\documentclass[11pt,a4paper,DIV=12,numbers=noenddot,sfdefaults=off,enabledeprecatedfontcommands]{scrartcl}
\pdfoutput=1

\usepackage[utf8]{inputenc}
\usepackage[T1]{fontenc}
\usepackage{lmodern}

\usepackage{graphicx}
\usepackage{amssymb}
\usepackage{amsmath}
\usepackage[dvipsnames]{xcolor}
\usepackage[normalem]{ulem}
\usepackage{slashed}

\usepackage{MnSymbol}

\usepackage{xspace}

\usepackage{booktabs}
\usepackage{makecell}

\usepackage{placeins}

\usepackage{pdflscape}

\usepackage{authblk}

\usepackage[format=plain,labelfont={bf},font={small}]{caption}

\usepackage[bookmarks=true,linktocpage,colorlinks=true,allbordercolors=white,allcolors=blue]{hyperref}

\usepackage[
  backend=bibtex,
  bibencoding=utf8,
  sorting=none,
  url=true,
  doi=false,
  style=phys,
  biblabel=brackets,
  subentry=false,
  eprint=true,
  maxbibnames=5
]{biblatex}
\numberwithin{equation}{section}

\newcommand{\email}[1]{\href{mailto:#1}{\nolinkurl{#1}}}

\newcommand{\s}{\newline \vspace*{-3.5mm}}
\newcommand{\beq}{\begin{eqnarray}}
\newcommand{\eeq}{\end{eqnarray}}
\newcommand{\CB}{{\mathrm{CB}}}
\newcommand{\CP}{{\mathrm{CP}}}
\newcommand{\BSMPT}{\texttt{BSMPT}}
\newcommand{\ScannerS}{\texttt{ScannerS}}

\newcommand{\sm}{{\ensuremath{\mathrm{SM}}}}

\title{A Comprehensive Analysis of the R2HDM Vacuum Evolution and the Induced GW and Collider Phenomenology}
\date{}

\titlehead{
\begin{flushright}
	KA-TP-20-2026 \\
    P3H-26-063
\end{flushright}
}

\author[1]{Lisa Biermann\thanks{\email{lisa.biermann@psi.ch}}}
\author[2]{Christoph Borschensky\thanks{\email{christoph.borschensky@kit.edu}}}
\author[2]{Rafael Boto\thanks{\email{rafael.boto@kit.edu}}}
\author[2]{Margarete Mühlleitner\thanks{\email{milada.muehlleitner@kit.edu}}}
\author[3,4]{Rui Santos\thanks{\email{rasantos@fc.ul.pt}}}
\author[3,4]{João Viana\thanks{\email{jfvvchico@hotmail.com}}}

\affil[1]{PSI Center for Neutron and Muon Sciences, 5232 Villigen PSI, Switzerland\vspace{2mm}}
\affil[2]{Institute for Theoretical Physics, Karlsruhe Institute of Technology,\authorcr\itshape Wolfgang-Gaede-Str.\ 1, 76131
Karlsruhe, Germany\vspace{2mm}}
\affil[3]{Departamento de Física, Faculdade de Ciências,\authorcr\itshape Universidade de Lisboa, 1749-016 Lisboa, Portugal\vspace{2mm}}
\affil[4]{Centro de Física Teórica e Computacional, Faculdade de Ciências,\authorcr\itshape Universidade de Lisboa, Campo Grande, Edifício C8 1749-016 Lisboa, Portugal\vspace{2mm}}

\begin{document}

\maketitle
\vspace*{-1cm}
\begin{abstract}
Extended Higgs sectors beyond the Standard Model (BSM) allow to dynamically generate the observed baryon asymmetry of the Universe through electroweak baryogenesis and thereby solve one of the most prominent open problems of the SM. The strong first-order phase transitions (PTs), required to preserve the generated asymmetry in the electroweak vacuum, source gravitational waves (GW) that can be tested at future experiments like \texttt{LISA}. Gravitational waves hence provide the exciting possibility to probe BSM physics through cosmological processes. In order to be able to eventually pin down the specific underlying physics, a good understanding of the evolution of the vacuum of the model under investigation is indispensable as well as of the uncertainties that are involved in the derivation of the GW spectrum. We use our code \texttt{BSMPTv3} that allows to reliably derive the finite temperature vacuum structure of extended Higgs sectors with multiple vacuum directions and calculates the GW spectrum of the found (multiple) strong first-order PTs, and we apply it to the real, i.e.~CP-conserving, 2-Higgs-Doublet Model. Taking into account all relevant theoretical and experimental constraints, we perform a thorough analysis of its vacuum evolution, the related collider and GW phenomenology and complement it by an uncertainty discussion.
\end{abstract}

\clearpage

\tableofcontents
\section{Introduction}
\label{s:introduction}
The Standard Model (SM) of particle physics has been a great success story. It was and is being tested with the highest accuracy and, with the discovery of the Higgs boson \cite{ATLAS:2012yve,CMS:2012qbp}, it has been structurally completed. There are, however, open questions which cannot be answered within the SM, among which the most prominent ones are those for the nature of Dark Matter and why there is more matter than antimatter in the universe \cite{WMAP:2012fli}. A dynamical mechanism for the generation of this baryon asymmetry is given by electroweak baryogenesis \cite{Kuzmin:1985mm,Cohen:1990it,Cohen:1993nk,Quiros:1994dr,Rubakov:1996vz,Funakubo:1996dw,Trodden:1998ym,Bernreuther:2002uj,Morrissey:2012db}, provided the three Sakharov conditions \cite{Sakharov:1967dj} are fulfilled, which are C and CP violation, the existence of baryon number violating processes, and departure from thermal equilibrium. The latter requires a strong first-order electroweak (EW) phase transition (PT). Although these ingredients are in principle present in the SM, its phenomenology is not compatible with the last condition, as for the measured value of the discovered Higgs boson mass of 125 GeV~\cite{ATLAS:2015yey} there is only a smooth cross-over \cite{Kajantie:1995kf,Kajantie:1996qd,Kajantie:1996mn,Karsch:1996yh,Gurtler:1997hr,Rummukainen:1998as,Csikor:1998eu,Aoki:1999fi,DOnofrio:2015gop}. We are therefore led to the investigation of extended Higgs sectors beyond-the-SM (BSM) in order to dynamically explain the observed matter-antimatter asymmetry. \s

Violent processes like strong first-order PTs can source a stochastic gravitational wave (GW) background. In 2016, the first direct observation of GWs was reported by the LIGO Collaboration \cite{LIGOScientific:2016aoc} and initiated a new era of multi-messenger astronomy. The increased sensitivity of future GW observatories thus provides us with the exciting possibility to probe BSM physics through the cosmological echo of a strong first-order PT. \s

At zero temperature, new physics extensions are constrained by numerous theoretical and experimental constraints, cf.~e.g.~\cite{Muhlleitner:2016mzt,Abouabid:2021yvw,Boto:2026bmz}. In particular, the discovered Higgs boson behaves very SM-like \cite{ATLAS:2022vkf,CMS:2022dwd}, and no direct sign of new physics has been established so far. This means that new physics is subtle and/or heavy. Its discovery will therefore be a challenge, and requires precise theoretical predictions as well as sophisticated experimental analyses. All available information from low-energy experiments, collider experiments, astrophysical observations, and cosmology should be taken into account, in order not to miss any sign of new physics. In case of a discovery, the extracted combined information will help to pin down the underlying parameters and corner new physics, leading to a better understanding of nature. \s

The challenge in inferring the underlying zero-temperature model from the measured GW spectra arises from the interplay of different areas of physics (field theory at zero and finite temperature, particle physics, non-equilibrium thermodynamics, perturbation theory, non-perturbative effects, \ldots) across a wide range of energy scales, as well as from conceptual issues such as the gauge dependence of the effective potential~\cite{Dolan:1973qd,Morrissey:2012db,Patel:2011th,Wainwright:2011qy,Garny:2012cg,Athron:2023xlk,Chiang:2017nmu}. Due to the complexity of the involved equations, solutions have to be derived numerically, and approximations are applied. For comprehensive reviews, cf.~e.g.~\cite{Athron:2023xlk,vandeVis:2025efm}. In order to  eventually be able to use the combined information from zero-temperature phenomenology and GW spectra from cosmological PTs, we have to have a very good understanding of which vacuum histories can be realised in specific BSM models, and uncertainty estimates for all the involved quantities along the way from the model to the GW spectrum have to be discussed and quantified. Only this allows us to derive from the GW signal meaningful predictions for zero-temperature phenomenology that can be tested at present and future colliders. \s

Numerous computational tools \cite{Masoumi:2017trx,Basler:2018cwe,Caprini:2019pxz,Basler:2020nrq,Basler:2024aaf,
Wainwright:2011kj,Camargo-Molina:2013qva,Camargo-Molina:2014pwa,Masoumi:2017trx,Hollik:2018wrr,Ferreira:2019iqb,Athron:2019nbd,Athron:2020sbe,Athron:2024xrh,Sato:2019wpo,Guada:2020xnz,Bardsley:2021lmq,Ekstedt:2022bff,Ekstedt:2023sqc,
Ekstedt:2024fyq,Ertas:2021xeh,Bringmann:2023iuz,Matuszak:2026xsz,Brdar:2025hxw,Costa:2025pew, BLOOP,PTPlot,PTtools,CosmoGW} have been developed that provide solutions for some or all of the steps along the road from the particle physics model to the GW spectra of strong first-order PTs, or to the baryon asymmetry~\cite{Barni:2025ifb,Muhlleitner:2026uqd}, which apply different approximations and algorithms. With \texttt{BSMPTv3} \cite{Basler:2024aaf}, we presented a tool that performs the whole chain from the particle physics model to the GW signal for arbitrary user-defined models with extended Higgs sectors. It is capable of treating multiple vacuum directions and multi-step PTs. It checks for discrete symmetries, flat directions, and electroweak symmetry restoration (EWSR). The nucleation, percolation, and completion temperatures are obtained from the solution of the bounce equation for the computation of the false vacuum decay probability. The GW spetrum is computed for bubble collisions and highly relativistic fluid shells, sound waves, and turbulence, and finally the signal-to-noise ratio (SNR) at \texttt{LISA} is provided. We use \texttt{BSMPTv3} together with our scanning tool \texttt{ScannerS} \cite{Coimbra:2013qq,Muhlleitner:2020wwk} to derive parameter samples of extended Higgs sector models that are compatible with all relevant theoretical and experimental constraints and that provide first-order PTs with detectable GW signals. \s

In this paper, we present a comprehensive and self-contained analysis of all possible phase histories in the Real 2-Higgs-Doublet Model (R2HDM). The R2HDM is a benchmark model that is extensively tested by experiment, and that provides the possibility of strong first-order PTs. It is a relatively simple extension of the SM, but at the same time complex enough to provide an interesting phenomenology that can be tested at colliders. For the found single and multi-step PT types we derive the collider phenomenology at zero temperature, the GW spectra, and the SNR for the \texttt{LISA} experiment after three years of data acquisition. We provide characteristic collider signatures related with the strong first-order PTs. To put our results in context, we discuss involved uncertainties and exemplarily show for some thermal parameters their impact on the GW spectrum. Our analysis goes beyond existing literature\footnote{For further work on the real 2HDM, cf.~e.g.~\cite{Fromme:2006cm,Dorsch:2014qja,Dorsch:2016nrg,Basler:2016obg,Dorsch:2017nza,Andersen:2017ika,Bernon:2017jgv,Kainulainen:2019kyp,Su:2020pjw,Aoki:2021oez,Goncalves:2021egx,Anisha:2022hgv,Biekotter:2022kgf,Aoki:2023lbz,Anisha:2023vvu,Biekotter:2023eil,Anisha:2025zbc,Aiko:2025tbk,Lee:2025hgb,Garbrecht:2025jgy,Bittar:2025lcr,Bhatnagar:2025jhh}.} by taking into account the most recent experimental constraints in our self-consistent tool chain and providing an overall picture of the possible PT histories of the R2HDM type 1, in particular also in the context of identifying differences and similarities between single- and multi-step PTs, and the relation between GW signals and collider phenomenology, critically addressing the existence of characteristic or, as they are sometimes called in the literature, ``smoking gun'' signatures. \s

The outline of the paper is as follows. In Sec.~\ref{s:models:r2hdm}, we introduce the R2HDM to set the notation. In Sec.~\ref{s:phasetransitions}, we briefly describe the main features of \texttt{BSMPTv3} relevant for this analysis. Section~\ref{s:numanalysis} summarises our scan ranges and applied constraints. In Secs.~\ref{s:PTsinR2HDM} and \ref{sec:pheno}, we present our numerical analysis. We describe the found PTs and investigate their phenomenology at colliders and GW experiments. We finish with the discussion of the uncertainties. Section~\ref{s:conclusions} summarises our results and contains our conclusions. In App.~\ref{app:band}, we describe the derivation of the \texttt{LISA} sensitivity band shown in our plots.
\section{The R2HDM}
\label{s:models:r2hdm}
In order to set the notation, we briefly introduce the CP-conserving 2HDM \cite{Lee:1973iz,Branco:2011iw}, called R2HDM in the following. The Higgs sector consists
of two $\mathrm{SU}(2)_L$ Higgs doublets $\Phi_1$ and $\Phi_2$, which transform as $\Phi_1\rightarrow \Phi_1,\ \Phi_2
\rightarrow -\Phi_2$ under a softly broken $\mathbb{Z}_2$ symmetry. The corresponding renormalisable tree-level potential is given by
\begin{align}
\begin{split}
V_{\text{tree}} &= m_{11}^2 \Phi_1^\dagger \Phi_1 + m_{22}^2
\Phi_2^\dagger \Phi_2 - \left[m_{12}^2 \Phi_1^\dagger \Phi_2 +
  \mathrm{h.c.} \right] + \frac{1}{2} \lambda_1 ( \Phi_1^\dagger
\Phi_1)^2 +\frac{1}{2} \lambda_2 (\Phi_2^\dagger \Phi_2)^2 \\
&\quad + \lambda_3 (\Phi_1^\dagger \Phi_1)(\Phi_2^\dagger\Phi_2) +
\lambda_4 (\Phi_1^\dagger \Phi_2)(\Phi_2^\dagger \Phi_1)
+ \left[ \frac{1}{2} \lambda_5 (\Phi_1^\dagger\Phi_2)^2  + \mathrm{h.c.} \right] \ ,
\end{split}\label{eq:model_1}
\end{align}
with the mass parameters $m_{11}^2$, $m_{22}^2$ and $m_{12}^2$ and the couplings
$\lambda_{1}\dots\lambda_{5}$ being real in the CP-conserving 2HDM. Applying $\mathrm{SU}(2)_L \times \mathrm{U}(1)_Y$ invariance, this potential allows for four vacuum expectation value (VEV) directions, given by the CP-even 
VEVs $\omega_{1,2}$ of the scalar components of the Higgs doublets,
the charge-breaking VEV $\omega_{\text{CB}}$, and the CP-breaking VEV
$\omega_{\text{CP}}$, so that the Higgs doublets can be parametrised in terms of the real fields $\rho_{i}$,
$\eta_i$, $\zeta_i$, and $\psi_i$ ($i=1,2$) as
\begin{align}\label{eq:2hdmdoublets}
    \Phi_1 = \frac{1}{\sqrt{2}}\begin{pmatrix}\rho_1+i\eta_1\\\zeta_1+\omega_1+i\psi_1\end{pmatrix}\,,\quad\Phi_2 = \frac{1}{\sqrt{2}}\begin{pmatrix}\rho_2+\omega_\text{CB}+i\eta_2\\\zeta_2+\omega_2+i\left(\psi_2+\omega_\text{CP}\right)\end{pmatrix}\,.
\end{align}
Compatibility with the zero-temperature Higgs phenomenology requires that
\begin{equation}
    \left.\{\omega_{\text{CB}},\,\omega_1,\,\omega_2,\,\omega_\text{CP}\}\right|_{T=0}
    = \{0,v_1,v_2,0 \}\,,
\end{equation}
with
\begin{equation}
    \left.\omega_{\text{EW}}\right|_{T=0} \equiv
    \left. \sqrt{\omega_1^2 + \omega_2^2 + \omega_{\text{CB}}^2 +
    \omega_{\text{CP}}^2}\right|_{T=0} =
    \sqrt{v_1^2 + v_2^2} \equiv v \approx 246 \mbox{ GeV} \;.
\end{equation}
The ratio of the zero-temperature CP-even VEVs is given by the mixing
angle $\beta$ as
\beq\label{eq:2hdmtanbeta}
\tan\beta = \frac{v_2}{v_1} \;.
\eeq
After electroweak symmetry breaking (EWSB), the Higgs spectrum consists of two scalar, $H_{1,2}$, and one
pseudoscalar, $A$, Higgs bosons, and a charged Higgs pair,
$H^\pm$. By convention, $H_1$ is taken as 
the lighter of the two CP-even Higgs bosons, i.e.~$m_{H_1} <
 m_{H_2}$. The mixing angles of the rotation matrix from the interaction to the mass basis are given by $\beta$ in the neutral CP-odd and charged Higgs sector, respectively, and in the neutral CP-even Higgs sector by the mixing angle $\alpha$.  They crucially determine the phenomenology of the Higgs sector. 
 \s 

The $\mathbb{Z}_2$ symmetry is extended to the Yukawa sector in order to avoid tree-level flavour-changing neutral currents, which then leads to four different types of the R2HDM. In this work, we concentrate on the R2HDM type 1, where the doublet $\Phi_2$
couples to all quarks and leptons.\footnote{In the 2HDM type 2, flavor constraints enforce an overall heavier Higgs spectrum \cite{Basler:2016obg,Anisha:2022hgv,Anisha:2025zbc}, which struggles to generate strong phase transitions.} In the following, we will thus not refer to any specific type anymore and implicitly consider always the 2HDM type 1. \s

The tree-level Higgs sector can be parametrised by eight input parameters which we choose as
\beq
v\,,\; \tan\beta\,, \; c_{H_{\text{non-SM}} VV} \,, \; m_{H_1} \,,\; m_{H_2} \,,\; m_{A} \,,\; m_{H^\pm} \,, \; m_{12}^2 \;.
\label{eq:2hdminput}
\eeq
For convenience, in the parameter scans performed for our numerical analysis, we replace the mixing angle $\alpha$ by the coupling modification factor $c_{H_{\text{non-SM}}VV}$ of the non-SM-like Higgs boson $H_{\text{non-SM}}$ coupling to the massive gauge bosons $(V \equiv Z,W^\pm)$ with respect to the corresponding Higgs coupling in the SM. Depending on the input parameter values, the non-SM-like Higgs boson can be the lighter or the heavier of the two CP-even Higgs bosons, i.e.~$H_1$ or $H_2$.

\section{Phase Transitions}
\label{s:phasetransitions}
In the numerical analysis of the R2HDM, we will use \texttt{BSMPTv3} for the derivation of the phase transitions, the thermal parameters, and the gravitational wave spectra. For self-consistency, we briefly summarise the most relevant features for our subsequent analysis and refer the reader for more details to \cite{Basler:2024aaf}. \s

In \texttt{BSMPT} \cite{Basler:2018cwe,Basler:2020nrq,Basler:2024aaf}, the effective potential is computed in four space-time dimensions. It is given by the sum of the tree-level potential $V_\text{tree}$, the one-loop corrected $\overline{\mathrm{MS}}$ renormalised Coleman-Weinberg potential $V_{\text{CW}}$ \cite{Coleman:1973jx}, the temperature-dependent UV-finite thermal contributions $V_\text{T}$ including the daisy-resummed thermal masses \cite{Carrington:1991hz}, and the finite counterterm potential $V_{\text{CT}}$, i.e.
\begin{eqnarray}
V (\omega, T) = V_{\text{tree}} (\omega) + V_{\text{CW}} (\omega) + V_{\text{T}} (\omega,T) + V_{\text{CT}} (\omega) \;,
\end{eqnarray}
where $\omega$ generically denotes the possible vacuum field directions. The renormalisation scale in the Coleman-Weinberg potential $V_{\text{CW}}$ is set to $\mu= v  = 246.22$~GeV. The finite counterterm potential  $V_{\text{CT}}$ is introduced to keep the VEV and the masses and mixing angles at zero temperature unchanged with respect to their tree-level values \cite{Basler:2016obg,Basler:2018cwe}. This on-shell (OS)-like renormalisation scheme allows theoretical and experimental constraints to be checked in parameter scans of the investigated models without the need for an iterative procedure. The daisy resummation can be chosen to be included in the effective potential either in the Arnold-Espinosa \cite{Arnold:1992rz} or the Parwani approach \cite{Parwani:1991gq}. In this work we use the Arnold-Espinosa scheme.\s

The code \texttt{BSMPTv3} performs the tracing of the minima of the effective potential as a function of the temperature. The code is able to trace several minimum directions and can treat multiple phase transitions. After determining the critical temperature $T_c$ of the found overlapping degenerate vacua, the transition rate is determined by solving the bounce equation. \texttt{BSMPTv3} also determines the nucleation, the percolation, and the completion temperature. It calculates the released latent heat $\alpha$ and the inverse duration of the PT, $\beta/H$, at the transition temperature, which is set by default equal to the percolation temperature $T_p$. The user can also choose, however, to use other temperatures, e.g.~the nucleation temperature $T_n$, as the transition temperature. Furthermore, all other relevant quantities are calculated for the derivation of the gravitational-wave spectrum. Note, however, that in \texttt{BSMPTv3} the wall velocity is taken as an input parameter given by the user. In the determination of the GW spectrum, the GWs originating from the collision of bubbles and highly relativistic fluid shells (``coll''), as well as from sound/shock waves (``sw'') and from magnetohydrodynamic turbulence (``turb'') are considered, and the spectra are implemented as given in \cite{Caprini:2024hue}. \s

We note finally that \texttt{BSMPT} is applicable to any extended Higgs sector. The user solely has to specify the coupling tensors, as explained in the manual \cite{Basler:2018cwe}. The code can be downloaded from the url: \url{https://github.com/BSMPT/BSMPT}.

\section{Constraints and Parameter Scan}
\label{s:numanalysis}
For the numerical analysis, we perform scans in the parameter space of the R2HDM and keep only those parameter sets that are compatible with the relevant theoretical and experimental constraints. The scans and checks are performed with the codes  \texttt{ScannerS}~\cite{Coimbra:2013qq,Muhlleitner:2020wwk} and \BSMPT{} version 3.3.1~\cite{Basler:2024aaf}. The program \ScannerS{} is used to generate parameter samples compatible with the most relevant experimental and theoretical constraints listed below. The generated points are subsequently fed into \texttt{BSMPT} to compute the phase transition histories as well as the gravitational-wave spectra. \s

In Table \ref{tab:r2hdmranges} we list the ranges of the R2HDM parameters over which we scan using the program \ScannerS{} for the two scans that we perform:  
We distinguish the cases \emph{normal hierarchy} (NH) where the SM-like Higgs, denoted by $H_{\text{SM}}$ from now on,  is the lighter CP-even Higgs state, $m_{H_1} = m_{H_\sm}$, and \emph{inverted hierarchy} (IH) where the SM-like Higgs is the heavier CP-even Higgs state, $m_{H_2} = m_{H_\sm}$. At non-zero temperature, we perform scans up to
\beq
T_{\text{high}} = 1 \; \mathrm{ TeV}\;.
\eeq
While there is no upper temperature limit, we impose an upper temperature cut in order to limit the amount of computation time. For the wall velocity $v_w$ we have chosen the default value in \BSMPT{},
\beq
v_w = 0.95 \;.
\eeq
The theoretical constraints tested by \ScannerS{} are the requirement that the R2HDM potential is bounded from below,
that perturbative unitarity holds, and that the electroweak vacuum is
the global minimum \cite{Branco:2011iw}. For the latter we use the
discriminant from \cite{Barroso:2013awa}. With \BSMPT{} we verify that the calculated next-to-leading order (NLO) global minimum is equal to the EW vacuum with $v \approx 246$~GeV at zero temperature. 
We furthermore check for EW-symmetry restoration, i.e.\ that in the high-temperature limit, the effective potential has its global minimum at the origin and is bounded from below. We discard all points that do not fulfill this criterion (unless specified otherwise). Additionally, we only keep points that lead to a calculable gravitational-wave signal, i.e.\ where the \texttt{status\_gw\_i} flags in the \BSMPT{\texttt{v3}} output for the corresponding transition denoted by the index \texttt{i} return the value \texttt{success}. \s

On the experimental side, we impose compatibility with the
electroweak precision data by demanding the computed values $S$,
$T$, and $U$ to be within $2\sigma$ of the SM fit \cite{Baak:2014ora}, where we take into account the full
correlation between the three parameters. We demand that one of the Higgs bosons has a mass of \cite{ATLAS:2015yey}
\beq
m_{H_\sm} = 125.09 \, \mbox{GeV} \,,
\eeq
and behaves SM-like. Its compatibility with the LHC Higgs data and the constraints on the non-SM Higgses from the experimental Higgs search limits are tested through the \ScannerS{} link to {\tt HiggsTools} \cite{Bechtle:2008jh,Bechtle:2013xfa,Bahl:2022igd,Bahl:2026yal}.
Consistency with flavour constraints is ensured by considering the global fit
of Ref.~\cite{Haller:2018nnx} in the
$m_{H^{\pm}}-\tan\beta$ plane that takes into consideration the exclusion regions based on neutral meson oscillations and $B$ meson decays~\cite{Haber:1999zh, Deschamps:2009rh,Mahmoudi:2009zx,Hermann:2012fc,Misiak:2015xwa, Misiak:2017bgg,Misiak:2020vlo,HFLAV:2016hnz,CMS:2014xfa,LHCb:2017rmj}.
\s

\renewcommand{\arraystretch}{1.4}
\begin{table}[tp]
\centering
\begin{tabular}{@{\hspace{1.0mm}}c|c@{\hspace{3.0mm}}c@{\hspace{3.0mm}}c@{\hspace{3.0mm}}c@{\hspace{3.0mm}}c@{\hspace{3.0mm}}c@{\hspace{3.0mm}}c@{\hspace{1.0mm}}}
\toprule
& $m_{H_1}$ [GeV] & $m_{H_2}$ [GeV] & $m_A$ [GeV] & $m_{H^\pm}$ [GeV] & $\tan\beta$ &
$c_{H_\text{non-SM} VV}$ & $m_{12}^2$ [GeV$^2$] \\ \midrule
\textbf{NH} & $125.09$ & $[130, 1500]$ & $[30, 1500]$ & $[30, 1500]$ & $[0.8, 250]$ & $[-0.5, 0.5]$ & $[10^{-3}, 10^6]$ \\ 
\textbf{IH} & $[30, 120]$ & $125.09$ & $[30, 1500]$ & $[30, 1500]$ & $[0.8, 120]$ & $[-0.5, 0.5]$ &
 $[10^{-3}, 10^6]$ \\ \bottomrule
\end{tabular}
\caption{Scan ranges of the R2HDM input parameters, cf.~Eq.~\eqref{eq:2hdminput}, where NH (IH) refers to the set-up where the lighter (heavier) of the two CP-even neutral Higgs bosons is the SM-like Higgs $H_\sm$, \emph{i.e.}~$H_1 (H_2) \equiv H_\sm$.}
\label{tab:r2hdmranges}
\end{table}
\renewcommand{\arraystretch}{1}

Our scan strategy is as follows. We start by performing an extensive random scan within the given parameter ranges of Table~\ref{tab:r2hdmranges} to obtain an initial selection of the different types of phase transition histories favouring different regions of the parameter space. The scan sample is subsequently improved by a machine-learning algorithm based on the Covariant Matrix Adaptation Evolutionary Strategy (CMA-ES) \cite{10.1162/106365601750190398,hansen2023cmaevolutionstrategytutorial} (for more details on the usage of this algorithm and its application to parameter scans in the 2HDM and 3HDM, cf.\ \cite{deSouza:2022uhk,Romao:2024gjx,deSouza:2025bpl,Boto:2025ovp,Boto:2026gzj,deSouza:2026sta}) which allows us to explore parameter regions with a low density of points and study outliers with specific phase histories. Finally, we perform multiple consecutive targeted Gaussian scans, i.e.~random scans with a Gaussian distribution centered around points of interest, to optimise our sample with respect to single- and multi-step transition histories as well as their transition strength and sizeable predictions for a gravitational-wave signal. In total, we generated about 1.6 million valid parameter configurations for each mass hierarchy which fulfill all theoretical and experimental constraints as checked by \ScannerS{}, and which we subsequently ran with the \BSMPT{\texttt{v3}} executable \texttt{CalcGW} followed by the analysis of its output.

\section{Phase Transitions in the R2HDM}
\label{s:PTsinR2HDM}
\subsection{Phase Transition Histories}
We use our obtained set of valid R2HDM configurations to first study what kind of phase transition histories can be realised in the R2HDM. In the following, we will denote first-order phase transitions as follows,
\begin{equation}
    (X \to Y_1 \to \ldots \to Y_n)_n\,,
\end{equation}
where the phases $X$, $Y_1$, \ldots, $Y_n$ of the $n$-step PT are labelled by\footnote{Since the numerical output of \BSMPT{} generally does not return exact zeros, we consider a VEV to be approximately zero if for its numerical value we have $|\omega_i| < 1$ GeV.}:
\begin{itemize}
    \item S: symmetric phase, $\omega_1 = \omega_2 = \omega_\CB = \omega_\CP = 0$,
    \item N: neutral electroweak (EW)-broken phase, $\omega_\CB = \omega_\CP = 0$ and either $\omega_1 \ne 0$ or $\omega_2 \ne 0$ or both $\omega_1, \omega_2 \ne 0$,
    \item EW: neutral EW-broken phase that ends at $T = 0$ in the physical EW vacuum with $v = 246$ GeV,
    \item CB: charge-breaking phase, $\omega_\CB \ne 0$,
    \item CP: CP-breaking phase, $\omega_\CP \ne 0$.
\end{itemize}
Here, $X$ denotes the phase in which the universe was at our highest scanning temperature $T_{\text{high}} = 1$~TeV. It thus potentially does not coincide with the global minimum for $T \to \infty$, and it is possible that for $T > T_{\text{high}}$, the universe underwent further PTs. However, as mentioned previously, we check explicitly with \BSMPT{} for symmetry restoration in the high-temperature limit. \s 

Furthermore, as we are mainly interested in strong first-order PTs, we classify the \emph{strength} of a phase transition by the quantity
\begin{equation}\label{eq:xipdefinition}
	\xi_* \equiv \frac{\sqrt{\displaystyle\sum_{i=1,2,\CB,\CP} \left[\omega^{\text{true}}_i(T_*) - \omega^{\text{false}}_i(T_*)\right]^2}}{T_*}\,.
\end{equation}
The labels ``true'' and ``false'' refer to the VEVs in the true and false minimum of the transition,
and $T_*$ denotes the transition temperature which can be set by the user. The default value set in \texttt{BSMPT} is the percolation temperature $T_p$. Alternatively, also the nucleation temperature $T_n$ can be chosen. We briefly remind the reader of the meaning of the various temperatures calculated by \texttt{BSMPTv3}. At the critical temperature, the false and the true minimum are degenerate. The nucleation temperature is the temperature at which the tunneling decay rate from the false to the true vacuum per Hubble volume matches the Hubble rate. If no nucleation temperature is found, the tunneling does not take place. At the percolation temperature, the probability of finding a point still in the false vacuum is 71\%. At this temperature,  there is a large connected structure of the true vacuum that spans the whole universe, and that is stable and cannot collapse back to the false vacuum. At the completion temperature, the probability of finding a point in the false vacuum is 1\%\footnote{We note that also the false vacuum fraction at the percolation and completion temperature are optional user inputs, cf.~\cite{Basler:2024aaf}.}. 
The sum inside the square root extends over all doublet VEVs, as they all play a role for the EW PT. Based on the conventional sphaleron suppression criterion $\xi_c \gtrsim 1$ (see e.g.\ \cite{Patel:2011th,Morrissey:2012db}), we will use in the following
\begin{equation}
	\xi_p > 1
\end{equation}
as the requirement for a strong first-order phase transition. In the following, when talking about PTs, we will usually refer to first-order PTs, unless specified otherwise. \s

A comment is at order here. The effective potential as it is used here is a gauge-dependent quantity, and so is the value $\xi_\star$ derived from it. For discussions on the gauge dependence of
$v_c/T_c$, see e.g.~\cite{Dolan:1973qd,Morrissey:2012db,Patel:2011th,Wainwright:2011qy,Garny:2012cg,Chiang:2017nmu,Athron:2023xlk}, and for recent developments, cf.~e.g.~\cite{Niemi:2020hto,Croon:2020cgk,Hirvonen:2021zej,Lofgren:2021ogg,Schicho:2022wty,Athron:2022jyi,Qin:2024dfp,Balui:2025kat,Balui:2025yvd,Balui:2026ghs}. As there has been no solution proposed yet for the 2HDM, we take the values of
$\xi_p > 1$ as rather indicative for a strong first order PT
\cite{Quiros:1994dr,Moore:1998swa}, and also check whether the released latent heat $\alpha$ supports the statement.

\subsection{Phase Transitions in the R2HDM}
From the analysis of our obtained parameter samples compatible with all relevant theoretical and experimental constraints, we find that in the R2HDM one-step, two-step, three-step, and even four-step PTs are possible, as well as exotic PTs involving intermediate CP- and CB-violating phases. Our findings are:
\begin{itemize}
\item[$\diamond$]
Starting from a symmetric vacuum at $T_{\text{high}}= 1$~TeV, we find one- and two-step phase transitions for both mass hierarchies, that end at $T=0$ in the physical EW vacuum $v=246$~GeV realised with both doublet VEVs being non-zero, i.e.~$\omega_1 (T=0)\equiv v_1 \ne 0$ and $\omega_2 (T=0) \equiv v_2 \ne 0$. All two-step PTs proceed via a neutral EW phase (N) with either both $\omega_1$ and $\omega_2$ non-zero or only one of them. We introduce the notation
\beq
(\text{S}\to \text{EW})_{1} \quad \text{and} \quad (\text{S}\to \text{N} \to\text{EW})_{2}
\eeq
for this kind of one-, respectively, two-step PTs. 
We also find (one- or two-step) phase histories that do not transition to the physical vacuum at $T=0$. They are \emph{trapped} in a wrong vacuum, which can be neutral or charge-breaking, as we found. Numerous among all of these scenarios do not restore EW symmetry at very high-energies. We reject both the non-symmetry-restoration (NSR) vacua and the trapped vacua as non-physical. 

\item[$\diamond$] We also find three-step phase transitions for the normal mass hierarchy, starting in a symmetric phase and transitioning through multiple neutral broken phases into the physical EW phase at $T = 0$,
\begin{equation}
    (\text{S} \to \text{N} \to \text{N} \to \text{EW})_3\,,
\end{equation}
with $\xi_p > 1$ for at least one of the transitions (which, in our case, is always the last one).  Furthermore, additional points with a three-step PT starting in a broken phase, therefore (possibly) exhibiting EW NSR, and then ending via an intermediate symmetry restoring phase in the physical EW vacuum also occurred in both mass hierarchies. The latter points will not be considered further here, as we restrict ourselves on symmetry-restoring scenarios. For the inverted hierarchy, we find three points with a three-step PT\footnote{Note that Ref.~\cite{Lee:2025hgb} found more points with a three-step PT, starting from a symmetric vacuum, and including a strong first-order PT for the inverted mass hierarchy. We explicitly checked these points and found that they were excluded by our applied constraints.}. We also find two points with a four-step PT for the normal hierarchy. Both the samples of three-step points in the inverted and four-step points in the normal hierarchy contain each only a single point for which one of the transitions has a $\xi_p$ slightly above 1. Due to the extremely small sample size, we therefore do not discuss these two points further.

\item[$\diamond$] Despite the absence of explicit CP-violating sources in the scalar potential at $T = 0$, the R2HDM also features phase histories with intermediate CP-violating vacua. Such a CP phase would not be subject to constraints from electric dipole measurements as they apply at zero temperature, where the CP-breaking VEV has transitioned back to zero again. For this spontaneous CP violation to arise, however, non-zero neutral VEVs are required, $\omega_{1/2} \ne 0$, such that at the one-loop level diagrams involving trilinear Higgs couplings can generate CP-violating effects. We thus find that the generation of the CP-breaking VEV actually does not occur directly via a first-order PT from the symmetric phase, but via a second-order PT, S$\stackrel{\text{1st}}{\longrightarrow}$(N$\stackrel{\text{2nd}}{\longrightarrow}$CP)$\stackrel{\text{1st}}{\longrightarrow}$EW, from a neutral broken phase with $\omega_{1/2} \ne 0$. The found points will therefore not be relevant for electroweak baryogenesis, as this would additionally require a first-order PT from the symmetric to the CP-breaking phase with non-zero $\omega_{1/2}$.

\item[$\diamond$] Also phase histories with intermediate charge-breaking (CB) vacua are possible. They have been found and analysed previously in \cite{Aoki:2023lbz}. Applying the latest experimental constraints, they are still allowed. All these scenarios, however, inevitably imply EW NSR at high temperature, as we discussed in detail in \cite{Aoki:2023lbz}. The phase histories involving CB vacua that we find predominantly occur via a two- or three-step phase transition involving an intermediate symmetry-restoring phase,
\begin{equation}
    (\text{CB} \to \text{S} \to \text{EW})_2 \quad \text{and} \quad (\text{CB} \to \text{N} \to \text{S} \to \text{EW})_3\,.
\end{equation}

\end{itemize}

\subsection{Properties of the Electroweak Phase Transitions of the R2HDM}
We now proceed to discuss the EW PTs in more detail. These are the 
first-order phase transitions (i.e.~$\xi_p > 1$) from the symmetric into an EW minimum. 

\subsubsection{Normal Hierarchy}
\begin{figure}[tp]
	\centering
    \includegraphics[width=\textwidth]{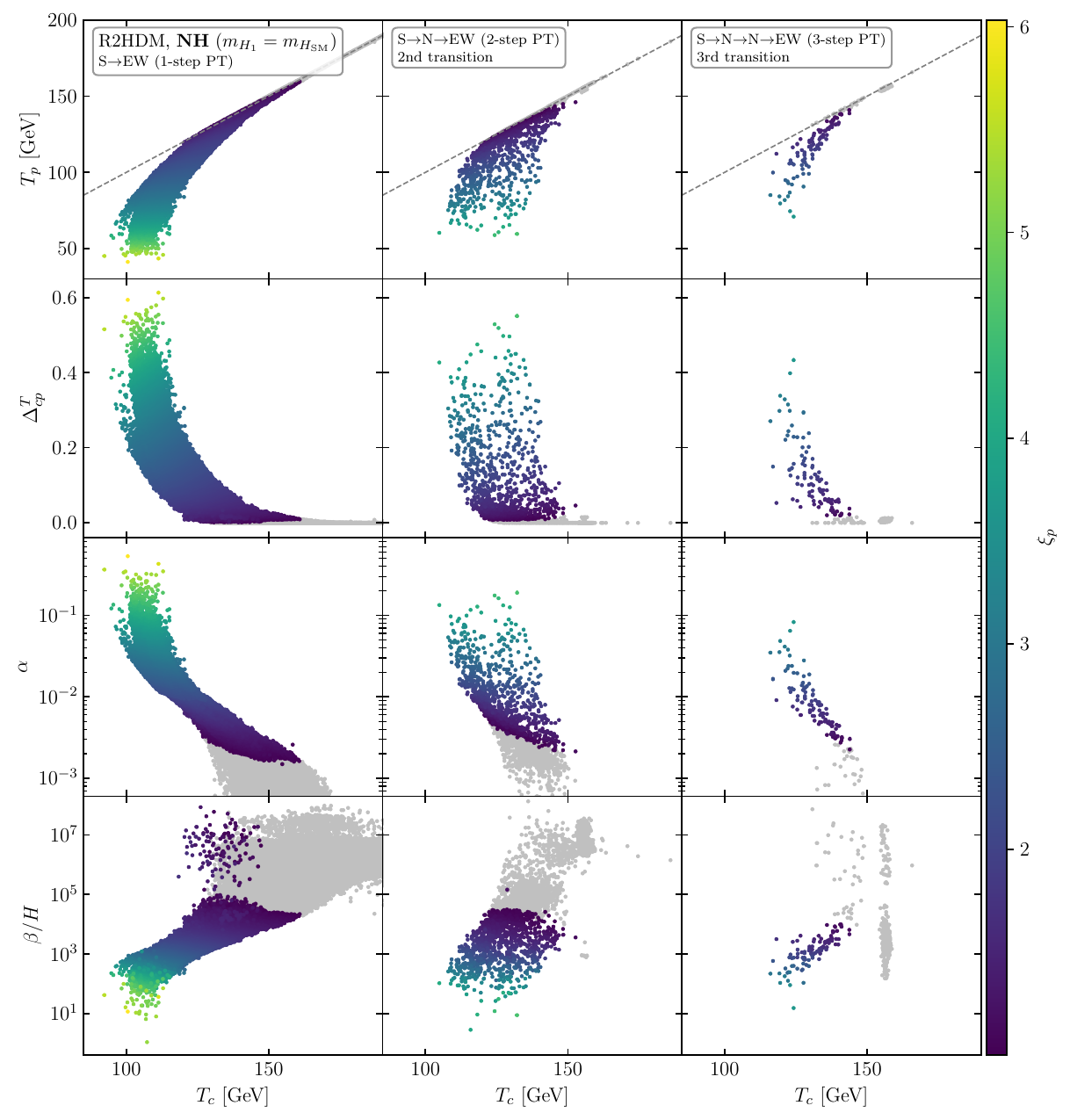}
	\caption{R2HDM normal hierarchy: Characteristic thermal quantities as a function of the critical temperature $T_c$. First row: percolation temperature $T_p$. Second row: relative difference between the critical and the percolation temperature,  $\Delta^T_{cp}$. Third row: released latent heat $\alpha$. Fourth row: inverse duration of the PT, $\beta/H$. The coloured (grey) points fulfill $\xi_p \ge 1$ ($\xi_p < 1$), with the colour bar denoting the size of $\xi_p$. The three columns correspond to one-step (left), two-step (middle), and three-step (right) PTs. For the multi-step PTs, only the strongest transition is shown (always corresponding to the last one from N$\to$EW).}
\label{fig:TpvsTcNormal}
\end{figure}

For the \textit{normal hierarchy}, where $m_{H_1} = m_{H_\sm}$, the ranges of the transition temperature $T_\star$, which we choose here and in the following, unless stated otherwise, as the percolation temperature $T_\star = T_p$, are found to be 
\begin{equation}
    \begin{split}
        (\text{S}\to \text{EW})_{1}^{\xi_p > 1}: \quad &T_* = [40, 160] \text{ GeV} \,,\\
        (\text{S} \to \text{N} \to \text{EW})_{2}^{\xi_p > 1}: \quad &T_*^{\text{1st}} = [165, 183] \text{ GeV}\,,~~ T_*^{\text{2nd}} = [58, 146] \text{ GeV} \,,\\
        (\text{S} \to \text{N} \to \text{N} \to \text{EW})_{3}^{\xi_p > 1}: \quad &T_*^{\text{3rd}} = [70, 141] \text{ GeV} \,.
    \end{split}
\end{equation}
For the multi-step phase transitions, we distinguish between the first, second, or third transition with the corresponding superscripts `1st', `2nd', and `3rd' (with the transitions being ordered in time, \emph{i.e.}\ the first transition happening earlier than the second transition, and so on). For the three-step PTs, values of $\xi_p > 1$ were found only for the last transition from N$\to$EW. The reason are the relatively small barrier heights separating the two minima in the first and the second transition, respectively. 
We note that for the one-step transitions, the critical temperature can be below $T_c = 100$~GeV, cf.~Fig.~\ref{fig:TpvsTcNormal} so that the high-temperature approximation applied in the daisy resummation may be questionable and the results have to be taken with caution for these scenarios. \s

In Fig.~\ref{fig:TpvsTcNormal} we plot characteristic thermal quantities as a function of the critical temperature for the three step-types of PTs found for the R2HDM, i.e.~the one- (left column), two- (middle column) and three-step PTs (right column). For the two- and three-step PTs we only plot the step with the strongest PTs which are the second and third step for the two- and three-step PT, respectively. In the first row, we plot the percolation temperature. In the second row, we show the relative difference between the critical temperature and the percolation temperature calculated as
\begin{equation}
    \Delta^T_{cp} \equiv \frac{T_c - T_p}{T_c}\,.
\end{equation}
It can be regarded as the duration of the phase transition with larger differences corresponding to a longer duration of the universe in the false vacuum. The third row displays the released latent heat $\alpha$ which is a measure for the strength of the PTs. The fourth and last row depicts $\beta/H$, where $H$ is the Hubble rate and 
\begin{equation}
    \beta = HT\frac{d}{dT}\left(\frac{S_3(T)}{T}\right),
\end{equation}
all evaluated at the transition temperature, and with $S_3(T)$ the three-dimensional Euclidean action for the O$(3)$-symmetric bounce solution, as appearing in the thermal tunneling rate (see also e.g.\ \cite{Ellis:2018mja}).
It quantifies the inverse duration of the PT. The colour code depicts the value of $\xi_p$ for values larger than 1, i.e.~strong first-order PTs. \s

For $\Delta^T_{cp}$, we find for the points with a strong first-order PT,
\begin{equation}
    \begin{split}
        (\text{S}\to \text{EW})_{1}^{\xi_p > 1}: \quad &\Delta^T_{cp} < 62\% \,,\\
        (\text{S} \to \text{N} \to \text{EW})_{2}^{\xi_p > 1}: \quad &\Delta^{T,\text{1st}}_{cp} < 28\%\,,~~ \Delta^{T,\text{2nd}}_{cp} < 55\% \,,\\
        (\text{S} \to \text{N} \to \text{N} \to \text{EW})_{3}^{\xi_p > 1}: \quad &\Delta^{T,\text{3rd}}_{cp} < 44\% \,.
    \end{split}
\end{equation}%
The differences in the critical temperature and in the percolation temperature can hence be quite substantial, as can also be inferred from the first two rows of Fig.~\ref{fig:TpvsTcNormal}.  
\s

Scenarios with smaller $T_c$ feature lower transition temperatures $T_p$ with a larger released latent heat $\alpha$ and hence stronger phase transitions with larger values of $\xi_p$. The values of $\alpha$ and $\xi_p$ for the single-, two-, and three-step PTs are found to be
\begin{equation}
    \begin{split}
        (\text{S}\to \text{EW})_{1}^{\xi_p > 1}: \quad &\alpha < 0.52 ~~(\xi_p < 6.1) \,,\\
        (\text{S} \to \text{N} \to \text{EW})_{2}^{\xi_p > 1}: \quad &\alpha^{\text{1st}} < 0.00057 ~~(\xi_p^{\text{1st}} < 1.1) \,,~~ \alpha^{\text{2nd}} < 0.19 ~~(\xi_p^{\text{2nd}} < 4.6) \,,\\
        (\text{S} \to \text{N} \to \text{N} \to \text{EW})_{3}^{\xi_p > 1}: \quad &\alpha^{\text{3rd}} < 0.081 ~~(\xi_p^{\text{3rd}} < 3.7) \,,
    \end{split}
    \label{ew:nhstrengths}
\end{equation}%
which is also shown in the third row of Fig.~\ref{fig:TpvsTcNormal}. We nicely see the correlation with larger values of $\Delta_{cp}^T$ (second row), implying larger barriers and hence stronger PTs,  reflected in larger values of $\alpha$ and $\xi_p$. \s

For the 
inverse durations of the phase transition, $\beta/H$, we find \begin{equation}
    \begin{split}
        (\text{S}\to \text{EW})_{1}^{\xi_p > 1}: \quad &\beta/H = [1.05, 8.42 \times 10^7] \,,\\
        (\text{S} \to \text{N} \to \text{EW})_{2}^{\xi_p > 1}: \quad &(\beta/H)^{\text{1st}} = [296, 1.37 \times 10^3] \,, \\ & (\beta/H)^{\text{2nd}} = [2.84, 1.41 \times 10^5] \,,\\
        (\text{S} \to \text{N} \to \text{N} \to \text{EW})_{3}^{\xi_p > 1}: \quad &(\beta/H)^{\text{3rd}} = [14.8, 9.58 \times 10^3] \,.
    \end{split}
\end{equation}%
The fourth row of Fig.~\ref{fig:TpvsTcNormal} confirms that stronger PTs (cf.~third row and colour code) go along with long durations (cf.~also second row). 

\begin{figure}[t!]
	\centering
\includegraphics[width=\textwidth]{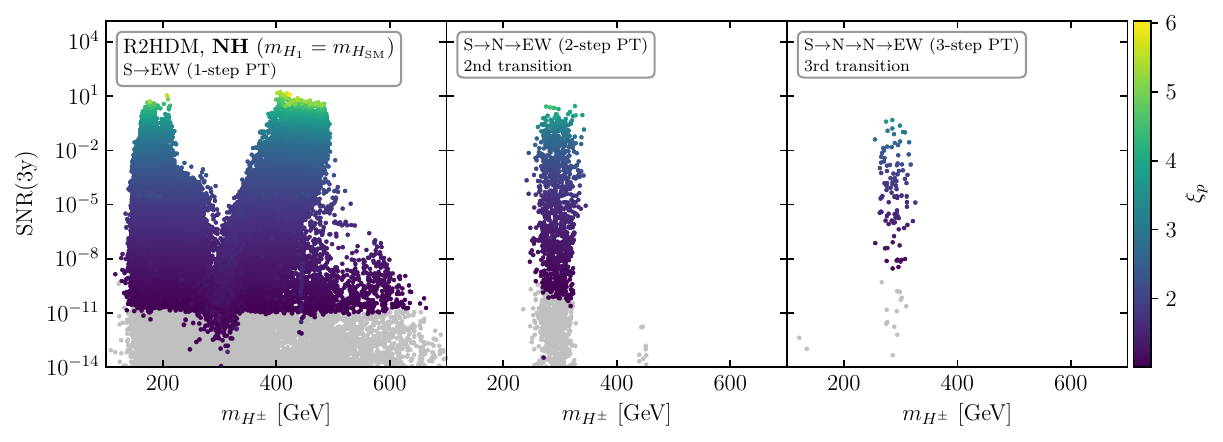}
	\caption{R2HDM normal hierarchy: Signal-to-noise ratio (SNR) for the \texttt{LISA} experiment after three years of data taking versus the charged Higgs mass. The columns as well as the colouring of the points are the same as in Fig.~\ref{fig:TpvsTcNormal}.}
\label{fig:SNRvsMHpmNormal}
\end{figure}
%
Next, we investigate the GW spectrum related to these strong first-order PTs. We used \texttt{BSMPTv3} to generate the GW spectra and the signal-to-noise ratio (SNR) at \texttt{LISA} assuming 3 years of data collection. The SNR is plotted in Fig.~\ref{fig:SNRvsMHpmNormal} as a function of the charged Higgs mass for the three step-types of PTs, where for the two- and three-step PTs we only show the transitions with the strongest PTs. The GW spectrum is dominantly sourced by sound waves with only a very small contribution to the SNR, typically less than 1\%, originating from magnetohydrodynamic turbulence. 
We are not yet in the regime where the bubble collisions play a role, as they become dominant only for {strongly supercooled scenarios with $\alpha \gg 1$. From Fig.~\ref{fig:TpvsTcNormal} (third row), it is clear that we do not reach such values in our scan. 
Overall, the signals are rather weak, despite $\xi_p$ values well above 1, and the SNR values barely reach 10. This is to be expected, as the found values of the released latent heat are rather low. Interestingly, we find that one-step PTs from our scan sample generate the largest signals with SNR values of up to around 17, while the two-step (three-step) PTs are weaker and reach SNR values of only 2.8 (0.44). From the distribution of points with respect to the charged Higgs mass, we see that, for one-step PTs, there are two regions with strong transitions centered around $m_{H^\pm} \approx 190$~GeV and $m_{H^\pm} \approx 420$~GeV. These regions  correspond to the two cases where the charged Higgs mass and either the heavy CP-even or the CP-odd Higgs masses are degenerate, $m_{H^\pm} \approx m_{H_2}$ or $m_{H^\pm} \approx m_A$, as will be discussed later. Furthermore, the region in between with $m_{H^\pm} \approx 300$~GeV appears to be filled with the points found with two- and three-step PTs, such that for almost any charged Higgs mass in the range of roughly $100~\text{GeV} < m_{H^\pm} < 500~\text{GeV}$, we have one- or multi-step PTs with strong transitions potentially leading to an observable GW signal with the \texttt{LISA} experiment. 

\subsubsection{Inverted Hierarchy}
For the \emph{inverted hierarchy}, we find both the one- and the two-step phase transition scenarios with a strong first-order PT. The corresponding ranges of the transition temperature are as follows:
\begin{equation}
    \begin{split}
        (\text{S}\to \text{EW})_{1}^{\xi_p > 1}: \quad &T_* = [50, 182] \text{ GeV} \,,\\
        (\text{S} \to \text{N} \to \text{EW})_{2}^{\xi_p > 1}: \quad &T_*^{\text{1st}} = [124, 247] \text{ GeV}\,,~~ T_*^{\text{2nd}} = [22, 145] \text{ GeV} \,.
    \end{split}
\end{equation}%
As in the case of the normal hierarchy, the superscripts `1st' and `2nd' refer to the sequence of the transitions. Here we now find for the two-step PTs points with a critical temperature below $T_c = 100$~GeV. \s

The transition temperature, which we take to be the percolation temperature, as well as further characteristic quantities are plotted in Fig.~\ref{fig:TpvsTcInverted} as a function of $T_c$. The left column shows the one-step PT and the right one the stronger transition of the two-step PT, which is the second one. The relative difference between the critical and transition temperature, shown in the second row, is found to be
\begin{equation}
    \begin{split}
        (\text{S}\to \text{EW})_{1}^{\xi_p > 1}: \quad &\Delta^T_{cp} < 55\% \,,\\
        (\text{S} \to \text{N} \to \text{EW})_{2}^{\xi_p > 1}: \quad &\Delta^{T,\text{1st}}_{cp} < 48\%\,,~~ \Delta^{T,\text{2nd}}_{cp} < 73\% \,.
    \end{split}
\end{equation}%
Overall, in the inverted mass hierarchy scenario, for the two-step PTs the trend is towards lower transition temperatures,  cf.~also~Fig.~\ref{fig:TpvsTcInverted} (upper row), whereas in the one-step PT they are higher compared to the normal hierarchy. Accordingly, we find for the released latent heat and the $\xi_p$ values smaller values for the one-step PTs, but larger values for the two-step PTs, namely 
\begin{equation}\label{eq:alphaPTInv}
\begin{split}
        (\text{S}\to \text{EW})_{1}^{\xi_p > 1}: \quad &\alpha < 0.21 ~~(\xi_p < 4.9) \,,\\
        (\text{S} \to \text{N} \to \text{EW})_{2}^{\xi_p > 1}: \quad &\alpha^{\text{1st}} < 0.0069 ~~(\xi_p^{\text{1st}} < 2.2) \,,~~ \alpha^{\text{2nd}} < 2.5 ~~(\xi_p^{\text{2nd}} < 11) \,,
\end{split}
\end{equation}%
as can also be inferred from Fig.~\ref{fig:TpvsTcInverted} (third row). \s

\begin{figure}[tp]
	\centering
    \includegraphics[width=.8\textwidth]{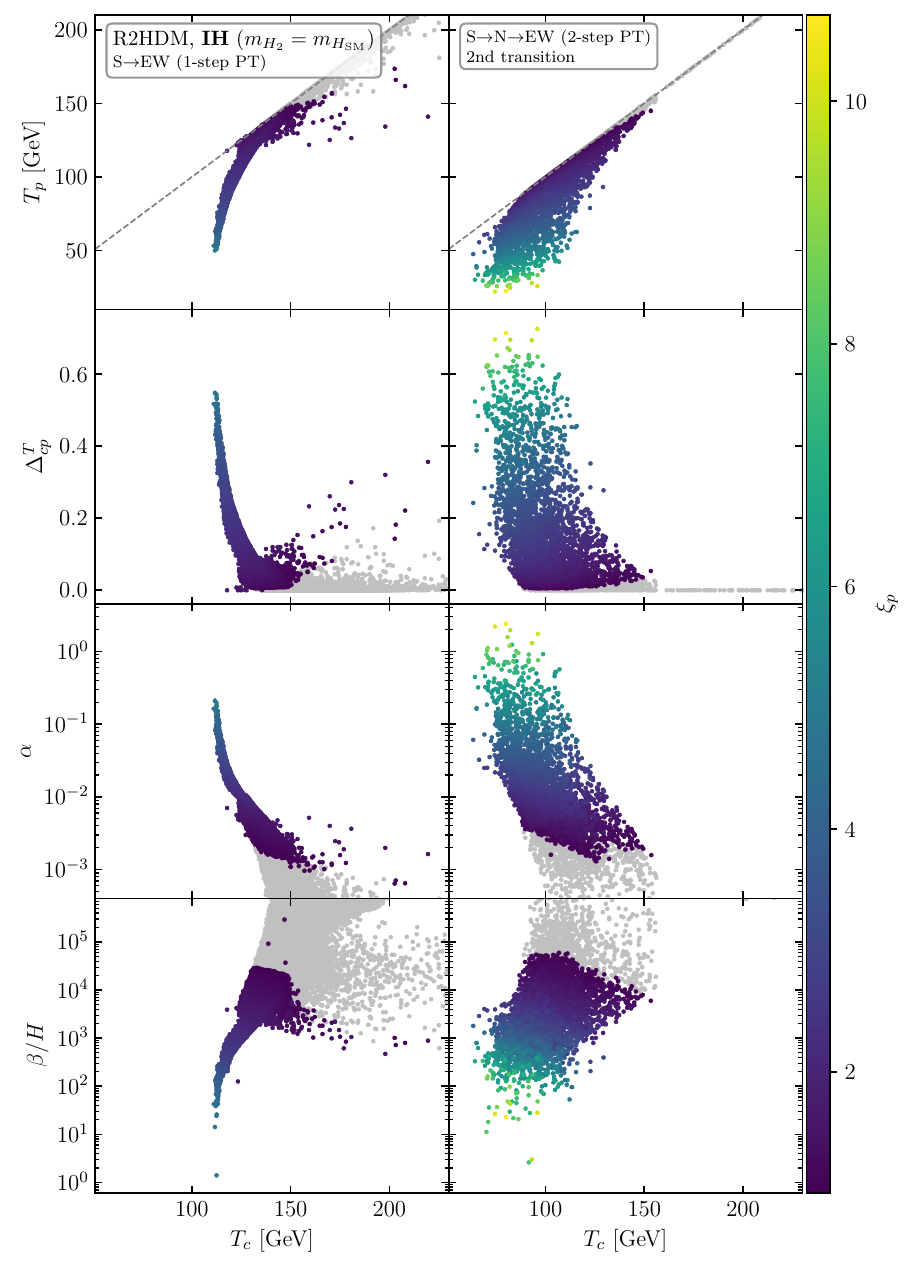}
	\caption{R2HDM inverted hierarchy: Characteristic thermal quantities as a function of the critical temperature $T_c$. First row: percolation temperature $T_p$. Second row: relative difference between the critical and the percolation temperature,  $\Delta^T_{cp}$. Third row: released latent heat $\alpha$. Fourth row: inverse duration of the PT, $\beta/H$. The coloured (grey) points fulfill $\xi_p \ge 1$ ($\xi_p < 1$), with the colour bar denoting the size of $\xi_p$. The two columns correspond to one-step (left) and two-step (right) PTs. For the two-step PTs, only the strongest transition is shown, corresponding to the second one from N$\to$EW.}
\label{fig:TpvsTcInverted}
\end{figure}%
The inverse durations of the phase transition $\beta/H$, shown in Fig.~\ref{fig:TpvsTcInverted} (fourth row), are in accordance with these findings and amount to 
\begin{equation}
    \begin{split}
        (\text{S}\to \text{EW})_{1}^{\xi_p > 1}: \quad &\beta/H = [1.40, 2.94 \times 10^5] \,,\\
        (\text{S} \to \text{N} \to \text{EW})_{2}^{\xi_p > 1}: \quad &(\beta/H)^{\text{1st}} = [46.9, 2.20 \times 10^7] \,, \\
        & (\beta/H)^{\text{2nd}} = [2.61, 5.97 \times 10^4] \,.
    \end{split}
\end{equation}%
The values of $\alpha$ found for the inverted hierarchy in the two-step phase transitions remain below 1 in the first transition, see Eq.~\eqref{eq:alphaPTInv}, and reach values above 1 in the second transition, however, for $T_c$ below 100~GeV only. Strong phase transitions correlate with relatively small $T_c$ compared to relatively large thermal VEVs at the critical temperature, $\omega_c = \omega(T_c)$. At the same time for large barriers, i.e.\ large $\omega_c$ between the false and the true vacuum, the universe remains long enough in the false vacuum so that the barrier has been damped sufficiently at lower temperature that the false vacuum decay can take place, which explains the larger $\Delta^T_{cp}$ values for these scenarios. \s

\begin{figure}[tp]
	\centering
    \includegraphics[width=.8\textwidth]{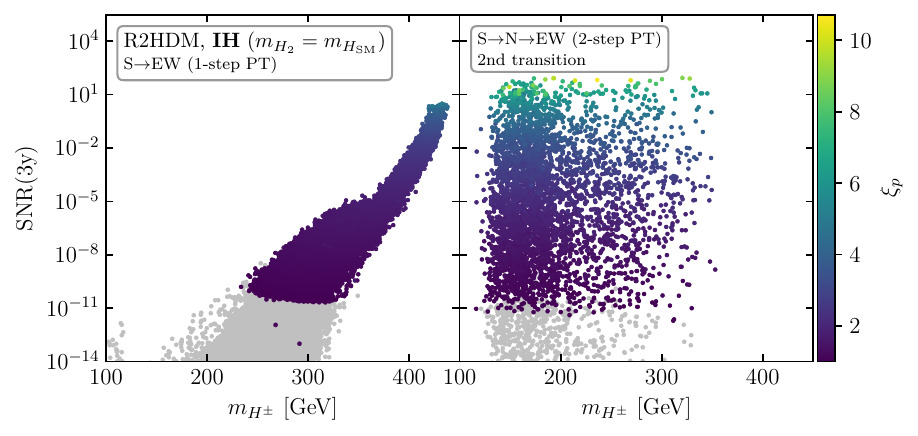}
	\caption{R2HDM inverted hierarchy: Signal-to-noise ratio (SNR) for the \texttt{LISA} experiment after three years of data taking versus the charged Higgs mass. The columns as well as the colouring of the points are the same as in Fig.~\ref{fig:TpvsTcInverted}.}
\label{fig:SNRvsMHpmInverted}
\end{figure}
In the inverted hierarchy, the values of $\alpha$ can be larger than 1 in some cases in the last PT of the two-step PTs. Accordingly, we find rather large SNRs at \texttt{LISA} with values of up to 88. The thermal quantities related to the one-step PT, $\alpha$ and $\xi_p$, are smaller than in the normal hierarchy, leading to smaller SNRs with values of maximally about 3. This can also be inferred from Fig.~\ref{fig:SNRvsMHpmInverted}, which shows the SNR for 3 years of data taking with the \texttt{LISA} experiment as a function of the charged Higgs mass. \s

In comparison to the normal hierarchy, we make two interesting observations.  Firstly, in the normal hierarchy, we found the largest values of $\xi_p$, $\alpha$, and the SNR for one-step transitions and decreasing maximum values with each additional transition. The inverted hierarchy, on the contrary, shows larger values of $\xi_p$, $\alpha$, and the SNR for two-step PTs. Secondly, similarly to the normal hierarchy, we find potentially detectable GW signals for almost the whole range of the charged Higgs mass between roughly $115~\text{GeV} < m_{H^\pm} < 440$~GeV, with one-step PTs dominating for $m_{H^\pm} > 400$~GeV and two-step PTs covering the range $115~\text{GeV} < m_{H^\pm} < 350$~GeV.

\section{Phenomenology of the R2HDM Phase Transitions \label{sec:pheno}}
We now turn to the discussion of the phenomenology related with the R2HDM electroweak phase transitions. We want to investigate whether there are specific phenomenological properties at $T=0$, i.e.~today, that are emerging from strong first-order PTs or multi-step PTs and that could be tested at colliders. 

\subsection{Normal Hierarchy}
\subsubsection{Mass Distributions}
In Figs.~\ref{fig:MassSpectrumNorm1Step}--\ref{fig:MassSpectrumNorm3Step},  we display for the normal hierarchy the mass spectrum in the parameter planes $m_A$ versus $m_{H^\pm}$, $m_A$ versus $m_{H_2}$, and $m_{H_2}$ versus $m_{H^\pm}$ for the one-step, two-step, and three-step phase transitions, respectively, starting from the symmetric vacuum at high temperatures and ending in the EW vacuum at zero temperature. For the two- and three-step PTs, these are hence the points related to the last transition, as this transition provides the strongest PTs among the possible multi-step PTs. The coloured points denote the  scenarios with strong first-order PTs and their corresponding $\xi_p$ values. \s

\begin{figure}[t!]
	\centering
    \includegraphics[width=.8\textwidth]{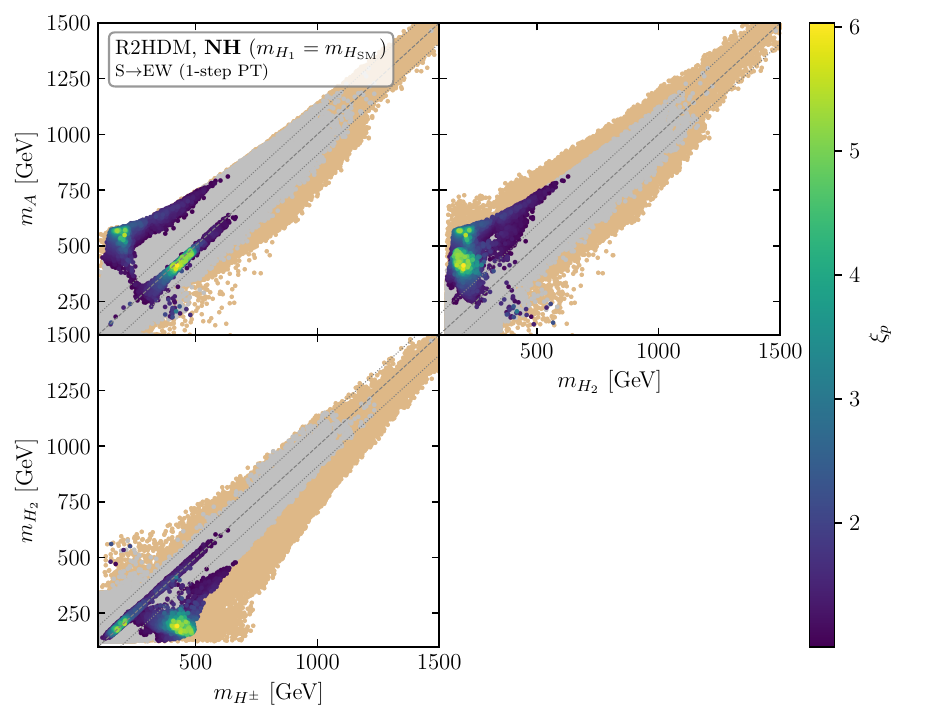}
	\caption{R2HDM normal hierarchy: Mass spectra of points with a one-step PT. Shown are the $m_A$ versus $m_{H^\pm}$ (top left), $m_{H_2}$ versus $m_A$ (top right), and $m_{H^\pm}$ versus $m_{H_2}$ (bottom) parameter planes. The coloured (grey) points fulfill $\xi_p \ge 1$ ($\xi_p < 1$), with the colour bar denoting the size of $\xi_p$. Light brown points denote the whole parameter point sample fulfilling all experimental and theoretical constraints as tested by \ScannerS{} (i.e.\ also points with no FOPTs or no EW symmetry restoration at high temperatures). The dashed gray line on the diagonal corresponds to the case where both masses on the $x$ and $y$ axes are degenerate, while the dotted gray lines indicate a distance of $m_Z$ away from the diagonal.}
    \label{fig:MassSpectrumNorm1Step}
\end{figure}
\begin{figure}[tp]
	\centering
    \includegraphics[width=.8\textwidth]{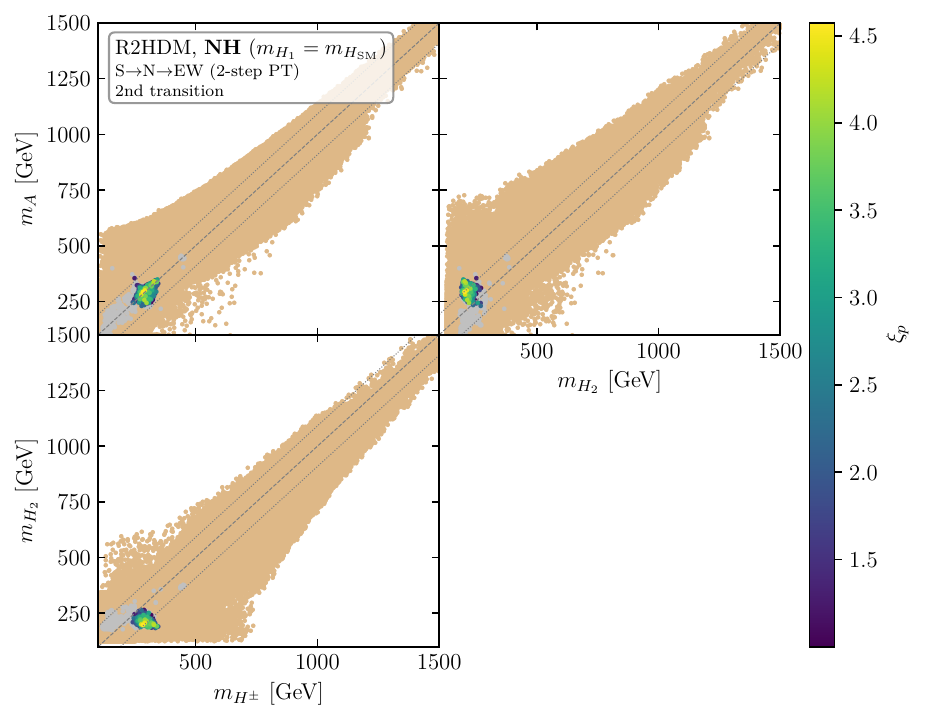}
	\caption{R2HDM normal hierarchy: Mass spectra of points with a two-step PT. The shown parameter planes and the colouring of the points are the same as in Fig.~\ref{fig:MassSpectrumNorm1Step}.}   \label{fig:MassSpectrumNorm2Step}
\end{figure}
\begin{figure}[tp]
	\centering
    \includegraphics[width=.8\textwidth]{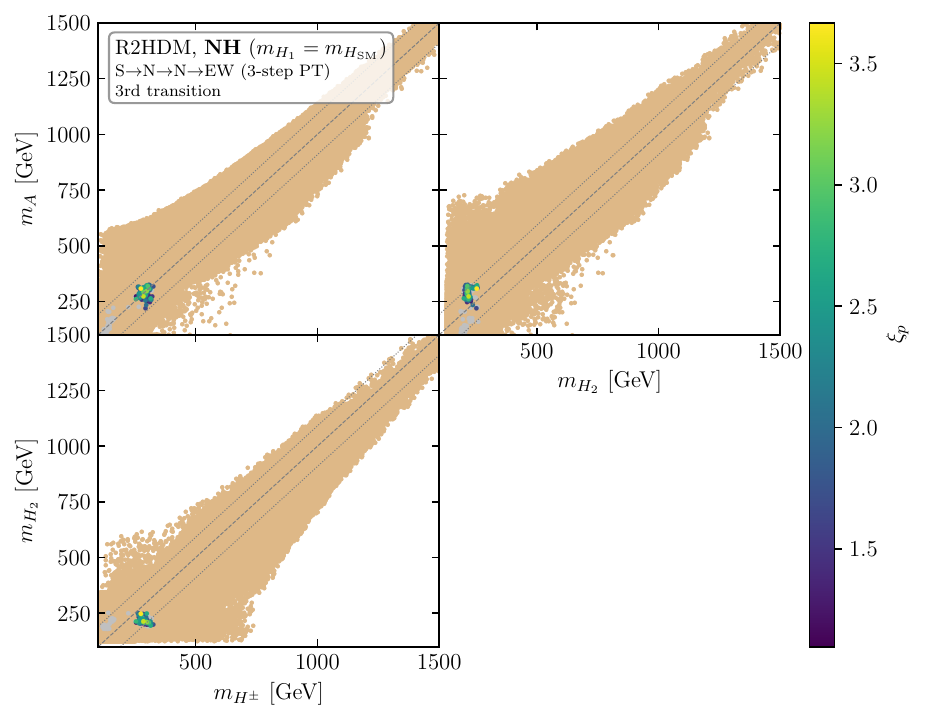}
	\caption{R2HDM normal hierarchy: Mass spectra of points with a three-step PT. The shown parameter planes and the colouring of the points are the same as in Fig.~\ref{fig:MassSpectrumNorm1Step}.}
\label{fig:MassSpectrumNorm3Step}
\end{figure}

In the one-step PTs, we can identify two specific mass regions centered around the largest $\xi_p$ values. They are characterised by (i) $m_{H_2} \approx m_{H^\pm} \approx 190$~GeV, and $m_A - m_{H\pm}/m_{H_2} \approx 360$~GeV and (ii) $m_A \approx m_{H^\pm} \approx 420$~GeV and $m_A/m_{H^\pm} - m_{H_2} \approx 220$~GeV for $\xi_p \gtrsim 3$, cf.~Fig.~\ref{fig:MassSpectrumNorm1Step}. At high-energy colliders, they would be characterised by $A \to Z H_{1,2}$ and $A\to W^\pm H^\mp$ decays for region (i) and by $A \to Z H_{1,2}$ and $H^\pm \to W^\pm H_{1,2}$ decays for region (ii).\footnote{The specific feature of the mass gap between $m_A$ and $m_{H_2}$ for strong first-order PTs, was also found previously in the literature, cf.~e.g.~\cite{Dorsch:2014qja,Dorsch:2016nrg,Basler:2016obg,Biekotter:2022kgf}.} Since less strong PTs can also appear for mass scenarios without these mass gaps, these decays would be a characteristic signature only for the strongest PTs, keeping in mind the caveat that the specific value of $\xi_p$ is subject to uncertainties, cf.~the comment above on gauge dependencies and the discussion on uncertainties below. It can furthermore clearly be observed that points with strong PTs come with rather light masses, i.e.~$m_A \lesssim 750$~GeV as well as $m_{H_2}, m_{H^\pm} \lesssim 650$~GeV, and $m_{H_2} \approx 190$~GeV for the strongest PTs. \s 

In the case of two-step PTs, the mass regions with strong first-order PTs shrink to one region centered around the largest $\xi_p$ values in the respective mass planes, cf.~Fig.~\ref{fig:MassSpectrumNorm2Step}. More specifically, region (i) of the one-step PTs disappears, and region (ii) shrinks considerably and shifts to smaller mass values, i.e.~$m_A \approx m_{H^\pm} \approx 300$~GeV and $m_{A}/m_{H^\pm} - m_{H_2} \approx m_Z$ for $\xi_p \gtrsim 3$. Specific collider signatures for the strongest decays would hence stem from $A \to ZH_{1,2}$ and $H^\pm \to W^\pm H_{1,2}$ decays. Note, however, that also here the mass gaps can disappear for less strong PTs, so that assuming these decays to be indicative for a strong first-order phase transition would require less uncertainties in the predictions of $\xi_p$. A striking feature, however, is the fact that for strong first-order two-step PTs $m_{A,H^\pm} \in [230, 340]$~GeV and $m_{H_2} \in [190, 270]$~GeV. \s

In the case of three-step PTs, the masses are located at the same regions in the mass planes for the largest $\xi_p$ values as in the two-step PTs, but the mass regions with $\xi_p > 1$ shrink further, cf.~Fig.~\ref{fig:MassSpectrumNorm3Step}. We find $m_{A,H^\pm} \in [230, 330]$~GeV and $m_{H_2} \in [240, 270]$~GeV. \s

We finish this discussion with a statement on the possible values of $\tan\beta$ compatible with strong first-order PTs. We found that strong one-step PTs occur for $\tan\beta$ values between 0.8 (our lower scan bound) and 65, whereas strong multi-step PTs reduce the $\tan\beta$ range to values between 20 and 50.  

\subsubsection{Trilinear Higgs Self-Couplings}
We proceed with analysing the impact of strong first-order PTs on the zero-temperature phenomenology focussing now on the trilinear Higgs self-coupling. In contrast to the couplings of the discovered 125~GeV Higgs boson to the SM particles, which are found to be very SM-like, the trilinear Higgs self-coupling is still poorly constrained. At the LHC, it can directly be determined from Higgs pair production \cite{Djouadi:1999rca,DiMicco:2019ngk}, where the gluon fusion is the dominant process \cite{Baglio:2012np,DiMicco:2019ngk}. The present limits on $\kappa_\lambda$, the deviation of the trilinear Higgs self-coupling from the SM value, assuming SM top-quark couplings are $-0.71 < \kappa_\lambda < 6.1$ at 95\% confidence level \cite{CMS:2026nuu}. Since the strength of the phase transition is determined by the evolution of the effective Higgs potential as a function of the temperature, there is an interplay between the Higgs potential parameters, masses and self-couplings, and the strength of the PT, which we have already seen and discussed for the Higgs masses. We now proceed to have a look at the effective trilinear Higgs self-couplings. Since we apply the on-shell-like renormalisation conditions only on the first and second derivatives of the Higgs potential, the trilinear Higgs self-couplings at NLO, derived from our effective Higgs potential, change with respect to the tree-level values. Furthermore, they will differ from the SM value, as we consider the R2HDM and demand the PT to be of strong first order. Note finally that the trilinear Higgs self-couplings derived from the effective potential neglect external momenta. A comparison with the Feynman diagrammatic approach taking into account finite momentum effects has shown that they play a minor role \cite{Arco:2025pgx} compared to the change with respect to the SM value. The SM value at NLO, which we apply here in our analysis is derived from the effective potential calculated in \texttt{BSMPT} using the ``SM'' option and amounts to $\lambda_{HHH}^{\text{SM,eff,NLO}} = 174.80$~GeV. \s

\begin{figure}[tp]
	\centering
    \includegraphics[width=\textwidth]{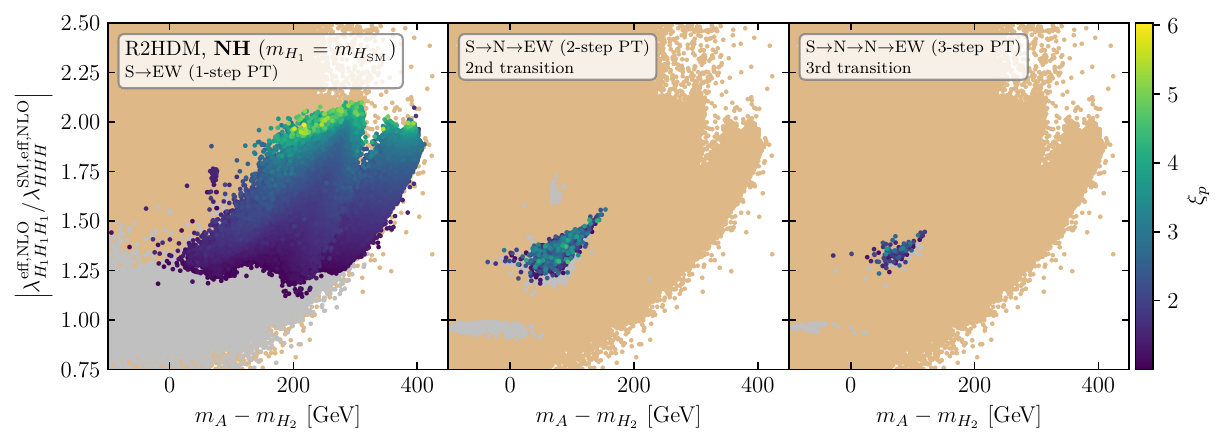}
	\caption{R2HDM normal hierarchy: Effective trilinear Higgs self-coupling of the SM-like Higgs $H_1$ at NLO normalised to the corresponding SM value as a function of the mass gap between the CP-odd and the heavy CP-even Higgs mass, $m_A - m_{H_2}$. The three panels correspond to one-step PT (left), second transition of the two-step PT (middle), and third transition of the three-step PT (right). The colouring of the points is the same as in Fig.~\ref{fig:MassSpectrumNorm1Step}.} \label{fig:MassSpectrumNormMassGap}
\end{figure}
In Fig.~\ref{fig:MassSpectrumNormMassGap} we display for the R2HDM points with a strong first-order PT the effective trilinear Higgs self-coupling of the SM-like Higgs ($H_1$ in the normal hierarchy) at NLO normalised to the corresponding SM value for the one-step PT (left), second transition of the two-step PT (middle), and the third transition of the three-step PT through the colour code denoting the related $\xi_p$ values, as a function of the mass gap $m_A - m_{H_2}$. 
For all three step-types of possible PTs in the R2HDM, we observe that the requirement of the strong first-order PT induces enhanced trilinear Higgs self-couplings compared to the SM case by at least 10\%. For the one-step PTs they are increased by a factor 1.1 to 2.1 where larger $\xi_p$ values lead to larger enhancements. For the two- and three-step PTs, however, we do not see such a correlation between the value of $\xi_p$, i.e.~the strength of the PT, and the relative enhancement. On the other hand, we see a correlation between the enhancement of the trilinear couplings and the mass gap, which increases with larger enhancements of the Higgs self-couplings. The enhancement of the trilinear couplings in these multi-step PTs is not as high as in the one-step case, reaching at most 1.6 for the two-step and 1.5 for the three-step PTs. \s

\begin{figure}[tp]
	\centering
	\includegraphics[width=\textwidth]{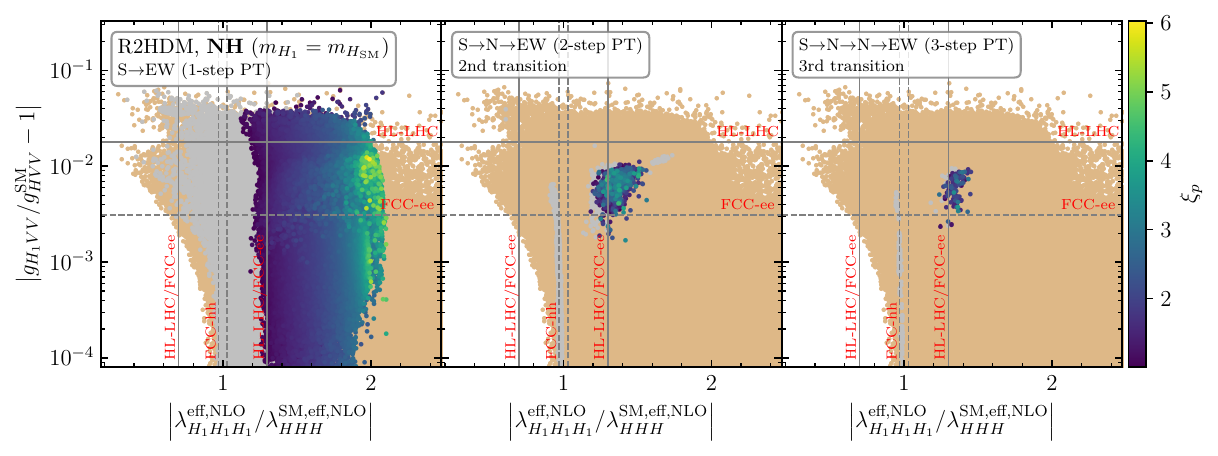}
	\caption{R2HDM normal hierarchy: Deviation of the SM-like Higgs $H_1$ coupling to massive gauge bosons ($V\equiv Z,W^\pm$) from the SM value versus the effective trilinear Higgs self-coupling of the SM-like Higgs at NLO normalised to the corresponding SM value. The three panels correspond to one-step PT (left), the second transition of the two-step PT (middle), and the third transition of the three-step PT (right).  The colour scheme of the points is the same as in Fig.~\ref{fig:MassSpectrumNorm1Step}. Also shown are the expected sensitivity bounds on the single- and di-Higgs couplings from future colliders, HL-LHC \cite{ATLAS:2025eii} (solid) and FCC-ee/-hh \cite{Maura:2025rcv,Selvaggi:2025kmd} (dashed).}
\label{fig:ghzvs3lnormal}
\end{figure}
Lastly, in order to put the found parameter regions in the context of current and future collider limits, we show (again for the three step-types of PTs) in Fig.~\ref{fig:ghzvs3lnormal} the relative deviation in the SM-like Higgs coupling to the massive gauge bosons ($V=Z,W^\pm$) from the SM value versus the effective trilinear Higgs self-coupling of the SM-like Higgs $H_1$ at NLO normalised to the corresponding SM value. We remind the reader that in \texttt{BSMPT}, the mixing angles are normalised to their tree-level values so that the loop-corrected effective Higgs couplings, i.e.~the loop-corrected Higgs couplings at zero external momentum, to the SM particles remain unchanged with respect to their tree-level values, in particular also the displayed coupling $g_{H_1VV}$. This means that after normalisation to the SM value, $g_{H_1VV}/g_{H_1VV}^{\text{SM}}$ is given by the coupling modification factor $\sin(\beta-\alpha)$, with the mixing angles renormalised in our OS-like scheme, which converges to 1 in the SM-limit. 
The 2HDM trilinear Higgs self-couplings on the other hand receive NLO corrections in our chosen OS-like renormalisation scheme and thereby change with respect to their tree-level values. \s

The horizontal lines show the limits on the Higgs couplings to the massive gauge bosons obtained at the High-Luminosity (HL)-LHC (solid) and the future FCC-ee (dashed). The vertical lines show the sensitivity to deviations from the SM trilinear Higgs self-coupling at the HL-LHC and the FCC-ee (solid) and at FCC-hh (dashed).\footnote{The projected HL-LHC limits are taken from \cite{ATLAS:2025eii} and the future  FCC-ee/-hh limits are extracted from \cite{Maura:2025rcv,Selvaggi:2025kmd}.} 
The left plot of Fig.~\ref{fig:ghzvs3lnormal} shows that in the R2HDM for one-step PTs we find no correlation between the single-Higgs coupling deviations and the di-Higgs coupling values.\footnote{In singlet models with only one mixing angle on the other hand there is a strong correlation between these couplings \cite{Bahl:2026aou}. This is also the case in the inert doublet model \cite{Bahl:2026aou}.} The increasing precision on the single Higgs couplings at future colliders will not further narrow down the trilinear Higgs self-coupling values associated with a strong first-order PT. The impact on observables can be different, however. As found in \cite{Anisha:2025zbc}, the combination of strong limits obtained from $S,T,U$ at the FCC-ee with the requirement of a strong first-order PT singles out parameter points leading to NLO deviations of the Higgs-strahlung cross section that will be testable at the FCC-ee.\footnote{We checked our found parameter points against the stricter FCC-ee bounds on the electroweak precision parameters $S$ and $T$ as was done in \cite{Anisha:2025zbc} and find that out of the approximately 90,000 points for the normal hierarchy with an SFOEWPT, 195 points pass the stricter bounds within a $2\sigma$ range, while out of the 195 points, 26 points are within $1\sigma$. For the set of approximately 30,000 strong first-order PT points for the inverted hierarchy, no points are found that pass the stricter bounds within the $2\sigma$ range.} \s

In the two- and three-step PTs with $\xi_p > 1$ on the other hand, where the parameter space for strong first-order PTs is much more reduced, the future precisions on the single Higgs couplings together with the constraints on the trilinear Higgs self-coupling from the HL-LHC and FCC-ee measurements will be able to exclude these scenarios, as can be inferred from Fig.~\ref{fig:ghzvs3lnormal} (middle) and (right). 

\subsection{Inverted Hierarchy}
\subsubsection{Mass Distributions}
\begin{figure}[t!]
	\centering
    \includegraphics[width=.8\textwidth]{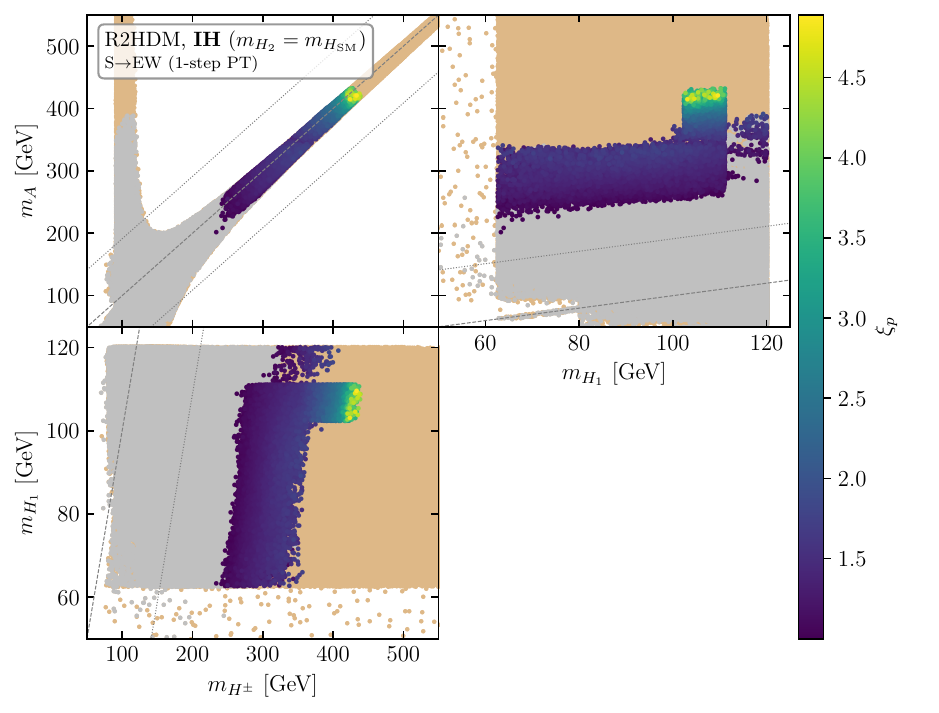}
	\caption{R2HDM inverted hierarchy: Mass spectra of points with a one-step PT. Shown are the $m_A$ versus $m_{H^\pm}$ (top left), $m_{H_2}$ versus $m_A$ (top right), and $m_{H^\pm}$ versus $m_{H_2}$ (bottom) parameter planes. The coloured (grey) points fulfill $\xi_p \ge 1$ ($\xi_p < 1$), with the colour bar denoting the size of $\xi_p$. The colouring as well as the meaning of the dashed and dotted lines is the same as in Fig.~\ref{fig:MassSpectrumNorm1Step}.}
    \label{fig:MassSpectrumInv1Step}
\end{figure}
We move on to discuss the mass spectra for the inverted hierarchy. In Fig.~\ref{fig:MassSpectrumInv1Step} we show the mass distributions for the one-step PTs, with the coloured points denoting strong transitions.
As here the neutral CP-even non-SM-like Higgs mass $H_1$ lies below $m_{H_2} = 125.09$~GeV,\footnote{We note that we enforced a mass gap between $H_1$ and $H_2$ of 5~GeV to avoid degenerate Higgs scenarios that require further special treatments when applying the LHC limits on the Higgs data.} the EW precision observables enforce the charged Higgs mass $m_{H^\pm}$ to align with either $m_{H_2}$ or $m_A$.  The requirement of a strong first-order PT, however, restricts the charged Higgs mass values to $m_{H^\pm} \approx m_{A} \in [220, 440]$~GeV. The induced large mass gaps imply the possibility of $A \to ZH_{1,2}$ and $H^\pm \to W^\pm H_{1,2}$ collider signatures, such that they can be considered a characteristic feature for this specific phase transition history in the inverted hierarchy.\footnote{The abrupt stop of strong first-order PT points at $m_{H_1} = 62.5$~GeV is a consequence of experimental constraints. The possible decay of the 125~GeV Higgs $H_2 \to H_1 H_1$ for $m_{H_1}\le 62.5$~GeV strongly reduces the number of allowed points due to the Higgs measurements, so that we did not find any strong first-order PTs in our parameter scan for this region, cf.~also~\cite{Lee:2025hgb}.} 
From Fig.~\ref{fig:MassSpectrumInv1Step} it can furthermore be inferred that the strength of the PTs increase with the heavier non-SM-like Higgs masses $m_A \approx m_{H^\pm}$. The region with larger $\xi_p$ values, $\xi_p \gtrsim 2.5$, requires $380~\text{GeV} \lesssim m_A \approx m_{H^\pm} \lesssim 440$~GeV and simultaneously $102~\text{GeV} \lesssim m_{H_1} \lesssim 112$~GeV. We checked that the particular shape of the allowed parameter space with a strong first-order PT relates to the requirement of EW symmetry restoration at high temperatures, see Fig.~\ref{fig:MassInvEWSR}, highlighting the interplay between the quartic couplings of the scalar potential and their influence on EW symmetry restoration and the Higgs masses. \s

\begin{figure}[tp]
	\centering
    \includegraphics[width=.8\textwidth]{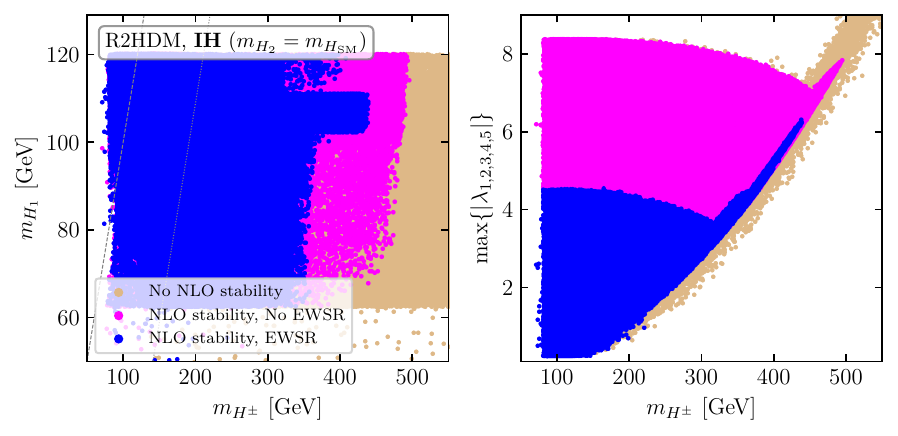}
	\caption{R2HDM inverted hierarchy: Non-SM-like light Higgs mass $m_{H_1}$ (left) and the maximum quartic coupling $|\lambda_{1,2,3,4,5}|$ (right) versus the charged Higgs mass $m_{H^\pm}$. All points  fulfill the constraints tested by \ScannerS{}. Light brown points are not NLO stable, i.e.\ the global minimum does not stay at $v \approx 246$~GeV when taking into account one-loop corrections to the effective potential and using the OS-like renormalisation scheme, magenta points fulfill NLO stability, but not EW symmetry restoration (EWSR) at high temperatures, while blue points fulfill both NLO stability and EWSR.}
    \label{fig:MassInvEWSR}
\end{figure}
The mass spectra for the two-step PTs are shown in Fig.~\ref{fig:MassSpectrumInv2Step}, where again we only plot the $\xi_p$ values for the second transition as it is the strongest of the two PTs. The points with strong PTs now extend to lower masses with $60$~GeV~$\lesssim m_A \lesssim 350$~GeV and $120~\text{GeV} \lesssim m_{H^\pm} \lesssim 350$~GeV. The charged Higgs masses now show a larger spread away from the $m_A = m_{H^\pm}$ line in particular for smaller masses, so that we now also have scenarios where $m_A \approx m_{H_1}$. Comparing with the results of \cite{Lee:2025hgb}, we find qualitative agreement in the range of the parameter regions for one- and two-step phase transitions. \s

\begin{figure}[tp]
	\centering
    \includegraphics[width=.8\textwidth]{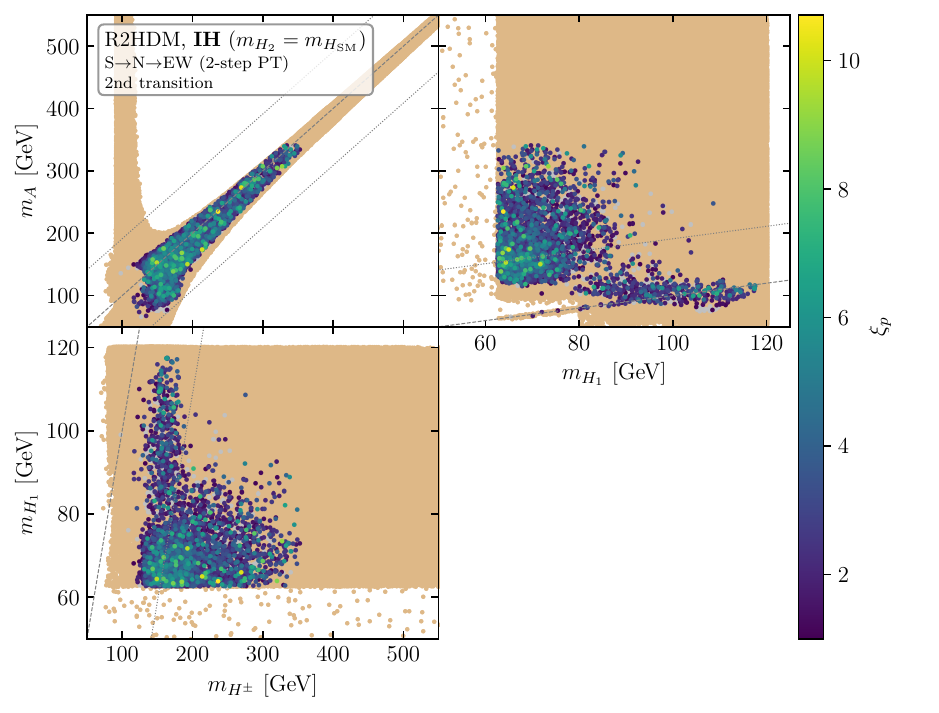}
	\caption{R2HDM inverted hierarchy: Mass spectra of points with a two-step PT. The shown parameter planes are the same as for Fig.~\ref{fig:MassSpectrumInv1Step} and the colouring of the points is the same as in Fig.~\ref{fig:MassSpectrumNorm1Step}.}
    \label{fig:MassSpectrumInv2Step}
\end{figure}

Finally, we find that for strong first-order PTs the $\tan\beta$ values are restricted to values below about 11 for one-step phase transitions, whereas for two-step phase transitions $\tan\beta$ can take values of up to about 34.

\subsubsection{Trilinear Higgs Self-Couplings}

\begin{figure}[tp]
	\centering
    \includegraphics[width=.8\textwidth]{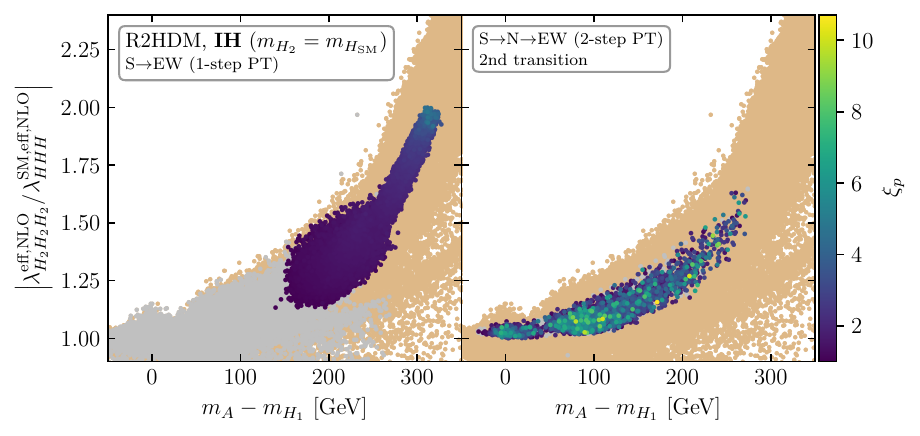}
	\caption{R2HDM inverted hierarchy: Effective trilinear Higgs self-coupling of the SM-like Higgs $H_2$ at NLO normalised to the corresponding SM value as a function of the mass gap between the CP-odd and the heavy CP-even Higgs mass, $m_A - m_{H_2}$. The two panels correspond to one-step PT (left) and the second transition of the two-step PT (right). The colouring of the points is the same as in Fig.~\ref{fig:MassSpectrumNorm1Step}.}
    \label{fig:MassSpectrumInvMassGap}
\end{figure}
Figure~\ref{fig:MassSpectrumInvMassGap} depicts the effective trilinear Higgs self-coupling of the SM-like Higgs ($H_2$ in the inverted hiearchy) at NLO normalised to the corresponding SM value for the one-step (left) and two-step PTs (right), as a function of the mass gap $m_A-m_{H_1}$. As in the normal hierarchy, we observe enhanced trilinear Higgs self-couplings for strong first-order PTs, lying between 1.1 up to now 2 times the corresponding SM value. We see an increase in the trilinear coupling and the mass gap for stronger PTs. The correlation between the trilinear coupling modification and the mass gap is stronger than in the normal hierarchy. \s 

For the two-step PTs, the points with a strong first-order PT range from a mass gap of negative values slightly below 0~GeV to almost 300~GeV, but are mostly limited to values of the trilinear coupling modifier below 1.7, with a large fraction of the points exhibiting very SM-like trilinear couplings. An enhanced trilinear Higgs self-coupling can hence not be taken as a requirement or consequence of a strong first-order PT any more in this case. Again, for the two-step PTs, the size of $\xi_p$ is not correlated to the mass gap. \s

\begin{figure}[tp]
	\centering
	\includegraphics[width=.8\textwidth]{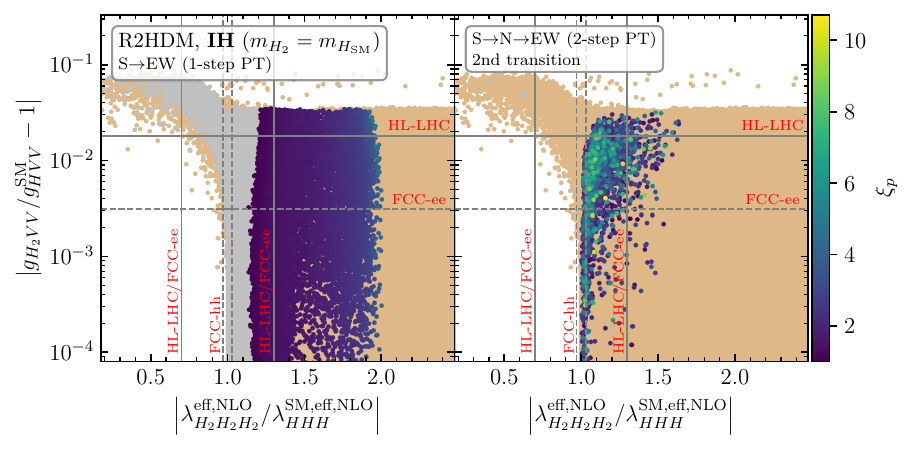}
	\caption{R2HDM inverted hierarchy: Deviation of the SM-like Higgs $H_2$ coupling to massive gauge bosons ($V\equiv Z,W^\pm$) from the SM value versus the effective trilinear Higgs self-coupling of the SM-like Higgs at NLO normalised to the corresponding SM value. The two panels correspond to one-step PT (left) and the second transition of the two-step PT (right).  The colour scheme of the points is the same as in Fig.~\ref{fig:MassSpectrumNorm1Step}. The shown future collider bounds are given as in  Fig.~\ref{fig:ghzvs3lnormal}.}
\label{fig:ghzvs3linv}
\end{figure}
The SM-like single- and di-Higgs coupling deviations are displayed in Fig.~\ref{fig:ghzvs3linv} for the one-step (left) and two-step phase transitions (right). In the case of one-step PTs, the increased precision on the single Higgs coupling to the massive gauge bosons at future colliders will not add further insights on the trilinear couplings and hence the relation to the strong first-order phase transition. This picture changes when we consider the two-step phase transitions where the trilinear couplings overall are closer to the SM value. They are now more correlated with the single Higgs couplings to the massive gauge bosons. Pushing these couplings more and more towards the SM values has the same effect on the SM-like trilinear Higgs self-coupling when combined with a strong phase transition, such that now single Higgs measurements can inform about the nature of the EWPT, keeping in mind, however, the uncertainty associated with the precise value of $\xi_p$.
The main difference between one- and two-step PTs is that for one-step PTs, we do not find first-order PTs with very SM-like trilinear self-couplings, while two-step PTs give us the possibility to have strong first-order PTs also for the trilinear coupling modifier close to 1, such that future collider experiments, which will constrain more strongly the region away from SM-like couplings, will eventually allow us to distinguish between the possibility of a single- or multi-step strong first-order PT in the R2HDM with an inverted mass hierarchy.

\subsection{Exotic Phase Transitions}
As discussed before, our scan sample also exhibits parameter configurations inducing phase histories involving exotic intermediate phases such as charge-breaking ($\omega_\CB \ne 0$) and CP-violating phases ($\omega_\CP \ne 0$). \s

\begin{figure}[tp]
	\centering
    \includegraphics[scale=.8]{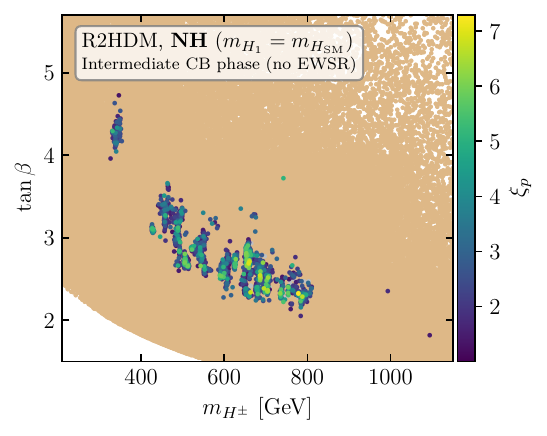}\includegraphics[scale=.8]{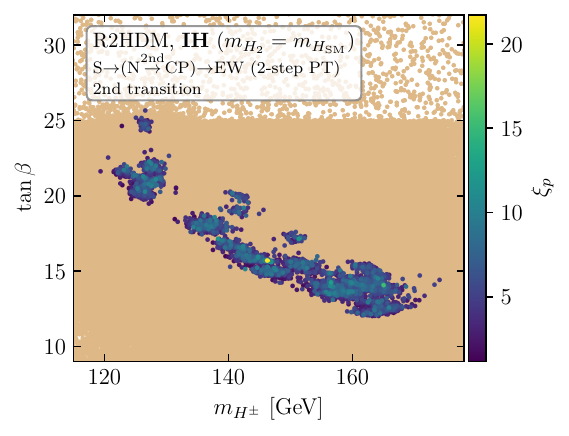}
    \caption{R2HDM exotic PTs in the $\tan\beta$ versus $m_{H^\pm}$ plane. Left:  transition histories with an intermediate charge-breaking phase for the normal hierarchy, no EWSR. Right: CP-breaking phase histories for inverted hierarchy. The colour scheme of the points is the same as in Fig.~\ref{fig:MassSpectrumNorm1Step}. For the left plot with intermediate CB phases, we show $\xi_p$ for the transition from the CB to a neutral or symmetric phase, which we typically found to be the strongest one. For the right plot with an intermediate CP-breaking phase, the colour scale denotes the strength of the second first-order transition from CP$\to$EW, as it is the stronger of the two first-order transitions.} 
    \label{fig:CBCPphases}
\end{figure}

In Fig.~\ref{fig:CBCPphases}~(left), we show in the $\tan\beta - m_{H\pm}$ plane for the normal hierarchy the part of the parameter space where multi-step PTs involve an intermediate CB phase. We find that for all CB points there is a mass gap between $m_A$ and $m_{H_2}$ with $m_{H_2} \approx m_{H^\pm}$. 
This allows for $A\to Z H_{1,2}$ and $A\to W^\pm H^\mp$ decays, as previously discussed in \cite{Aoki:2023lbz}. Furthermore, all CB points lie in the region for small $\tan\beta < 5$. While in \cite{Aoki:2023lbz}, we focussed mainly on studying the low $m_{H^\pm}$ region, as a notable difference, we now also find CB points here for intermediate to heavy $300~\text{GeV} \lesssim m_{H^\pm} \lesssim 1000$~GeV. As before, we however do not find CB points which allow for EW symmetry restoration at high $T$. This is related to the opposing requirements of small quartic Higgs potential couplings for EW symmetry restoration at high $T$ and large quartic couplings for the occurrence of an intermediate CB phase. \s

In the case of the inverted hierarchy, we find points with an intermediate CP-violating phase in the parameter region shown in Fig.~\ref{fig:CBCPphases}~(right), namely for $m_{H^\pm} \lesssim 175$~GeV and intermediate to high $\tan\beta$, $12 \lesssim \tan\beta \lesssim 26$. Notably, we find that the CP-violating phase does not arise directly via a first-order PT from the symmetric phase, which would give a relevant contribution to baryogenesis. Instead, the universe first undergoes a first-order PT from the symmetric to a neutral-broken phase without CP violation. A CP-violating VEV then develops via a second-order PT from the neutral-broken phase, and eventually, the universe transitions into the EW-broken phase. This is to be expected, as there is no explicit CP violation in the tree-level Lagrangian of the R2HDM, so spontaneous CP violation at $T \ne 0$ (i.e.\ $\omega_\CP \ne 0$) can only arise as a loop effect. This requires, however, scalar loops with trilinear Higgs self-couplings that are non-zero only for non-zero neutral VEVs $\omega_{1,2} \ne 0$. An intermediate CP-violating PT starting from a R2HDM at $T=0$ hence at least requires an intermediate phase with $\omega_{1,2} \ne 0$ to spontanteously develop a CP-violating VEV.

\subsection{Gravitational-Wave Signals and Uncertainties}
With the discovery of the GWs, we now have an additional observable that can inform us about cosmological vacuum evolution and ultimately, about the underlying new physics model. For a meaningful parameter derivation, however, we need precise predictions and quantitative information on the involved uncertainties, cf.~e.g.~\cite{Croon:2020cgk}. In the derivation for the gravitational wave signals there are multiple sources for uncertainties. They stem, to name a few of them, e.g.~from the uncertainty in the measured SM input parameters (parametric uncertainties), they  are related to the missing higher-order corrections in the effective Higgs potential (perturbative uncertainties), they can stem from the way the infrared uncertainties are treated (Linde problem), they arise from the gauge dependence of the Higgs potential itself, they are due to approximations made in the description of the PT (e.g.~which transition temperature is applied or uncertainties in the prediction of the transition rate), they source from the way the wall velocity is obtained, which feeds in the derivation of the transition temperatures and ultimately in the GWs, or they can stem from the description of the GW spectrum itself. For comprehensive overviews and discussions, cf.~e.g.~\cite{Athron:2023xlk,vandeVis:2025efm}.\s

Our approach uses the four-dimensional effective potential, which is gauge-dependent and sensitive to the renormalisation scale. Moreover, the Linde problem \cite{Linde:1980ts} challenges the thermodynamic description in theories with non-Abelian gauge fields. The problem is related to non-perturbative effects that arise from massless vector bosons. They can be treated by daisy resummation \cite{Arnold:1992rz,Espinosa:1995se}, or in the improved approach of optimised partial dressing \cite{Curtin:2016urg,Curtin:2022ovx,Bahl:2024ykv,Bittar:2025lcr}. The problem of gauge dependence was considered e.g.~in \cite{Chiang:2017nmu,Patel:2011th,Hirvonen:2021zej,Hirvonen:2021zej}, in the context of simple benchmark models. We are not aware, however, of a gauge-independent approach in the 2HDM nor in a model-independent way, and also not in the OS-like renormalisation scheme which we employ in our calculation. Alternatively, effective descriptions are suggested, in particular the three-dimensional effective field theory (EFT) approach. In dimensional reduction \cite{Ginsparg:1980ef,Appelquist:1981vg} the hierarchy of thermal scales appearing in the thermal plasma is treated by integrating out the contributions from the heavy thermal scale to the parameters of the low-energy EFT. This approach improves the perturbative convergence of the effective potential and  reduces its gauge dependence, cf.~e.g.~\cite{Kainulainen:2019kyp}. With \texttt{DRalgo} \cite{Ekstedt:2022bff} an algorithm has been presented that constructs an effective dimensionally reduced, high-temperature field theory for generic models. An implementation in \texttt{BSMPTv3} is in principle possible, would require, however, for its correct application the examination of the hierarchy of the involved scales for each studied parameter point of the model under investigation at each temperature step taken in the tracing of the vacuum phases. This would therefore require a major upgrade and is beyond the scope of the current paper. We therefore treat the computed value of $\xi_p$ according to Eq.~\eqref{eq:xipdefinition} as a proxy for the (gauge-independent) sphaleron energy in the broken phase. The requirement of $\xi_p > 1$ is taken as an approximate minimal condition that is necessary for electroweak baryogenesis to be realised. Additionally, we compare our calculated $\xi_p$ values with the derived released latent heat to further support this statement. \s

\begin{figure}[tp]
	\centering
\includegraphics[width=\textwidth]{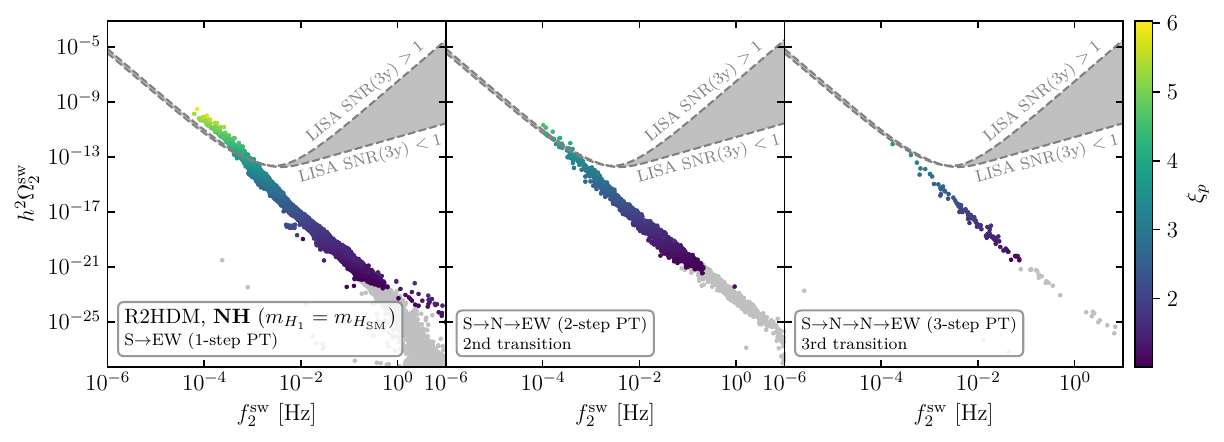}
	\caption{R2HDM normal hierarchy: GW amplitude $h^2\Omega_2^\text{sw}$ for the sound-wave power spectrum (``sw'') versus the characteristic frequency $f_2^\text{sw}$. The columns as well as the colouring of the points are the same as in Fig.~\ref{fig:TpvsTcNormal}. The grey band denotes the sensitivity of the \texttt{LISA} experiment after 3 years of data taking: points above (below) the band have a SNR $> 1$ ($< 1$), while points lying within the grey area can have a SNR $> 1$ or $< 1$ depending on their value of $f_1^\text{sw}$. We refer the reader to App.~\ref{app:band} for more details.} 
\label{fig:omegaGWnorm}
\end{figure}

\subsubsection{Gravitational Wave Spectra}
While a full uncertainty analysis of the computation of the gravitational wave spectrum is far beyond the scope of this paper, we will analyse the impact of the variation of a few parameters entering the effective potential and the GW computation  in more detail. We start by showing the expected gravitational-wave spectra for the strong first-order PT points from our scan. As the sound-wave contribution is dominant for all  parameter points of our sample, we show in the following the relevant amplitudes of the sound-wave power spectrum depending on the frequency. In \BSMPT{}, the sound-wave power spectrum is implemented following \cite{Caprini:2024hue} as a double broken power law with characteristic frequency breaks  $f_1^\text{sw}$ and $f_2^\text{sw}$, and the amplitude $h^2 \Omega_2^\text{sw}$, where $h = 0.674 \pm 0.005$ is the reduced Hubble constant \cite{Planck:2018vyg}.\footnote{For more details on the computation of the GW spectrum in \texttt{BSMPTv3}, we refer to \cite{Basler:2024aaf}.} In Fig.~\ref{fig:omegaGWnorm}, we show for the R2HDM points with normal mass hierarchy $h^2 \Omega_2^\text{sw}$ versus $f_2^\text{sw}$. The corresponding plots for the inverted hierarchy are shown  in Fig.~\ref{fig:omegaGWinv}. \s
	
\begin{figure}[tp]
	\centering    \includegraphics[width=.8\textwidth]{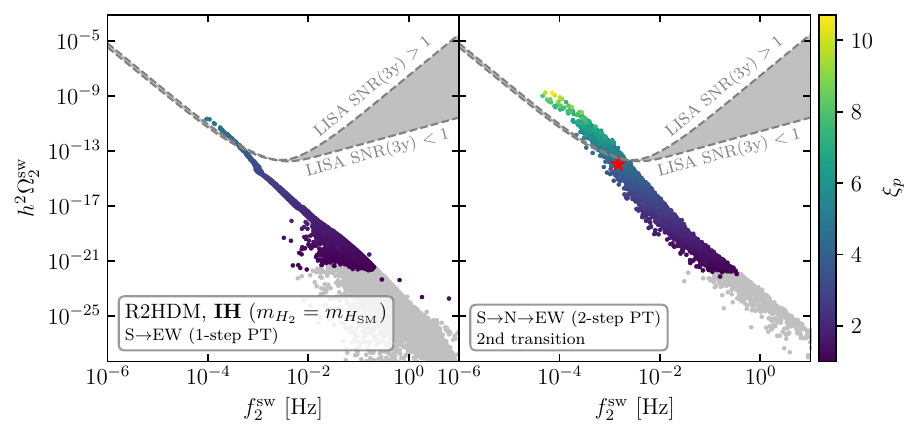}
	\caption{R2HDM inverted hierarchy: GW amplitude $h^2\Omega_2^\text{sw}$ for the sound-wave power spectrum (``sw'') versus the characteristic frequency $f_2^\text{sw}$. The columns as well as the colouring of the points are the same as in Fig.~\ref{fig:TpvsTcInverted}. The meaning of the grey \texttt{LISA} sensitivity band is described in the caption of Fig.~\ref{fig:omegaGWnorm} and App.~\ref{app:band}. The red star $\textcolor{red}{\filledstar}$ denotes the location of the benchmark point BP$\filledstar$ discussed below.}
\label{fig:omegaGWinv}
\end{figure}

As can be inferred from Figs.~\ref{fig:omegaGWnorm} and \ref{fig:omegaGWinv} and as was already shown in Figs.~\ref{fig:SNRvsMHpmNormal} and \ref{fig:SNRvsMHpmInverted}, larger values of $\xi_p$ lead to stronger GW signals and thus larger values of $h^2\Omega_2^\text{sw}$. The latter reaches values of up to $10^{-9}$ or slightly above. The strong first-order PT points with $\xi_p > 1$ are found in a frequency range of $10^{-4}\,\text{Hz} \lesssim f_2^\text{sw} \lesssim 1\,\text{Hz}$. For most of the points we see that smaller values of $f_2^\text{sw}$ lead to larger values of $h^2\Omega_2^\text{sw}$ and vice versa. 
The difference between the normal and inverted mass hierarchy is that for the normal hierarchy we  obtain the strongest GW signals and largest amplitudes for points with a one-step transition, and consequently smaller amplitudes for two-step and even smaller ones for three-step PTs. For the inverted hierarchy on the other hand, the largest amplitudes and GW signals are found for the two-step transitions, with one-step transition points above and slightly below the \texttt{LISA} sensitivity band being constrained to a very thin line. This is explained by the fact that the strength of the PTs in the normal hierarchy decreases from the one-step PT to the second step of the two-step PT to the third step of the three-step PT, cf.~Eq.~(\ref{ew:nhstrengths}), whereas the strength of the PT in the inverted hierarchy increases from the one-step PT to the second step of the two-step PT, cf.~Eq.~(\ref{eq:alphaPTInv}). \s 

\begin{figure}[tp]
	\centering
	\includegraphics[width=\textwidth]{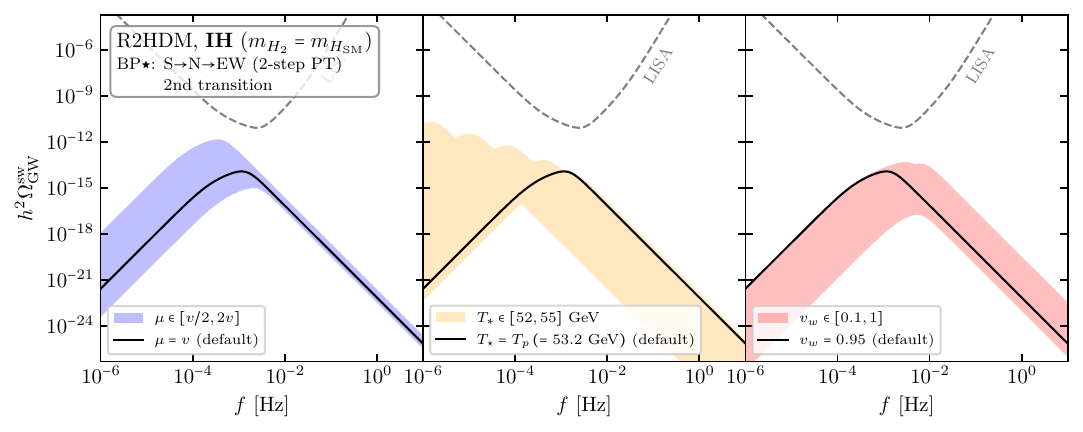}
	\caption{R2HDM inverted hierarchy, BP$\filledstar$: Power spectrum of the sound-wave contribution according to the double broken power law template, showing $h^2\Omega_\text{GW}^\text{sw}$ as a function of the frequency $f$ for the benchmark point BP$\filledstar$. The three plots show each a variation of the renormalisation scale $\mu$ (left panel, blue band), the transition temperature $T_\star$ (middle panel, yellow band), and the bubble wall velocity $v_w$ (right panel, red band), respectively, around their default values (black line). The grey dashed curve corresponds to the nominal sensitivity of the \texttt{LISA} experiment, cf.~e.g.~Eq.~(3.87) of \cite{Basler:2024aaf}.}
\label{fig:omegaGWinvBPParamVar}
\end{figure}

\subsubsection{Uncertainty Discussion}
Out of the sample of points with a two-step PT transition of the inverted hierarchy, we study one benchmark point, denoted as BP$\filledstar$ in Fig.~\ref{fig:omegaGWinv}, in more detail. The benchmark point corresponds to a transition history of the type S$\to$N$\to$EW, where the second transition from N$\to$EW is the stronger of the two with $\xi_p^\text{2nd} = 4.49$ as compared to $\xi_p^\text{1st} = 0.34$ of the first transition from S$\to$N. We will consider the stronger second transition in the following. The input parameters for BP$\filledstar$, together with the SM Higgs mass $m_{H_2} = 125.09$~GeV and the EW VEV $v = 246.22$~GeV, are given as,
\begin{equation}
    \begin{gathered}
        m_{H_1} = 64~\text{GeV},~~ m_A = 153~\text{GeV},~~ m_{H^\pm} = 136~\text{GeV},~~ m_{12}^2 = 1496~\text{GeV}^2, \\
        \tan\beta = 3.4,~~ \cos(\alpha - \beta) = 0.99651\,.
    \end{gathered}
\end{equation}
The calculated PT and GW parameters related to the second transition are,
\begin{equation}
    \begin{gathered}
        T_c^\text{2nd} = 81.15~\text{GeV},~~ T_n^\text{2nd} = 55.09~\text{GeV},~~ T_p^\text{2nd} = 53.20~\text{GeV},~~ \xi_p^\text{2nd} = 4.491, \\
        \alpha^\text{2nd} = 0.07,~~ (\beta/H)^\text{2nd} = 561,~~ f_1^\text{sw,2nd} = 0.24~\text{mHz}, \\
        f_2^\text{sw,2nd} = 1.48~\text{mHz},~~ h^2\Omega_2^\text{sw,2nd} = 2.48\times 10^{-14},~~ \text{SNR(3y)}^\text{2nd} = 0.33\,.
    \end{gathered}
\end{equation}
In Fig.~\ref{fig:omegaGWinvBPParamVar} we show the full power spectrum of the sound-wave contribution, $h^2 \Omega^{\text{sw}}_{\text{GW}}$ as a function of the frequency $f$ for BP$\filledstar$. From left to right, we vary three parameters that enter the effective potential and GW calculations: the renormalisation scale $\mu$,  the transition temperature $T_\star$ of the second transition, and the bubble wall velocity $v_w$. The default values for these parameters in \BSMPT{} are $\mu^\text{def} = v$, $T_\star^\text{def} = T_p$, $v_w^\text{def} = 0.95$, and we vary them in the ranges $\mu \in [v/2, 2v]$, $T_\star \in [52, 55]$~GeV with the default value for the second transition, $T_p^\text{2nd} = 53.20$~GeV, lying within this interval, and $v_w \in [0.1, 1]$. We note that a change of $\mu$ and also $v_w$ implicitly changes the percolation temperature and thus the transition temperature in the bands of the left and right panels of Fig.~\ref{fig:omegaGWinvBPParamVar}.\footnote{The percolation temperature changes between $[39.84,61.30]$~GeV when varying $\mu \in [v/2, 2v]$, and between $[52.56, 53.22]$~GeV when varying $v_w \in [0.1, 1]$.} As obvious from the plots, a variation of any of these parameters within the given range can significantly change the predicted strength and frequency of the GW signal.\s

\paragraph{Impact of the Renormalisation Scale}
We start by discussing the change of the renormalisation scale $\mu$ and its impact on the GW spectrum. Note, that despite the usage of an on-shell-like renormalisation scheme at zero temperature for the first and second derivatives of the effective potential, i.e.~its VEV and mass values, we still have a renormalisation scale dependence in the zero-temperature trilinear Higgs self-couplings derived from the effective potential. At non-zero temperatures all parameters derived from the effective potential become scale-dependent, as the on-shell-like renormalisation conditions are applied at the zero-temperature minimum of the Higgs potential. They do not hold any more at the minima of the vacuum at non-zero temperature, which differ from the zero-temperature value. Since the renormalisation scale dependence is logarithmic, at $T=0$ it has only a moderate impact when varying the scale from 1/2 to 2 times its value  \cite{Biekotter:2022kgf,Biekotter:2025npc}. At non-zero temperature, depending on the change of the VEV values and the spread of the scanned scales, the impact on the quantities derived from the effective potential can be larger. This then propagates into the determination of the GW spectra. Here, a possible induced shift of the peak frequency away from the sensitive region of \texttt{LISA} can then have dramatic consequences on the SNR. \s

\begin{figure}[tp]
	\centering
	\includegraphics[width=0.6\textwidth]{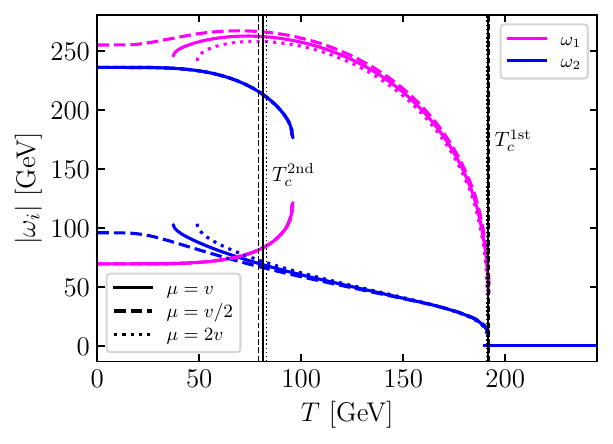}
	\caption{R2HDM inverted hierarchy, BP$\filledstar$: Evolution of the neutral EW VEVs $\omega_1$ (magenta) and $\omega_2$ (blue) as a function of the temperature for three different renormalisation scales, $\mu = \mu^{\text{def}} = v$ (full line), $\mu = v/2$ (dashed), and $\mu=2v$ (dotted). The vertical lines show the three corresponding critical temperatures for the first PT and for the second PT, respectively. The charge and CP-breaking VEVs remain zero throughout the traced temperature interval.}
\label{fig:omegaTmuvar}
\end{figure}

In Fig.~\ref{fig:omegaTmuvar}, we display for BP$\filledstar$ the evolution of the neutral EW VEVs $\omega_1$ (magenta) and $\omega_2$ (blue) as a function of the temperature for the three chosen $\mu$ values, $\mu=v$ (solid), $\mu = v/2$ (dashed), and $\mu=2v$ (dotted). The charge- and CP-breaking VEVs remain zero throughout the whole traced temperature interval. The vertical lines show the three corresponding critical temperatures for the first PT and for the second PT, respectively. As can be inferred from the figure, the spread in the critical temperature $T_c$ due to the change of the renormalisation scale $\mu$ is negligible at the first PT. At the second PT, it is larger, but still rather small, with a relative change of at most $-2.5$\%. However, while the last phase, i.e.\ the EW phase, is basically the same for all three renormalisation scales and both $\omega_{1,2}$, the intermediate phase differs when the renormalisation scale is changed. This induces a more important difference in the percolation temperatures, that differ from the default scale by $-25$\% for $\mu=v/2$ and $+15$\% for $\mu=2 v$. This has dramatic consequences for the thermal parameters $\alpha$ and $\beta/H$ and eventually on the SNR(3y), i.e.\ the SNR after three years of data acquisition. The values can be read off from Tab.~\ref{tab:BPstarScaleVariation}. \s

As can be inferred from Fig.~\ref{fig:omegaGWinvBPParamVar} (left), the variation of $\mu$ has a similar effect on the GW spectrum as the change of the PT strength $\xi_p$: a decrease of $\mu$ moves the peak of the spectrum to higher values and to lower frequencies, and vice versa for an increase of $\mu$, as already seen in Figs.~\ref{fig:omegaGWnorm} and \ref{fig:omegaGWinv}. In the end, the range of values for the SNR induced by the variation of the renormalisation scale lies between 0.045 and 5.7, with a SNR of 0.33 at our default scale. This example shows that the change of the $T_c$ due to the renormalisation scale change may be small, however, the phase evolution can change and severely affect the transition temperature $T_\star=T_p$ which then propagates through to the GW spectra inducing large changes. Had we taken instead a temperature closer to $T_c$ as transition temperature, the changes in SNR would have been much less dramatic. This also shows the importance of the discussion on the proper choice of the transition temperature. \s

The dependence of the effective potential on the parameters of the chosen point is large, however, such that even a slight shift in the input parameters can drive back the GW spectra again to the starting (the default) point, cf.~e.g.~the discussion in \cite{Biekotter:2022kgf}. Since we perform large scans over the entire R2HDM parameter space we are therefore confident to give an overall realistic picture of which GW signals can be expected in the R2HDM. \s

\renewcommand{\arraystretch}{1.4}
\begin{table}[t!]
    \centering
    \begin{tabular}{c|ccc}
        \toprule
        & & \multicolumn{2}{c}{Relative error} \\
        & $\mu = v$ & $\mu = v/2$ & $\mu = 2v$ \\ \midrule
        $\Delta \omega(T_c)$~[GeV] & $230.3$ & $+3.1\%$ & $-3.1\%$ \\
        $\Delta \omega(T_n)$~[GeV] & $239.6$ & $+1.9\%$ & $-2.7\%$ \\
        $\Delta \omega(T_p)$~[GeV] & $238.9$ & $+1.7\%$ & $-2.6\%$ \\
        $T_c$~[GeV] & $81.15$ & $-2.5\%$ & $+2.3\%$ \\
        $T_n$~[GeV] & $55.09$ & $-22\%$ & $+14\%$ \\
        $T_p$~[GeV] & $53.20$ & $-25\%$ & $+15\%$ \\
        $\Delta_{cp}^T$ & $0.3444$ & $+44\%$ & $-24\%$ \\
        $\alpha$ & $0.07101$ & $+219\%$ & $-44\%$ \\
        $\beta/H$ & $560.9$ & $-60\%$ & $+57\%$ \\
        SNR(3y) & $0.3299$ & $+1630\%$ & $-86\%$ \\
        \bottomrule
    \end{tabular}
    \caption{Thermal parameters for the second phase transition of BP$\filledstar$ from N$\to$EW and related SNR after three years of data taking, when varying the renormalisation scale $\mu$. The second column denotes the absolute values for the central scale $\mu = v$, while the third and fourth columns show the relative variation in percent when changing the scale to $\mu = v/2$ and $\mu = 2v$, respectively. The VEV difference $\Delta\omega$ between the true and false vacua at the characteristic temperatures is defined as $\Delta\omega(T_x) = \sqrt{\sum_{i} \left[\omega^{\text{true}}_i(T_x) - \omega^{\text{false}}_i(T_x)\right]^2}$ with the sum extending over all allowed directions $i = 1, 2, \CB, \CP$ and $x=c,n,p$.}
\label{tab:BPstarScaleVariation}
\end{table}
\renewcommand{\arraystretch}{1}

\paragraph{Variation of the Transition Temperature}
The change of the renormalisation scale among others affects the value of the derived transition temperature. In Fig.~\ref{fig:omegaGWinvBPParamVar} (middle) we show the change of the GW spectrum upon a variation of the transition temperature between 52 and 55 GeV, with $T_* \equiv T_p = 53.2$~GeV being the BP$\filledstar$ value. The variation of the transition temperature $T_\star$ changes both the height and the position of the peak, with higher $T_\star$ moving the peak towards smaller frequencies while increasing its height, and vice versa for smaller $T_\star$. The value of the SNR consequently changes between $2.5\times 10^{-5}$ and $0.36$. These results should nevertheless be taken with care, as the transition temperature was varied manually, despite being interdependent with the vacuum evolution, the derived VEV values, the computed thermal quantities, etc., as clearly illustrated by the discussion in the previous paragraph.
While we computed the thermal quantities at the changed transition temperature, the interdependence with the vacuum evolution is obviously not taken into account. Since, the (consistently derived) nucleation temperature $T_n=55.1$~GeV ranges at the upper bound of our variation interval, however, and since there is also some discussion on which transition temperature should be used, this variation gives us insight on the related uncertainties. \s

\paragraph{Impact of the Wall Velocity}
Finally, we show in Fig.~\ref{fig:omegaGWinvBPParamVar} (right) the impact of the wall velocity. The variation of the wall velocity $v_w$ changes both the position of the peak and its height, with the GW spectrum peaking at the highest (lowest) frequencies found for $v_w \approx 0.6$ ($v_w \approx 1$). We furthermore found for this BP$\filledstar$ that towards smaller $v_w < 0.6$, the height of the peak is continuously decreasing, such that within the varied range of $v_w$, we obtain SNR values between $6.2\times 10^{-4}$ and 2.5. This behaviour has to be taken with caution, however, as we set the wall velocity by hand in the input file and do not consistently solve the coupled system
of scalar field equations, hydrodynamic conservation laws for the plasma, and the 
Boltzmann equations for the distribution functions of all relevant species, nor do we consider any out-of-equilibrium corrections. \s

In summary, we find that the strength and the frequency of the GW spectrum strongly depend on the thermal parameters. Their values are sensitive to the effective potential used in the computation. However, also a self-consistent derivation of the thermal parameters taking into account the plasma properties and out-of-equilibrium effects is essential. As mentioned above, due to our extensive parameter scan across the whole allowed model space, we still get an overall realistic picture of the GWs in the R2HDM. For future comparisons with data from space-based GW observatories such as the \texttt{LISA} experiment and conclusions drawn on the specific values of the input parameters and hence the underlying model, however, it is of the utmost importance to reduce the uncertainties, refine the assumptions on early universe dynamics and derive the thermal parameters in a self-consistent way taking into account the plasma dynamics, together with providing a realistic derivation of the remaining uncertainties.\footnote{Ab initio calculations on the lattice take this approach, require, however, a lot of computational time and are hence not of practical use for large parameter scans across a variety of BSM models, such that we have an interest in improving the predictions applying perturbation theory as much as possible.} We note that a self-consistent derivation of the wall velocity, cf.~e.g.~\cite{Ekstedt:2024fyq,Branchina:2025jou},  taking into account out-of-equilibrium effects \cite{DeCurtis:2022hlx,DeCurtis:2023hil,DeCurtis:2024hvh,Branchina:2025adj} has recently been implemented in \texttt{BSMPTv3} and will be published soon. 
\section{Summary and Conclusions}
\label{s:conclusions}
\renewcommand{\arraystretch}{1.4}

The understanding of the Higgs vacuum is of the utmost importance for our understanding of the mass generation of particles, of the evolution of the universe, and to answer the question why there is more matter than antimatter. The dynamical generation of the matter-antimatter asymmetry during the electroweak epoch requires strong first-order PTs which can only be realised in extended Higgs sectors beyond the SM. The GW spectra generated in such PTs not only provide us with a cosmological tool giving  evidence for BSM physics, eventually they may also allow us to draw conclusions on the underlying physics model. The challenge is to make solid predictions for the GW spectrum starting out from zero-temperature physics of today. In this contribution, we presented a thorough investigation of the well-motivated and experimentally tested R2HDM with in principle four possible vacuum directions at non-zero temperature. Applying all relevant theoretical and experimental constraints, we derived the phase histories with strong first-order PTs that can be realised in this model. We investigated in detail the related phenomenology that can be tested at present and future high-energy colliders and the GW spectra generated by the strong PTs. We discussed the uncertainties involved in the derivation of the GWs and examined the impact of the uncertainties related to some key thermal parameters. The analysis presented in this paper  gives us a global picture of what phase histories and GW signals we can expect to find  in case the R2HDM is realised, and how we should design our collider investigations in order to learn more about the cosmological evolution of the Higgs vacuum and the related new physics model. We summarise in the following our key observations. \s

Taking into account all relevant theoretical and experimental constraints, we derived for the R2HDM the possible vacuum evolution from the early universe to today and the related GW spectrum that can be induced by strong first-order PTs. We found PTs that involve normal EW vacua ($\omega_{1,2}\ne 0$), but also intermediate charge-breaking ($\omega_{\text{CB}}\ne 0$) or CP-violating ($\omega_{\text{CP}}\ne 0$) phases at non-zero temperature. Charge-breaking vacua were found to be realised only for the normal hierarchy ($m_{H_\sm} = m_{H_1} < m_{H_2}$) and always involve non-symmetry restoration at high temperature. CP-violating phases were only realised for the inverted hierarchy ($m_{H_\sm} = m_{H_2} > m_{H_1}$). We found that the phase histories that can be realised for strong first-order PTs, i.e.~$\xi_p > 1$, via CP-even neutral vacua 
($\omega_{1,2}$) starting from a symmetric vacuum in the early universe and ending in the electroweak minimum of today can be given by one-step but also multi-step PTs, both for the normal hierarchy and for the inverted hierarchy. For the multi-step PTs, the strongest PTs are always realised in the last transition. 
For strong first-order PT histories with $\omega_{1,2} \ne 0$, we analysed the related scalar mass spectra and trilinear Higgs self-couplings, in order to deduce, whenever possible, characteristic observables and signatures that could be tested at present and future colliders. We derived the GW spectra and the SNR expected at \texttt{LISA}. 
\s

More specifically, we found for the \emph{normal hierarchy} one-, two-, and three-step PTs. The values of the thermal parameters released latent heat ($\alpha$) and PT strength at the percolation temperature ($\xi_p$) decrease with increasing step number. The largest values are hence given by the one-step PTs. The largest values for the released latent heat and PT strength are found to be $\alpha = 0.52$ and $\xi_p=6.1$, respectively. The transition temperature can here go down to $T_\star \equiv T_p = 40$~GeV. We also found very few points with four-step PTs, which we did not discuss further. 
In the \emph{inverted hierarchy}, we found one- and two-step PTs. Here, the thermal parameters $\alpha$ and $\xi_p$ increase with increasing step number. We find the largest values of the released latent heat and PT strength in the second transition of the two-step PTs given by $\alpha=2.5$, $\xi_p=11$. Here, the transition temperature can be as low as $T_\star=22$~GeV. 
We also found very few points with three-step PTs, which we did not discuss further. \s

The characteristics of the mass distributions and $\tan\beta$ values of these PTs are:

\vspace*{-0.2cm}
\begin{itemize}
\item[--] Normal hierarchy: \\
\emph{One-step PTs:} Points with a strong first-order PTs have rather light masses, i.e.~$m_A \lesssim 750$~GeV as well as $m_{H_2}, m_{H^\pm} \lesssim 650$~GeV, and $m_{H_2} \approx 190$~GeV for the strongest PTs. 
We find two specific mass regions centered around $\xi_p \gtrsim 3$: \\[0.1cm]
$(i)$ $m_{H_2} \approx m_{H^\pm} \approx 190$~GeV, and $m_A - m_{H\pm}/m_{H_2} \approx 360$~GeV. This region is characterised by $A \to Z H_{1,2}$ and $A\to W^\pm H^\mp$ decays. \\[0.1cm]
$(ii)$ $m_A \approx m_{H^\pm} \approx 420$~GeV and $m_A/m_{H^\pm} - m_{H_2} \approx 220$~GeV. Here, $A \to Z H_{1,2}$ and $H^\pm \to W^\pm H_{1,2}$ decays are characteristic. \\[0.2cm]
\emph{Two-step PTs:} Strong first-order PTs are found for 
$m_{A,H^\pm} \in [230, 340]$~GeV and $m_{H_2} \in [190, 270]$~GeV.
We find one specific mass region centered around the largest $\xi_p \gtrsim 3$: \\[0.1cm]
$m_A \approx m_{H^\pm} \approx 300$~GeV and $m_{A}/m_{H^\pm} - m_{H_2} \approx m_Z$. The possible decays characteristic for this region are $A \to ZH_{1,2}$ and $H^\pm \to W^\pm H_{1,2}$ decays.  
\\[0.2cm]
\emph{Three-step PTs:} Here, the masses are distributed as $m_{A,H^\pm} \in [230, 330]$~GeV and $m_{H_2} \in [240, 270]$~GeV. \\[0.2cm]
The values of $\tan\beta$ for strong first-order one-step PTs are $\tan\beta \in [0.8, 65]$, and for multi-step PTs they lie in the interval $\tan\beta \in [20, 50]$.
\item[--] \emph{Inverted hierarchy:} \\
\emph{One-step PTs:} 
Regions with strong first-order PTs are characterised by $m_{H^\pm} \approx m_{A} \in [220, 440]$~GeV, and the PT strength increases with $m_A \approx m_{H^\pm}$. Due to the existing mass gaps to $m_{H_{1,2}} \lesssim 125$~GeV, the decays $A \to ZH_{1,2}$ and $H^\pm \to W^\pm H_{1,2}$ are truly characteristic features for this phase history. 
\\[0.1cm]
\emph{Two-step PTs:} Points with strong PTs now extend to lower masses with $60$~GeV~$\lesssim m_A \lesssim 350$~GeV and $120$~GeV~$\lesssim m_{H^\pm} \lesssim 350$~GeV, and we also have scenarios with $m_A \approx m_{H_1}$. Possible decay signatures are $A\to Z H_{1,2}$, $A \to W^\pm H^\mp$ and $H^\pm \to W^\pm H_{1,2}$.\\[0.1cm]
For strong first-order PTs the $\tan\beta$ values are restricted to values below about 11 for one-step phase transitions, whereas for two-step phase transitions $\tan\beta$ can take values of up to about 34.
\end{itemize}
As for the trilinear Higgs self-couplings, we found that in the \emph{normal hierarchy} the SM-like trilinear Higgs self-coupling is enhanced for all three step-types of PTs. More specifically, we have $1.1 \lesssim \kappa_\lambda \lesssim 2.1$ for one-step PTs, with larger $\xi_p$ values leading to larger enhancements.  In the two- and three-step PTs no correlation between the enhancement factor and $\xi_p$ was observed. In the two-step PTs we have $1.15 \lesssim \kappa_\lambda \lesssim 1.6$ and in the three-step PTs,
$1.2 \lesssim \kappa_\lambda \lesssim 1.5$. In all three cases, larger mass gaps $m_A - m_{H_2}$ are correlated with larger trilinear Higgs self-coupling enhancements. 
Investigating the potential of future colliders, to constrain these scenarios further through coupling measurements, we found that increased precision in the Higgs-to-$VV$ ($V\equiv Z,W^\pm$) coupling does not further constrain the trilinear Higgs self-coupling associated with a strong first-order one-step PT. In the two- and three-step PTs the future increased precision in Higgs-to-$VV$ and in the trilinear Higgs self-coupling has the potential to exclude these scenarios. \s

In the \emph{inverted hierarchy}, one-step PTs are characterised by $1.1 \lesssim \kappa_\lambda \lesssim 2$. Larger enhancements of the trilinear Higgs self-couplings are correlated with larger mass gaps $m_A - m_{H_1}$. This is also the case for the two-step PTs. However, here the mass gap can also become zero. We found $1.0 \lesssim \kappa_\lambda \lesssim 1.7$. This means that we can have strong PTs also for SM-like trilinear Higgs self-couplings, which is in striking contrast to all other analysed PTs. Hence, we will be able to distinguish between one- and two-step PTs with increased precision on the trilinear Higgs self-coupling. 
In one-step PTs, the future increased precision in the Higgs-to-$VV$ coupling will not further constrain the trilinear Higgs self-coupling associated with a strong first-order PT. In the two-step PTs, 
there is also a slight correlation between the Higgs-to-$VV$ coupling and the trilinear Higgs self-couplings of the strong first-order PTs such that the increase in precision in the future single Higgs measurements will be able to inform about the nature of the PT. \s

As for the GW spectra, we found that they are all dominated by the sound waves. In the normal hierarchy, the largest signal-to-noise ratio at \texttt{LISA} after three years of data taking is found to be as large as SNR(3y)~$= 17$ for the one-step PTs. In the inverted hierarchy it can go up to SNR(3y)~$= 88$ for the second step in the two-step PTs.
\s

The derivation of the GW spectra related to strong PTs in the vacuum evolution of extended Higgs sectors, starting from a specific BSM model at $T=0$, is accompanied by numerous uncertainties and approximations. In order to assess the potential of deriving the underlying model parameters from the GW signal, the uncertainties must be discussed in detail and quantified. We investigated the impact of the renormalisation scale in the effective potential, of the choice of the transition temperature and of the wall velocity on the GW spectra. Small induced variations by the change of these parameters of the derived thermal parameters get amplified in the GW spectra leading to significant changes in the SNR due to shifts in the peak amplitude and peak frequency. As the change of the input parameters such as the physical masses can significantly alter the GW spectra, the performed scan across the whole parameter space of the R2HDM gives a realistic picture, however, of what signals can be expected. To be able to pin down the specific related underlying parameter configuration or to be able to distinguish between different BSM scenarios, however, requires improvement on all steps along the GW spectra derivation starting from the zero-temperature model, and crucially, the discussion and quantification of the involved uncertainties, as well as the combination with all available information from high-energy physics, astrophysics, and cosmology.  


\section*{Acknowledgements}

The research of L.B.\ is supported by the Swiss National Science Foundation (SNSF). The research of C.B., R.B., and M.M.\ is supported by the Deutsche Forschungsgemeinschaft (DFG, German Research Foundation) under grant 396021762 - TRR 257.
J.V.\ and R.S.\ acknowledge financial support from the Portuguese Foundation for Science and Technology (FCT) under the contracts: UID/PRR2/00618/2025 (\nolinkurl{https://doi.org/10.54499/UID/PRR2/00618/2025}), UID/PRR/00618/2025 (\nolinkurl{https://doi.org/10.54499/UID/PRR/00618/2025}), and \linebreak UID/00618/2025 (\nolinkurl{https://doi.org/10.54499/UID/00618/2025}). 
J.V.\ is also supported by FCT under contract PRT/BD/154191/2022 (\nolinkurl{https://doi.org/10.54499/PRT/BD/154191/2022}).


\appendix

\section{Determination of the \texttt{LISA} Sensitivity Band}\label{app:band}
We specify in the following how the grey \texttt{LISA} sensitivity bands shown in Figs.~\ref{fig:omegaGWnorm} and \ref{fig:omegaGWinv} are obtained. The SNR with a data acquisition time $\mathcal T$ is generally defined as the overlap between the expected GW signal power spectrum $h^2\Omega_\text{GW}(f)$ for a specific frequency $f$ and the experimental sensitivity,
\begin{equation}\label{eq:SNRGWs}
    \text{SNR} = \sqrt{\mathcal{T}\int_{f_\text{min}}^{f_\text{max}} df \left[\frac{h^2\Omega_\text{GW}(f)}{h^2\Omega_\text{Sens}(f)}\right]^2} \,,
\end{equation}
with the interval $[f_\text{min}, f_\text{max}]$ denoting the frequency range in which the experiment is sensitive to a signal, and the GW power spectrum $h^2\Omega_\text{GW}(f)$ being dominated by sound-wave contributions for our parameter sample, $h^2\Omega_\text{GW}(f) \approx h^2\Omega_\text{sw}(f)$, the latter of which are described by a double broken power law \cite{Caprini:2024hue} as
\begin{equation}\label{eq:DBPLswspectrum}
    h^2\Omega_\text{sw}(f) = h^2\Omega_2^\text{sw} \times S_2(f; f_1^\text{sw}, f_2^\text{sw})\,.
\end{equation}
$S_2(f; f_1^\text{sw}, f_2^\text{sw})$ denotes the shape function (see e.g.~\cite{Basler:2024aaf} for its definition) which is normalised such that 
\begin{equation}
    S_2(f = f_2^\text{sw}; f_1^\text{sw}, f_2^\text{sw}) = 1 \quad \mbox{and consequently} \quad h^2\Omega_\text{sw}(f = f_2^\text{sw}) = h^2\Omega_2^\text{sw}\;.
\end{equation}
The characteristic frequency breaks $f_1^\text{sw}$ and $f_2^\text{sw}$ as well as the characteristic amplitude $h^2\Omega_2^\text{sw}$ are calculated by \texttt{BSMPTv3} from the thermal parameters.
We now want to visualise in a two-dimensional plot of $h^2\Omega_2^\text{sw}$ versus $f_2^\text{sw}$ for which parameter pairs $(f_2^\text{sw}, h^2\Omega_2^\text{sw})$ we obtain a SNR larger than some threshold SNR$_\text{thresh}$. We thus insert Eq.~\eqref{eq:DBPLswspectrum} into Eq.~\eqref{eq:SNRGWs} and solve for $h^2\Omega_2^\text{sw}$:
\begin{equation}\label{eq:omega2threshold}
    \text{SNR} \ge \text{SNR}_\text{thresh} \quad \Leftrightarrow \quad h^2\Omega_2^\text{sw} \ge \frac{\text{SNR}_\text{thresh}}{\sqrt{\mathcal{T}\int_{f_\text{min}}^{f_\text{max}} df \left[\frac{S_2(f; f_1^\text{sw}, f_2^\text{sw})}{h^2\Omega_\text{Sens}(f)}\right]^2}}\,.
\end{equation}
We denote the smallest possible value of $h^2\Omega_2^\text{sw}$ which is needed to obtain the threshold value SNR$_\text{thresh}$ as $h^2\Omega_{2,\text{thresh}}^\text{sw}$, corresponding to the equality in Eq.~\eqref{eq:omega2threshold}. 
Since the right-hand side of Eq.~\eqref{eq:omega2threshold}  still depends on $f_1^\text{sw}$ and $f_2^\text{sw}$, we can consider the threshold $h^2\Omega_{2,\text{thresh}}^\text{sw} = h^2\Omega_{2,\text{thresh}}^\text{sw}(f_1^\text{sw}, f_2^\text{sw})$ as a function of these two frequency breaks.
Thus, since we are interested in plotting only the $f_2^\text{sw}$ dependence of the parameter points, we do the following:
\begin{enumerate}
    \item We vary $f_2^\text{sw}$ in the range $[f_\text{min}, f_\text{max}] = [10^{-6}, 10]$~Hz.
    \item We evaluate, for each value of $f_2^\text{sw}$, Eq.~\eqref{eq:omega2threshold} for sufficiently many values of $f_1^\text{sw}$ in its allowed range (in practice, we choose the range as $f_1^\text{sw} \in [10^{-20}~\text{Hz}, f_2^\text{sw}]$, since $f_1^\text{sw} < f_2^\text{sw}$).
    \item We determine, for each value of $f_2^\text{sw}$, the minimum and maximum values of $h^2\Omega_{2,\text{thresh}}^\text{sw}$ as obtained from the variation of $f_1^\text{sw}$ in step 2.
\end{enumerate}
The last step results in a sensitivity band depending on the value of $f_2^\text{sw}$ which can be interpreted as follows. If the $(f_2^\text{sw}, h^2\Omega_2^\text{sw})$ pair of a parameter point lies above (below) the band, that is above the maximum value (below the minimum value) of $h^2\Omega_{2,\text{thresh}}^\text{sw}$, then the point is ensured to have an  SNR~$>$~SNR$_\text{thresh}$ ($<$~SNR$_\text{thresh}$). In case the value pair is inside the band, whether the SNR is above or below the threshold depends on the precise value of $f_1^\text{sw}$ of the parameter point and cannot directly be read off from the corresponding plot.
\s

Since most of the points in our scan lie in the region where the band is very narrow, see Figs.~\ref{fig:omegaGWnorm} and \ref{fig:omegaGWinv}, it can be considered as a quick quantifier if a given point has a sufficiently large SNR. For concreteness, in the sensitivity band shown in Figs.~\ref{fig:omegaGWnorm} and \ref{fig:omegaGWinv}, we choose a SNR threshold value of SNR$_\text{thresh} = 1$ and a data acquisition time of $\mathcal T = 3$~years.


\printbibliography

@article{Aiko:2025tbk,
    author = "Aiko, Masashi and Endo, Motoi and Kanemura, Shinya and Mura, Yushi",
    title = "{Electroweak baryogenesis in 2HDM without EDM cancellation}",
    eprint = "2504.07705",
    archivePrefix = "arXiv",
    primaryClass = "hep-ph",
    reportNumber = "KEK-TH-2695, OU-HET-1268",
    doi = "10.1007/JHEP07(2025)236",
    journal = "JHEP",
    volume = "07",
    pages = "236",
    year = "2025"
}

@article{Garbrecht:2025jgy,
    author = {Garbrecht, Bj{\"o}rn and Wang, Edward},
    title = "{Baryogenesis and EDMs in the 2HDM+CS}",
    eprint = "2512.17695",
    archivePrefix = "arXiv",
    primaryClass = "hep-ph",
    month = "12",
    year = "2025"
}

@article{Biekotter:2023eil,
    author = {Biek{\"o}tter, Thomas and Heinemeyer, Sven and No, Jose Miguel and Radchenko, Kateryna and Romacho, Mar{\'\i}a Olalla Olea and Weiglein, Georg},
    title = "{First shot of the smoking gun: probing the electroweak phase transition in the 2HDM with novel searches for A {\textrightarrow} ZH in $ {\ell}^{+}{\ell}^{-}t\overline{t} $ and $ \nu \nu b\overline{b} $ final states}",
    eprint = "2309.17431",
    archivePrefix = "arXiv",
    primaryClass = "hep-ph",
    reportNumber = "DESY-23-144, KA-TP-19-2023, IFT-CSIC-120",
    doi = "10.1007/JHEP01(2024)107",
    journal = "JHEP",
    volume = "01",
    pages = "107",
    year = "2024"
}

@article{Matuszak:2026xsz,
    author = "Matuszak, Jonas and Tasillo, Carlo",
    title = "{TransitionListener v2.0 -- Robust gravitational wave predictions for cosmological phase transitions}",
    eprint = "2605.15259",
    archivePrefix = "arXiv",
    primaryClass = "hep-ph",
    month = "5",
    year = "2026"
}

@article{Barni:2025ifb,
    author = "Barni, Giulio",
    title = "{Electroweak Baryogenesis with BARYONET: a self-contained review of the WKB approach}",
    eprint = "2510.21915",
    archivePrefix = "arXiv",
    primaryClass = "hep-ph",
    doi = "10.21468/SciPostPhys.21.2.032",
    journal = "SciPost Phys.",
    volume = "21",
    pages = "032",
    year = "2026"
}

@article{Athron:2024xrh,
    author = "Athron, Peter and Balazs, Csaba and Fowlie, Andrew and Morris, Lachlan and Searle, William and Xiao, Yang and Zhang, Yang",
    title = "{PhaseTracer2: from the effective potential to gravitational waves}",
    eprint = "2412.04881",
    archivePrefix = "arXiv",
    primaryClass = "astro-ph.CO",
    doi = "10.1140/epjc/s10052-025-14258-y",
    journal = "Eur. Phys. J. C",
    volume = "85",
    number = "5",
    pages = "559",
    year = "2025"
}

@article{Boto:2026bmz,
    author = {Boto, Rafael and Dao, Thi Nhung and Egle, Felix and Elyaouti, Karim and Gabelmann, Martin and M{\"u}hlleitner, Margarete and Plotnikov, Johann},
    title = "{NMSSMScanner: Efficient Scans in the NMSSM Parameter Space Proof of Concept}",
    eprint = "2604.25009",
    archivePrefix = "arXiv",
    primaryClass = "hep-ph",
    reportNumber = "DESY-26-056, FR-PHENO-2026-008, KA-TP-09-2026",
    month = "4",
    year = "2026"
}

@article{Muhlleitner:2016mzt,
    author = "Muhlleitner, Margarete and Sampaio, Marco O. P. and Santos, Rui and Wittbrodt, Jonas",
    title = "{The N2HDM under Theoretical and Experimental Scrutiny}",
    eprint = "1612.01309",
    archivePrefix = "arXiv",
    primaryClass = "hep-ph",
    doi = "10.1007/JHEP03(2017)094",
    journal = "JHEP",
    volume = "03",
    pages = "094",
    year = "2017"
}

@article{CMS:2022dwd,
    author = "Tumasyan, Armen and others",
    collaboration = "CMS",
    title = "{A portrait of the Higgs boson by the CMS experiment ten years after the discovery.}",
    eprint = "2207.00043",
    archivePrefix = "arXiv",
    primaryClass = "hep-ex",
    reportNumber = "CMS-HIG-22-001, CERN-EP-2022-039",
    doi = "10.1038/s41586-022-04892-x",
    journal = "Nature",
    volume = "607",
    number = "7917",
    pages = "60--68",
    year = "2022",
    note = "[Erratum: Nature 623, (2023)]"
}

@article{ATLAS:2022vkf,
    author = "Aad, Georges and others",
    collaboration = "ATLAS",
    title = "{A detailed map of Higgs boson interactions by the ATLAS experiment ten years after the discovery}",
    eprint = "2207.00092",
    archivePrefix = "arXiv",
    primaryClass = "hep-ex",
    reportNumber = "CERN-EP-2022-057",
    doi = "10.1038/s41586-022-04893-w",
    journal = "Nature",
    volume = "607",
    number = "7917",
    pages = "52--59",
    year = "2022",
    note = "[Erratum: Nature 612, E24 (2022)]"
}

@article{Dorsch:2016nrg,
    author = "Dorsch, G. C. and Huber, S. J. and Konstandin, T. and No, J. M.",
    title = "{A Second Higgs Doublet in the Early Universe: Baryogenesis and Gravitational Waves}",
    eprint = "1611.05874",
    archivePrefix = "arXiv",
    primaryClass = "hep-ph",
    reportNumber = "DESY-16-213",
    doi = "10.1088/1475-7516/2017/05/052",
    journal = "JCAP",
    volume = "05",
    pages = "052",
    year = "2017"
}

@article{Dorsch:2014qja,
    author = "Dorsch, G. C. and Huber, S. J. and Mimasu, K. and No, J. M.",
    title = "{Echoes of the Electroweak Phase Transition: Discovering a second Higgs doublet through $A_0 \rightarrow ZH_0$}",
    eprint = "1405.5537",
    archivePrefix = "arXiv",
    primaryClass = "hep-ph",
    doi = "10.1103/PhysRevLett.113.211802",
    journal = "Phys. Rev. Lett.",
    volume = "113",
    number = "21",
    pages = "211802",
    year = "2014"
}

@article{Curtin:2022ovx,
    author = "Curtin, David and Roy, Jyotirmoy and White, Graham",
    title = "{Gravitational waves and tadpole resummation: Efficient and easy convergence of finite temperature QFT}",
    eprint = "2211.08218",
    archivePrefix = "arXiv",
    primaryClass = "hep-ph",
    doi = "10.1103/PhysRevD.109.116001",
    journal = "Phys. Rev. D",
    volume = "109",
    number = "11",
    pages = "116001",
    year = "2024"
}

@article{Curtin:2016urg,
    author = "Curtin, David and Meade, Patrick and Ramani, Harikrishnan",
    title = "{Thermal Resummation and Phase Transitions}",
    eprint = "1612.00466",
    archivePrefix = "arXiv",
    primaryClass = "hep-ph",
    reportNumber = "YITP-2016-48",
    doi = "10.1140/epjc/s10052-018-6268-0",
    journal = "Eur. Phys. J. C",
    volume = "78",
    number = "9",
    pages = "787",
    year = "2018"
}

@article{Coleman:1973jx,
    author = "Coleman, Sidney R. and Weinberg, Erick J.",
    title = "{Radiative Corrections as the Origin of Spontaneous Symmetry Breaking}",
    doi = "10.1103/PhysRevD.7.1888",
    journal = "Phys. Rev. D",
    volume = "7",
    pages = "1888--1910",
    year = "1973"
}

@article{Carrington:1991hz,
    author = "Carrington, M. E.",
    title = "{The Effective potential at finite temperature in the Standard Model}",
    reportNumber = "TPI-MINN-91-48-T-REV, TPI-MINN-91-48-T",
    doi = "10.1103/PhysRevD.45.2933",
    journal = "Phys. Rev. D",
    volume = "45",
    pages = "2933--2944",
    year = "1992"
}

@article{Caprini:2024hue,
    author = "Caprini, Chiara and Jinno, Ryusuke and Lewicki, Marek and Madge, Eric and Merchand, Marco and Nardini, Germano and Pieroni, Mauro and Roper Pol, Alberto and Vaskonen, Ville",
    collaboration = "LISA Cosmology Working Group",
    title = "{Gravitational waves from first-order phase transitions in LISA: reconstruction pipeline and physics interpretation}",
    eprint = "2403.03723",
    archivePrefix = "arXiv",
    primaryClass = "astro-ph.CO",
    reportNumber = "LISA-COSWG-24-01, CERN-TH-2024-029",
    doi = "10.1088/1475-7516/2024/10/020",
    journal = "JCAP",
    volume = "10",
    pages = "020",
    year = "2024"
}

@article{Basler:2016obg,
    author = "Basler, P. and Krause, M. and Muhlleitner, M. and Wittbrodt, J. and Wlotzka, A.",
    title = "{Strong First Order Electroweak Phase Transition in the CP-Conserving 2HDM Revisited}",
    eprint = "1612.04086",
    archivePrefix = "arXiv",
    primaryClass = "hep-ph",
    doi = "10.1007/JHEP02(2017)121",
    journal = "JHEP",
    volume = "02",
    pages = "121",
    year = "2017"
}

@article{Basler:2024aaf,
    author = {Basler, Philipp and Biermann, Lisa and M{\"u}hlleitner, Margarete and M{\"u}ller, Jonas and Santos, Rui and Viana, Jo{\~a}o},
    title = "{BSMPT v3 a tool for phase transitions and primordial gravitational waves in extended Higgs sectors}",
    eprint = "2404.19037",
    archivePrefix = "arXiv",
    primaryClass = "hep-ph",
    reportNumber = "KA-TP-08-2024",
    doi = "10.1016/j.cpc.2025.109766",
    journal = "Comput. Phys. Commun.",
    volume = "316",
    pages = "109766",
    year = "2025"
}

@article{Linde:1980ts,
    author = "Linde, Andrei D.",
    title = "{Infrared Problem in Thermodynamics of the Yang-Mills Gas}",
    reportNumber = "LEBEDEV-80-106",
    doi = "10.1016/0370-2693(80)90769-8",
    journal = "Phys. Lett. B",
    volume = "96",
    pages = "289--292",
    year = "1980"
}

@article{vandeVis:2025efm,
    author = "van de Vis, Jorinde and de Vries, Jordy and Postma, Marieke",
    title = "{Bubble trouble: A review on electroweak baryogenesis}",
    eprint = "2508.09989",
    archivePrefix = "arXiv",
    primaryClass = "hep-ph",
    reportNumber = "CERN-TH-2025-161, Nikhef 2025-012",
    doi = "10.1016/j.ppnp.2026.104244",
    journal = "Prog. Part. Nucl. Phys.",
    volume = "150",
    pages = "104244",
    year = "2026"
}

@article{Bahl:2024ykv,
    author = "Bahl, Henning and Carena, Marcela and Ireland, Aurora and Wagner, Carlos E. M.",
    title = "{Improved thermal resummation for multi-field potentials}",
    eprint = "2404.12439",
    archivePrefix = "arXiv",
    primaryClass = "hep-ph",
    reportNumber = "EFI 24-2, FERMILAB-PUB-24-0040-T",
    doi = "10.1007/JHEP09(2024)153",
    journal = "JHEP",
    volume = "09",
    pages = "153",
    year = "2024"
}

@article{Biekotter:2025npc,
    author = {Biek{\"o}tter, Thomas and Dashko, Andrii and L{\"o}schner, Maximilian and Weiglein, Georg},
    title = "{Perturbative aspects of the electroweak phase transition with a complex singlet and implications for gravitational wave predictions}",
    eprint = "2511.14831",
    archivePrefix = "arXiv",
    primaryClass = "hep-ph",
    reportNumber = "DESY-25-131, IFT-UAM/CSIC-25-104",
    month = "11",
    year = "2025"
}

@article{Bittar:2025lcr,
    author = "Bittar, Pedro and Roy, Subhojit and Wagner, Carlos E. M.",
    title = "{Self consistent thermal resummation: a case study of the phase transition in 2HDM}",
    eprint = "2504.02024",
    archivePrefix = "arXiv",
    primaryClass = "hep-ph",
    reportNumber = "EFI-25-3",
    doi = "10.1007/JHEP12(2025)021",
    journal = "JHEP",
    volume = "12",
    pages = "021",
    year = "2025"
}

@article{Kainulainen:2019kyp,
    author = "Kainulainen, Kimmo and Keus, Venus and Niemi, Lauri and Rummukainen, Kari and Tenkanen, Tuomas V. I. and Vaskonen, Ville",
    title = "{On the validity of perturbative studies of the electroweak phase transition in the Two Higgs Doublet model}",
    eprint = "1904.01329",
    archivePrefix = "arXiv",
    primaryClass = "hep-ph",
    doi = "10.1007/JHEP06(2019)075",
    journal = "JHEP",
    volume = "06",
    pages = "075",
    year = "2019"
}

@article{Ginsparg:1980ef,
    author = "Ginsparg, Paul H.",
    title = "{First Order and Second Order Phase Transitions in Gauge Theories at Finite Temperature}",
    reportNumber = "SACLAY-DPh-T 80/27",
    doi = "10.1016/0550-3213(80)90418-6",
    journal = "Nucl. Phys. B",
    volume = "170",
    pages = "388--408",
    year = "1980"
}

@article{Appelquist:1981vg,
    author = "Appelquist, Thomas and Pisarski, Robert D.",
    title = "{High-Temperature Yang-Mills Theories and Three-Dimensional Quantum Chromodynamics}",
    reportNumber = "Print-81-0020 (YALE), YTP-81-01, COO-3075-203",
    doi = "10.1103/PhysRevD.23.2305",
    journal = "Phys. Rev. D",
    volume = "23",
    pages = "2305",
    year = "1981"
}

@article{Chiang:2017nmu,
    author = "Chiang, Cheng-Wei and Ramsey-Musolf, Michael J. and Senaha, Eibun",
    title = "{Standard Model with a Complex Scalar Singlet: Cosmological Implications and Theoretical Considerations}",
    eprint = "1707.09960",
    archivePrefix = "arXiv",
    primaryClass = "hep-ph",
    reportNumber = "NCTS-PH-1724, ACFI-T17-16",
    doi = "10.1103/PhysRevD.97.015005",
    journal = "Phys. Rev. D",
    volume = "97",
    number = "1",
    pages = "015005",
    year = "2018"
}

@article{Hirvonen:2021zej,
    author = {Hirvonen, Joonas and L{\"o}fgren, Johan and Ramsey-Musolf, Michael J. and Schicho, Philipp and Tenkanen, Tuomas V. I.},
    title = "{Computing the gauge-invariant bubble nucleation rate in finite temperature effective field theory}",
    eprint = "2112.08912",
    archivePrefix = "arXiv",
    primaryClass = "hep-ph",
    reportNumber = "ACFI-T21-16, HIP-2021-45/TH, NORDITA 2021-111",
    doi = "10.1007/JHEP07(2022)135",
    journal = "JHEP",
    volume = "07",
    pages = "135",
    year = "2022"
}

@article{Arco:2025pgx,
    author = {Arco, F. and Heinemeyer, S. and M{\"u}hlleitner, M.},
    title = "{Large one-loop effects of BSM triple Higgs couplings on double Higgs production at e$^{+}$e$^{-}$ colliders}",
    eprint = "2505.02947",
    archivePrefix = "arXiv",
    primaryClass = "hep-ph",
    reportNumber = "DESY-25-073, IFT--UAM/CSIC-25-016, KA-TP-13-2025",
    doi = "10.1007/JHEP01(2026)160",
    journal = "JHEP",
    volume = "01",
    pages = "160",
    year = "2026"
}

@article{Baglio:2012np,
    author = {Baglio, J. and Djouadi, A. and Gr{\"o}ber, R. and M{\"u}hlleitner, M. M. and Quevillon, J. and Spira, M.},
    title = "{The measurement of the Higgs self-coupling at the LHC: theoretical status}",
    eprint = "1212.5581",
    archivePrefix = "arXiv",
    primaryClass = "hep-ph",
    reportNumber = "KA-TP-44-2012, SFB-CPP-12-102, LPT-ORSAY-12-124, PSI-PR-12-10",
    doi = "10.1007/JHEP04(2013)151",
    journal = "JHEP",
    volume = "04",
    pages = "151",
    year = "2013"
}

@article{DiMicco:2019ngk,
    author = "Alison, J. and others",
    editor = "Di Micco, Biagio and Gouzevitch, Maxime and Mazzitelli, Javier and Vernieri, Caterina",
    title = "{Higgs boson potential at colliders: Status and perspectives}",
    eprint = "1910.00012",
    archivePrefix = "arXiv",
    primaryClass = "hep-ph",
    reportNumber = "FERMILAB-CONF-19-468-E-T, LHCXSWG-2019-005",
    doi = "10.1016/j.revip.2020.100045",
    journal = "Rev. Phys.",
    volume = "5",
    pages = "100045",
    year = "2020"
}

@article{Djouadi:1999rca,
    author = "Djouadi, A. and Kilian, W. and Muhlleitner, M. and Zerwas, P. M.",
    title = "{Production of neutral Higgs boson pairs at LHC}",
    eprint = "hep-ph/9904287",
    archivePrefix = "arXiv",
    reportNumber = "DESY-99-033, TTP-99-17, PM-99-21",
    doi = "10.1007/s100529900083",
    journal = "Eur. Phys. J. C",
    volume = "10",
    pages = "45--49",
    year = "1999"
}

@article{Dolan:1973qd,
    author = "Dolan, L. and Jackiw, R.",
    title = "{Symmetry Behavior at Finite Temperature}",
    reportNumber = "MIT-CTP-406",
    doi = "10.1103/PhysRevD.9.3320",
    journal = "Phys. Rev. D",
    volume = "9",
    pages = "3320--3341",
    year = "1974"
}

@article{Patel:2011th,
    author = "Patel, Hiren H. and Ramsey-Musolf, Michael J.",
    title = "{Baryon Washout, Electroweak Phase Transition, and Perturbation Theory}",
    eprint = "1101.4665",
    archivePrefix = "arXiv",
    primaryClass = "hep-ph",
    doi = "10.1007/JHEP07(2011)029",
    journal = "JHEP",
    volume = "07",
    pages = "029",
    year = "2011"
}

@article{Wainwright:2011qy,
    author = "Wainwright, Carroll and Profumo, Stefano and Ramsey-Musolf, Michael J.",
    title = "{Gravity Waves from a Cosmological Phase Transition: Gauge Artifacts and Daisy Resummations}",
    eprint = "1104.5487",
    archivePrefix = "arXiv",
    primaryClass = "hep-ph",
    reportNumber = "NPAC-11-04",
    doi = "10.1103/PhysRevD.84.023521",
    journal = "Phys. Rev. D",
    volume = "84",
    pages = "023521",
    year = "2011"
}

@article{Garny:2012cg,
    author = "Garny, Mathias and Konstandin, Thomas",
    title = "{On the gauge dependence of vacuum transitions at finite temperature}",
    eprint = "1205.3392",
    archivePrefix = "arXiv",
    primaryClass = "hep-ph",
    reportNumber = "DESY-12-073, CERN-PH-TH-2012-127",
    doi = "10.1007/JHEP07(2012)189",
    journal = "JHEP",
    volume = "07",
    pages = "189",
    year = "2012"
}

@article{Niemi:2020hto,
    author = "Niemi, Lauri and Ramsey-Musolf, Michael J. and Tenkanen, Tuomas V. I. and Weir, David J.",
    title = "{Thermodynamics of a Two-Step Electroweak Phase Transition}",
    eprint = "2005.11332",
    archivePrefix = "arXiv",
    primaryClass = "hep-ph",
    reportNumber = "HIP-2020-11/TH, ACFI-T20-05",
    doi = "10.1103/PhysRevLett.126.171802",
    journal = "Phys. Rev. Lett.",
    volume = "126",
    number = "17",
    pages = "171802",
    year = "2021"
}

@article{Croon:2020cgk,
    author = "Croon, Djuna and Gould, Oliver and Schicho, Philipp and Tenkanen, Tuomas V. I. and White, Graham",
    title = "{Theoretical uncertainties for cosmological first-order phase transitions}",
    eprint = "2009.10080",
    archivePrefix = "arXiv",
    primaryClass = "hep-ph",
    reportNumber = "HIP-2020-26/TH",
    doi = "10.1007/JHEP04(2021)055",
    journal = "JHEP",
    volume = "04",
    pages = "055",
    year = "2021"
}

@article{Lofgren:2021ogg,
    author = {L\"ofgren, Johan and Ramsey-Musolf, Michael J. and Schicho, Philipp and Tenkanen, Tuomas V. I.},
    title = "{Nucleation at Finite Temperature: A Gauge-Invariant Perturbative Framework}",
    eprint = "2112.05472",
    archivePrefix = "arXiv",
    primaryClass = "hep-ph",
    reportNumber = "ACFI-T21-15, HIP-2021-44/TH, NORDITA 2021-110",
    doi = "10.1103/PhysRevLett.130.251801",
    journal = "Phys. Rev. Lett.",
    volume = "130",
    number = "25",
    pages = "251801",
    year = "2023"
}

@article{Athron:2022jyi,
    author = "Athron, Peter and Balazs, Csaba and Fowlie, Andrew and Morris, Lachlan and White, Graham and Zhang, Yang",
    title = "{How arbitrary are perturbative calculations of the electroweak phase transition?}",
    eprint = "2208.01319",
    archivePrefix = "arXiv",
    primaryClass = "hep-ph",
    doi = "10.1007/JHEP01(2023)050",
    journal = "JHEP",
    volume = "01",
    pages = "050",
    year = "2023"
}

@article{Schicho:2022wty,
    author = "Schicho, Philipp and Tenkanen, Tuomas V. I. and White, Graham",
    title = "{Combining thermal resummation and gauge invariance for electroweak phase transition}",
    eprint = "2203.04284",
    archivePrefix = "arXiv",
    primaryClass = "hep-ph",
    reportNumber = "HIP-2022-2/TH, NORDITA 2022-009",
    doi = "10.1007/JHEP11(2022)047",
    journal = "JHEP",
    volume = "11",
    pages = "047",
    year = "2022"
}

@article{Qin:2024dfp,
    author = "Qin, Renhui and Bian, Ligong",
    title = "{First-order Electroweak phase transition with Gauge-invariant approach}",
    eprint = "2408.09677",
    archivePrefix = "arXiv",
    primaryClass = "hep-ph",
    month = "8",
    year = "2024"
}

@article{Muhlleitner:2020wwk,
    author = {M{\"u}hlleitner, Margarete and Sampaio, Marco O. P. and Santos, Rui and Wittbrodt, Jonas},
    title = "{ScannerS: parameter scans in extended scalar sectors}",
    eprint = "2007.02985",
    archivePrefix = "arXiv",
    primaryClass = "hep-ph",
    reportNumber = "KA-TP-05-2020, LU TP 20-38",
    doi = "10.1140/epjc/s10052-022-10139-w",
    journal = "Eur. Phys. J. C",
    volume = "82",
    number = "3",
    pages = "198",
    year = "2022"
}

@article{ATLAS:2015yey,
    author = "Aad, Georges and others",
    collaboration = "ATLAS, CMS",
    title = "{Combined Measurement of the Higgs Boson Mass in $pp$ Collisions at $\sqrt{s}=7$ and 8 TeV with the ATLAS and CMS Experiments}",
    eprint = "1503.07589",
    archivePrefix = "arXiv",
    primaryClass = "hep-ex",
    reportNumber = "ATLAS-HIGG-2014-14, CMS-HIG-14-042, CERN-PH-EP-2015-075",
    doi = "10.1103/PhysRevLett.114.191803",
    journal = "Phys. Rev. Lett.",
    volume = "114",
    pages = "191803",
    year = "2015"
}

@article{Barroso:2013awa,
    author = "Barroso, A. and Ferreira, P. M. and Ivanov, I. P. and Santos, Rui",
    title = "{Metastability bounds on the two Higgs doublet model}",
    eprint = "1303.5098",
    archivePrefix = "arXiv",
    primaryClass = "hep-ph",
    doi = "10.1007/JHEP06(2013)045",
    journal = "JHEP",
    volume = "06",
    pages = "045",
    year = "2013"
}

@article{Haber:1999zh,
    author = "Haber, Howard E. and Logan, Heather E.",
    title = "{Radiative corrections to the Z b anti-b vertex and constraints on extended Higgs sectors}",
    eprint = "hep-ph/9909335",
    archivePrefix = "arXiv",
    reportNumber = "SCIPP-98-47",
    doi = "10.1103/PhysRevD.62.015011",
    journal = "Phys. Rev. D",
    volume = "62",
    pages = "015011",
    year = "2000"
}

@article{Deschamps:2009rh,
    author = "Deschamps, O. and Descotes-Genon, S. and Monteil, S. and Niess, V. and T'Jampens, S. and Tisserand, V.",
    title = "{The Two Higgs Doublet of Type II facing flavour physics data}",
    eprint = "0907.5135",
    archivePrefix = "arXiv",
    primaryClass = "hep-ph",
    reportNumber = "LPC-CF-2009-05, LPT-ORSAY-2009-43, LAPP-EXP-2009-03",
    doi = "10.1103/PhysRevD.82.073012",
    journal = "Phys. Rev. D",
    volume = "82",
    pages = "073012",
    year = "2010"
}

@article{Mahmoudi:2009zx,
    author = "Mahmoudi, Farvah and Stal, Oscar",
    title = "{Flavor constraints on the two-Higgs-doublet model with general Yukawa couplings}",
    eprint = "0907.1791",
    archivePrefix = "arXiv",
    primaryClass = "hep-ph",
    doi = "10.1103/PhysRevD.81.035016",
    journal = "Phys. Rev. D",
    volume = "81",
    pages = "035016",
    year = "2010"
}

@article{Hermann:2012fc,
    author = "Hermann, Thomas and Misiak, Mikolaj and Steinhauser, Matthias",
    title = "{$\bar{B}\to X_s \gamma$ in the Two Higgs Doublet Model up to Next-to-Next-to-Leading Order in QCD}",
    eprint = "1208.2788",
    archivePrefix = "arXiv",
    primaryClass = "hep-ph",
    reportNumber = "SFB-CPP-12-60, TTP12-29, IFT-5-2012",
    doi = "10.1007/JHEP11(2012)036",
    journal = "JHEP",
    volume = "11",
    pages = "036",
    year = "2012"
}

@article{Misiak:2015xwa,
    author = "Misiak, M. and others",
    title = "{Updated NNLO QCD predictions for the weak radiative B-meson decays}",
    eprint = "1503.01789",
    archivePrefix = "arXiv",
    primaryClass = "hep-ph",
    reportNumber = "TTP15-007, SFB-CPP-14-121, SI-HEP-2015-08, QFET-2015-09, TTK-15-09, IFT-2-2015",
    doi = "10.1103/PhysRevLett.114.221801",
    journal = "Phys. Rev. Lett.",
    volume = "114",
    number = "22",
    pages = "221801",
    year = "2015"
}

@article{Misiak:2017bgg,
    author = "Misiak, Mikolaj and Steinhauser, Matthias",
    title = "{Weak radiative decays of the B meson and bounds on $M_{H^\pm }$ in the Two-Higgs-Doublet Model}",
    eprint = "1702.04571",
    archivePrefix = "arXiv",
    primaryClass = "hep-ph",
    reportNumber = "TTP17-004, IFT-1-2017",
    doi = "10.1140/epjc/s10052-017-4776-y",
    journal = "Eur. Phys. J. C",
    volume = "77",
    number = "3",
    pages = "201",
    year = "2017"
}

@article{Misiak:2020vlo,
    author = "Misiak, M. and Rehman, Abdur and Steinhauser, Matthias",
    title = "{Towards $ \overline{B}\to {X}_s\gamma $ at the NNLO in QCD without interpolation in m$_{c}$}",
    eprint = "2002.01548",
    archivePrefix = "arXiv",
    primaryClass = "hep-ph",
    reportNumber = "TTP20-001, P3H-20-005, IFT-01/2020",
    doi = "10.1007/JHEP06(2020)175",
    journal = "JHEP",
    volume = "06",
    pages = "175",
    year = "2020"
}

@article{Lee:2025hgb,
    author = "Lee, Soojin and Kim, Dongjoo and Cho, Jin-Hwan and Kim, Jinheung and Song, Jeonghyeon",
    title = "{Multistep strong first-order electroweak phase transitions in the inverted type-I 2HDM: Parameter space, gravitational waves, and collider phenomenology}",
    eprint = "2506.03260",
    archivePrefix = "arXiv",
    primaryClass = "hep-ph",
    reportNumber = "KIAS-P25030",
    doi = "10.1103/cbgr-w9cb",
    journal = "Phys. Rev. D",
    volume = "112",
    number = "5",
    pages = "055035",
    year = "2025"
}

@article{Baak:2014ora,
    author = {Baak, M. and C\'uth, J. and Haller, J. and Hoecker, A. and Kogler, R. and M\"onig, K. and Schott, M. and Stelzer, J.},
    collaboration = "Gfitter Group",
    title = "{The global electroweak fit at NNLO and prospects for the LHC and ILC}",
    eprint = "1407.3792",
    archivePrefix = "arXiv",
    primaryClass = "hep-ph",
    reportNumber = "DESY-14-124",
    doi = "10.1140/epjc/s10052-014-3046-5",
    journal = "Eur. Phys. J. C",
    volume = "74",
    pages = "3046",
    year = "2014"
}

@article{ATLAS:2012yve,
    author = "Aad, Georges and others",
    collaboration = "ATLAS",
    title = "{Observation of a new particle in the search for the Standard Model Higgs boson with the ATLAS detector at the LHC}",
    eprint = "1207.7214",
    archivePrefix = "arXiv",
    primaryClass = "hep-ex",
    reportNumber = "CERN-PH-EP-2012-218",
    doi = "10.1016/j.physletb.2012.08.020",
    journal = "Phys. Lett. B",
    volume = "716",
    pages = "1--29",
    year = "2012"
}

@article{CMS:2012qbp,
    author = "Chatrchyan, Serguei and others",
    collaboration = "CMS",
    title = "{Observation of a New Boson at a Mass of 125 GeV with the CMS Experiment at the LHC}",
    eprint = "1207.7235",
    archivePrefix = "arXiv",
    primaryClass = "hep-ex",
    reportNumber = "CMS-HIG-12-028, CERN-PH-EP-2012-220",
    doi = "10.1016/j.physletb.2012.08.021",
    journal = "Phys. Lett. B",
    volume = "716",
    pages = "30--61",
    year = "2012"
}

@article{LIGOScientific:2016aoc,
    author = "Abbott, B. P. and others",
    collaboration = "LIGO Scientific, Virgo",
    title = "{Observation of Gravitational Waves from a Binary Black Hole Merger}",
    eprint = "1602.03837",
    archivePrefix = "arXiv",
    primaryClass = "gr-qc",
    reportNumber = "LIGO-P150914",
    doi = "10.1103/PhysRevLett.116.061102",
    journal = "Phys. Rev. Lett.",
    volume = "116",
    number = "6",
    pages = "061102",
    year = "2016"
}

@article{WMAP:2012fli,
    author = "Bennett, C. L. and others",
    collaboration = "WMAP",
    title = "{Nine-Year Wilkinson Microwave Anisotropy Probe (WMAP) Observations: Final Maps and Results}",
    eprint = "1212.5225",
    archivePrefix = "arXiv",
    primaryClass = "astro-ph.CO",
    doi = "10.1088/0067-0049/208/2/20",
    journal = "Astrophys. J. Suppl.",
    volume = "208",
    pages = "20",
    year = "2013"
}

@article{Kuzmin:1985mm,
      author         = "Kuzmin, V. A. and Rubakov, V. A. and Shaposhnikov, M. E.",
      title          = "{On the Anomalous Electroweak Baryon Number
                        Nonconservation in the Early Universe}",
      journal        = "Phys. Lett.",
      volume         = "155B",
      year           = "1985",
      pages          = "36",
      doi            = "10.1016/0370-2693(85)91028-7",
      reportNumber   = "IC/85/8",
      SLACcitation   = "%%CITATION = PHLTA,155B,36;%%"
}

@article{Cohen:1990it,
      author         = "Cohen, Andrew G. and Kaplan, David B. and Nelson, Ann E.",
      title          = "{Baryogenesis at the weak phase transition}",
      journal        = "Nucl. Phys.",
      volume         = "B349",
      year           = "1991",
      pages          = "727-742",
      doi            = "10.1016/0550-3213(91)90395-E",
      reportNumber   = "NSF-ITP-90-85, UCSD-PTH-90-09, BUHEP-90-15",
      SLACcitation   = "%%CITATION = NUPHA,B349,727;%%"
}

@article{Cohen:1993nk,
      author         = "Cohen, Andrew G. and Kaplan, D. B. and Nelson, A. E.",
      title          = "{Progress in electroweak baryogenesis}",
      journal        = "Ann. Rev. Nucl. Part. Sci.",
      volume         = "43",
      year           = "1993",
      pages          = "27-70",
      doi            = "10.1146/annurev.ns.43.120193.000331",
      eprint         = "hep-ph/9302210",
      archivePrefix  = "arXiv",
      primaryClass   = "hep-ph",
      reportNumber   = "UCSD-PTH-93-02, BUHEP-93-4",
      SLACcitation   = "%%CITATION = HEP-PH/9302210;%%"
}

@article{Quiros:1994dr,
      author         = "Quiros, M.",
      title          = "{Field theory at finite temperature and phase
                        transitions}",
      journal        = "Helv. Phys. Acta",
      volume         = "67",
      year           = "1994",
      pages          = "451-583",
      SLACcitation   = "%%CITATION = HPACA,67,451;%%"
}

@article{Rubakov:1996vz,
      author         = "Rubakov, V. A. and Shaposhnikov, M. E.",
      title          = "{Electroweak baryon number nonconservation in the early
                        universe and in high-energy collisions}",
      journal        = "Usp. Fiz. Nauk",
      volume         = "166",
      year           = "1996",
      pages          = "493-537",
      doi            = "10.1070/PU1996v039n05ABEH000145",
      note           = "[Phys. Usp.39,461(1996)]",
      eprint         = "hep-ph/9603208",
      archivePrefix  = "arXiv",
      primaryClass   = "hep-ph",
      reportNumber   = "CERN-TH-96-13, INR-0913-96",
      SLACcitation   = "%%CITATION = HEP-PH/9603208;%%"
}

@article{Funakubo:1996dw,
      author         = "Funakubo, Koichi",
      title          = "{CP violation and baryogenesis at the electroweak phase
                        transition}",
      journal        = "Prog. Theor. Phys.",
      volume         = "96",
      year           = "1996",
      pages          = "475-520",
      doi            = "10.1143/PTP.96.475",
      eprint         = "hep-ph/9608358",
      archivePrefix  = "arXiv",
      primaryClass   = "hep-ph",
      reportNumber   = "SAGA-HE-108",
      SLACcitation   = "%%CITATION = HEP-PH/9608358;%%"
}

@article{Trodden:1998ym,
      author         = "Trodden, Mark",
      title          = "{Electroweak baryogenesis}",
      journal        = "Rev. Mod. Phys.",
      volume         = "71",
      year           = "1999",
      pages          = "1463-1500",
      doi            = "10.1103/RevModPhys.71.1463",
      eprint         = "hep-ph/9803479",
      archivePrefix  = "arXiv",
      primaryClass   = "hep-ph",
      reportNumber   = "CWRU-P6-98",
      SLACcitation   = "%%CITATION = HEP-PH/9803479;%%"
}

@article{Bernreuther:2002uj,
      author         = "Bernreuther, Werner",
      title          = "{CP violation and baryogenesis}",
      booktitle      = "{Workshop of the Graduate College of Elementary Particle
                        Physics Berlin, Germany, April 2-5, 2001}",
      journal        = "Lect. Notes Phys.",
      volume         = "591",
      year           = "2002",
      pages          = "237-293",
      note           = "[,237(2002)]",
      eprint         = "hep-ph/0205279",
      archivePrefix  = "arXiv",
      primaryClass   = "hep-ph",
      reportNumber   = "PITHA-02-08",
      SLACcitation   = "%%CITATION = HEP-PH/0205279;%%"
}

@article{Morrissey:2012db,
      author         = "Morrissey, David E. and Ramsey-Musolf, Michael J.",
      title          = "{Electroweak baryogenesis}",
      journal        = "New J. Phys.",
      volume         = "14",
      year           = "2012",
      pages          = "125003",
      doi            = "10.1088/1367-2630/14/12/125003",
      eprint         = "1206.2942",
      archivePrefix  = "arXiv",
      primaryClass   = "hep-ph",
      reportNumber   = "NPAC-12-08",
      SLACcitation   = "%%CITATION = ARXIV:1206.2942;%%"
}

@article{Sakharov:1967dj,
      author         = "Sakharov, A. D.",
      title          = "{Violation of CP Invariance, C asymmetry, and baryon
                        asymmetry of the universe}",
      journal        = "Pisma Zh. Eksp. Teor. Fiz.",
      volume         = "5",
      year           = "1967",
      pages          = "32-35",
      doi            = "10.1070/PU1991v034n05ABEH002497",
      note           = "[Usp. Fiz. Nauk161,no.5,61(1991)]",
      SLACcitation   = "%%CITATION = ZFPRA,5,32;%%"
}

@article{Ellis:2018mja,
    author = "Ellis, John and Lewicki, Marek and No, Jos\'e Miguel",
    title = "{On the Maximal Strength of a First-Order Electroweak Phase Transition and its Gravitational Wave Signal}",
    eprint = "1809.08242",
    archivePrefix = "arXiv",
    primaryClass = "hep-ph",
    reportNumber = "KCL-PH-TH/2018-46, CERN-TH/2018-197, IFT-UAM/CSIC-18-94, CERN-TH-2018-197",
    doi = "10.1088/1475-7516/2019/04/003",
    journal = "JCAP",
    volume = "04",
    pages = "003",
    year = "2019"
}

@software{PTPlot,
    author = {Weir, David},
    title = {{PTPlot}},
    url = {https://www.ptplot.org/ptplot/}
}

@software{PTtools,
    author = {Hindmarsh, Mark and M\"aki, Mika"},
    title = {{PTtools}},
    url = {https://github.com/CFT-HY/pttools}
}

@software{CosmoGW,
    author = {Roper Pol, Alberto},
    title = {{CosmoGW}},
    url = {https://github.com/CosmoGW/CosmoGW}
}

@software{BLOOP,
    author = {Keus, Venus and Lewitt, Lucy and Thomson-Cooke, Jasmine},
    title = {{BLOOP}},
    url = {https://github.com/MOREHIGGS/Bloop}
}

@article{Basler:2018cwe,
    author = {Basler, Philipp and M\"uhlleitner, Margarete},
    title = "{BSMPT (Beyond the Standard Model Phase Transitions): A tool for the electroweak phase transition in extended Higgs sectors}",
    eprint = "1803.02846",
    archivePrefix = "arXiv",
    primaryClass = "hep-ph",
    doi = "10.1016/j.cpc.2018.11.006",
    journal = "Comput. Phys. Commun.",
    volume = "237",
    pages = "62--85",
    year = "2019"
}

@article{Basler:2020nrq,
    author = {Basler, Philipp and M\"uhlleitner, Margarete and M\"uller, Jonas},
    title = "{BSMPT v2 a tool for the electroweak phase transition and the baryon asymmetry of the universe in extended Higgs Sectors}",
    eprint = "2007.01725",
    archivePrefix = "arXiv",
    primaryClass = "hep-ph",
    doi = "10.1016/j.cpc.2021.108124",
    journal = "Comput. Phys. Commun.",
    volume = "269",
    pages = "108124",
    year = "2021"
}

@article{Bahl:2022igd,
    author = {Bahl, Henning and Biek\"otter, Thomas and Heinemeyer, Sven and Li, Cheng and Paasch, Steven and Weiglein, Georg and Wittbrodt, Jonas},
    title = "{HiggsTools: BSM scalar phenomenology with new versions of HiggsBounds and HiggsSignals}",
    eprint = "2210.09332",
    archivePrefix = "arXiv",
    primaryClass = "hep-ph",
    doi = "10.1016/j.cpc.2023.108803",
    journal = "Comput. Phys. Commun.",
    volume = "291",
    pages = "108803",
    year = "2023"
}

@article{Bahl:2026yal,
    author = {Bahl, Henning and Biek{\"o}tter, Thomas and Heinemeyer, Sven and Radchenko Serdula, Kateryna and Weiglein, Georg},
    title = "{HiggsTools for LHC Run 3 and Beyond}",
    eprint = "2608.05401",
    archivePrefix = "arXiv",
    primaryClass = "hep-ph",
    reportNumber = "DESY-26-011 DESY-26-011, IFT--UAM/CSIC-25-072",
    month = "8",
    year = "2026"
}

@article{Su:2020pjw,
    author = "Su, Wei and Williams, Anthony G. and Zhang, Mengchao",
    title = "{Strong first order electroweak phase transition in 2HDM confronting future Z \& Higgs factories}",
    eprint = "2011.04540",
    archivePrefix = "arXiv",
    primaryClass = "hep-ph",
    reportNumber = "ADP-20-31/T1141",
    doi = "10.1007/JHEP04(2021)219",
    journal = "JHEP",
    volume = "04",
    pages = "219",
    year = "2021"
}

@article{Anisha:2022hgv,
    author = {Anisha and Biermann, Lisa and Englert, Christoph and M\"uhlleitner, Margarete},
    title = "{Two Higgs doublets, effective interactions and a strong first-order electroweak phase transition}",
    eprint = "2204.06966",
    archivePrefix = "arXiv",
    primaryClass = "hep-ph",
    doi = "10.1007/JHEP08(2022)091",
    journal = "JHEP",
    volume = "08",
    pages = "091",
    year = "2022"
}

@article{Anisha:2023vvu,
    author = {Anisha and Azevedo, Duarte and Biermann, Lisa and Englert, Christoph and M\"uhlleitner, Margarete},
    title = "{Effective 2HDM Yukawa interactions and a strong first-order electroweak phase transition}",
    eprint = "2311.06353",
    archivePrefix = "arXiv",
    primaryClass = "hep-ph",
    doi = "10.1007/JHEP02(2024)045",
    journal = "JHEP",
    volume = "02",
    pages = "045",
    year = "2024"
}

@article{Camargo-Molina:2013qva,
    author = "Camargo-Molina, J. E. and O'Leary, B. and Porod, W. and Staub, F.",
    title = "{$\mathbf{Vevacious}$: A Tool For Finding The Global Minima Of One-Loop Effective Potentials With Many Scalars}",
    eprint = "1307.1477",
    archivePrefix = "arXiv",
    primaryClass = "hep-ph",
    doi = "10.1140/epjc/s10052-013-2588-2",
    journal = "Eur. Phys. J. C",
    volume = "73",
    number = "10",
    pages = "2588",
    year = "2013"
}

@article{Camargo-Molina:2014pwa,
    author = "Camargo-Molina, J. E. and Garbrecht, B. and O'Leary, B. and Porod, W. and Staub, F.",
    title = "{Constraining the Natural MSSM through tunneling to color-breaking vacua at zero and non-zero temperature}",
    eprint = "1405.7376",
    archivePrefix = "arXiv",
    primaryClass = "hep-ph",
    doi = "10.1016/j.physletb.2014.08.036",
    journal = "Phys. Lett. B",
    volume = "737",
    pages = "156--161",
    year = "2014"
}

@article{Espinosa:1995se,
    author = "Espinosa, J. R. and Quiros, M.",
    title = "{Improved metastability bounds on the standard model Higgs mass}",
    eprint = "hep-ph/9504241",
    archivePrefix = "arXiv",
    reportNumber = "CERN-TH-95-18, CERN-TH-95-018, DESY-95-039, IEM-FT-97-95",
    doi = "10.1016/0370-2693(95)00572-3",
    journal = "Phys. Lett. B",
    volume = "353",
    pages = "257--266",
    year = "1995"
}

@article{Kajantie:1996mn,
    author = "Kajantie, K. and Laine, M. and Rummukainen, K. and Shaposhnikov, Mikhail E.",
    title = "{Is there a~ hot electroweak phase transition at $m_H \gtrsim m_W$?}",
    eprint = "hep-ph/9605288",
    archivePrefix = "arXiv",
    reportNumber = "CERN-TH-96-126, HD-THEP-96-15, IUHET-333",
    doi = "10.1103/PhysRevLett.77.2887",
    journal = "Phys. Rev. Lett.",
    volume = "77",
    pages = "2887--2890",
    year = "1996"
}

@article{Csikor:1998eu,
    author = "Csikor, F. and Fodor, Z. and Heitger, J.",
    title = "{Endpoint of the hot electroweak phase transition}",
    eprint = "hep-ph/9809291",
    archivePrefix = "arXiv",
    reportNumber = "ITP-BUDAPEST-541, KEK-TH-580, KEK-PREPRINT-98-160, MS-TPI-98-16",
    doi = "10.1103/PhysRevLett.82.21",
    journal = "Phys. Rev. Lett.",
    volume = "82",
    pages = "21--24",
    year = "1999"
}

@article{Aoki:2023lbz,
    author = {Aoki, Mayumi and Biermann, Lisa and Borschensky, Christoph and Ivanov, Igor P. and M\"uhlleitner, Margarete and Shibuya, Hiroto},
    title = "{Intermediate charge-breaking phases and symmetry non-restoration in the 2-Higgs-Doublet Model}",
    eprint = "2308.04141",
    archivePrefix = "arXiv",
    primaryClass = "hep-ph",
    reportNumber = "KANAZAWA-23-08, KA-TP-16-2023",
    doi = "10.1007/JHEP02(2024)232",
    journal = "JHEP",
    volume = "02",
    pages = "232",
    year = "2024"
}

@article{Hollik:2018wrr,
    author = "Hollik, Wolfgang G. and Weiglein, Georg and Wittbrodt, Jonas",
    title = "{Impact of Vacuum Stability Constraints on the Phenomenology of Supersymmetric Models}",
    eprint = "1812.04644",
    archivePrefix = "arXiv",
    primaryClass = "hep-ph",
    reportNumber = "DESY 18-148, DESY-18-148, TTP 18-036",
    doi = "10.1007/JHEP03(2019)109",
    journal = "JHEP",
    volume = "03",
    pages = "109",
    year = "2019"
}

@article{Ferreira:2019iqb,
    author = {Ferreira, P. M. and M\"uhlleitner, Margarete and Santos, Rui and Weiglein, Georg and Wittbrodt, Jonas},
    title = "{Vacuum Instabilities in the N2HDM}",
    eprint = "1905.10234",
    archivePrefix = "arXiv",
    primaryClass = "hep-ph",
    reportNumber = "DESY-19-085, DESY 19-085, KA-TP-07-2019",
    doi = "10.1007/JHEP09(2019)006",
    journal = "JHEP",
    volume = "09",
    pages = "006",
    year = "2019"
}

@misc{LISA,
      title={Laser Interferometer Space Antenna}, 
      author={Pau Amaro-Seoane and Heather Audley and Stanislav Babak and John Baker and Enrico Barausse and Peter Bender and Emanuele Berti and Pierre Binetruy and Michael Born and Daniele Bortoluzzi and Jordan Camp and Chiara Caprini and Vitor Cardoso and Monica Colpi and John Conklin and Neil Cornish and Curt Cutler and Karsten Danzmann and Rita Dolesi and Luigi Ferraioli and Valerio Ferroni and Ewan Fitzsimons and Jonathan Gair and Lluis Gesa Bote and Domenico Giardini and Ferran Gibert and Catia Grimani and Hubert Halloin and Gerhard Heinzel and Thomas Hertog and Martin Hewitson and Kelly Holley-Bockelmann and Daniel Hollington and Mauro Hueller and Henri Inchauspe and Philippe Jetzer and Nikos Karnesis and Christian Killow and Antoine Klein and Bill Klipstein and Natalia Korsakova and Shane L Larson and Jeffrey Livas and Ivan Lloro and Nary Man and Davor Mance and Joseph Martino and Ignacio Mateos and Kirk McKenzie and Sean T McWilliams and Cole Miller and Guido Mueller and Germano Nardini and Gijs Nelemans and Miquel Nofrarias and Antoine Petiteau and Paolo Pivato and Eric Plagnol and Ed Porter and Jens Reiche and David Robertson and Norna Robertson and Elena Rossi and Giuliana Russano and Bernard Schutz and Alberto Sesana and David Shoemaker and Jacob Slutsky and Carlos F. Sopuerta and Tim Sumner and Nicola Tamanini and Ira Thorpe and Michael Troebs and Michele Vallisneri and Alberto Vecchio and Daniele Vetrugno and Stefano Vitale and Marta Volonteri and Gudrun Wanner and Harry Ward and Peter Wass and William Weber and John Ziemer and Peter Zweifel},
      year={2017},
      eprint={1702.00786},
      archivePrefix={arXiv},
      primaryClass={astro-ph.IM}
}

@article{Aoki:2021oez,
    author = "Aoki, Mayumi and Komatsu, Takatoshi and Shibuya, Hiroto",
    title = "{Possibility of a multi-step electroweak phase transition in the two-Higgs doublet models}",
    eprint = "2106.03439",
    archivePrefix = "arXiv",
    primaryClass = "hep-ph",
    reportNumber = "KANAZAWA-21-08",
    doi = "10.1093/ptep/ptac068",
    journal = "PTEP",
    volume = "2022",
    number = "6",
    pages = "063B05",
    year = "2022"
}

@article{Wainwright:2011kj,
    author = "Wainwright, Carroll L.",
    title = "{CosmoTransitions: Computing Cosmological Phase Transition Temperatures and Bubble Profiles with Multiple Fields}",
    eprint = "1109.4189",
    archivePrefix = "arXiv",
    primaryClass = "hep-ph",
    doi = "10.1016/j.cpc.2012.04.004",
    journal = "Comput. Phys. Commun.",
    volume = "183",
    pages = "2006--2013",
    year = "2012"
}

@article{Athron:2023xlk,
    author = "Athron, Peter and Bal\'azs, Csaba and Fowlie, Andrew and Morris, Lachlan and Wu, Lei",
    title = "{Cosmological phase transitions: From perturbative particle physics to gravitational waves}",
    eprint = "2305.02357",
    archivePrefix = "arXiv",
    primaryClass = "hep-ph",
    doi = "10.1016/j.ppnp.2023.104094",
    journal = "Prog. Part. Nucl. Phys.",
    volume = "135",
    pages = "104094",
    year = "2024"
}

@article{Masoumi:2017trx,
    author = "Masoumi, Ali and Olum, Ken D. and Wachter, Jeremy M.",
    title = "{Approximating tunneling rates in multi-dimensional field spaces}",
    eprint = "1702.00356",
    archivePrefix = "arXiv",
    primaryClass = "gr-qc",
    doi = "10.1088/1475-7516/2017/10/022",
    journal = "JCAP",
    volume = "10",
    pages = "022",
    year = "2017",
    note = "[Erratum: JCAP 05, E01 (2023)]"
}

@article{Athron:2019nbd,
    author = "Athron, Peter and Bal\'azs, Csaba and Bardsley, Michael and Fowlie, Andrew and Harries, Dylan and White, Graham",
    title = "{BubbleProfiler: finding the field profile and action for cosmological phase transitions}",
    eprint = "1901.03714",
    archivePrefix = "arXiv",
    primaryClass = "hep-ph",
    reportNumber = "CoEPP-MN-19-1",
    doi = "10.1016/j.cpc.2019.05.017",
    journal = "Comput. Phys. Commun.",
    volume = "244",
    pages = "448--468",
    year = "2019"
}

@article{Guada:2020xnz,
    author = "Guada, Victor and Nemev\v{s}ek, Miha and Pintar, Matev\v{z}",
    title = "{FindBounce: Package for multi-field bounce actions}",
    eprint = "2002.00881",
    archivePrefix = "arXiv",
    primaryClass = "hep-ph",
    doi = "10.1016/j.cpc.2020.107480",
    journal = "Comput. Phys. Commun.",
    volume = "256",
    pages = "107480",
    year = "2020"
}

@article{Sato:2019wpo,
    author = "Sato, Ryosuke",
    title = "{SimpleBounce : a simple package for the false vacuum decay}",
    eprint = "1908.10868",
    archivePrefix = "arXiv",
    primaryClass = "hep-ph",
    reportNumber = "DESY 19-148, DESY-19-148",
    doi = "10.1016/j.cpc.2020.107566",
    journal = "Comput. Phys. Commun.",
    volume = "258",
    pages = "107566",
    year = "2021"
}

@article{Ertas:2021xeh,
    author = "Ertas, Fatih and Kahlhoefer, Felix and Tasillo, Carlo",
    title = "{Turn up the volume: listening to phase transitions in hot dark sectors}",
    eprint = "2109.06208",
    archivePrefix = "arXiv",
    primaryClass = "astro-ph.CO",
    reportNumber = "TTK-21-36, DESY-22-014",
    doi = "10.1088/1475-7516/2022/02/014",
    journal = "JCAP",
    volume = "02",
    number = "02",
    pages = "014",
    year = "2022"
}

@article{Bringmann:2023iuz,
    author = "Bringmann, Torsten and Gonzalo, Tom\'as E. and Kahlhoefer, Felix and Matuszak, Jonas and Tasillo, Carlo",
    title = "{Hunting WIMPs with LISA: correlating dark matter and gravitational wave signals}",
    eprint = "2311.06346",
    archivePrefix = "arXiv",
    primaryClass = "astro-ph.CO",
    doi = "10.1088/1475-7516/2024/05/065",
    journal = "JCAP",
    volume = "05",
    pages = "065",
    year = "2024"
}

@article{Bardsley:2021lmq,
    author = "Bardsley, Michael",
    title = "{An optimisation based algorithm for finding the nucleation temperature of cosmological phase transitions}",
    eprint = "2103.01985",
    archivePrefix = "arXiv",
    primaryClass = "astro-ph.CO",
    doi = "10.1016/j.cpc.2021.108252",
    journal = "Comput. Phys. Commun.",
    volume = "273",
    pages = "108252",
    year = "2022"
}

@article{Athron:2020sbe,
    author = "Athron, Peter and Bal\'azs, Csaba and Fowlie, Andrew and Zhang, Yang",
    title = "{PhaseTracer: tracing cosmological phases and calculating transition properties}",
    eprint = "2003.02859",
    archivePrefix = "arXiv",
    primaryClass = "hep-ph",
    reportNumber = "CoEPP-MN-20-3",
    doi = "10.1140/epjc/s10052-020-8035-2",
    journal = "Eur. Phys. J. C",
    volume = "80",
    number = "6",
    pages = "567",
    year = "2020"
}

@article{Fromme:2006cm,
    author = "Fromme, Lars and Huber, Stephan J. and Seniuch, Michael",
    title = "{Baryogenesis in the two-Higgs doublet model}",
    eprint = "hep-ph/0605242",
    archivePrefix = "arXiv",
    reportNumber = "CERN-PH-TH-2006-094, BI-TP-2006-18",
    doi = "10.1088/1126-6708/2006/11/038",
    journal = "JHEP",
    volume = "11",
    pages = "038",
    year = "2006"
}

@article{Moore:1998swa,
    author = "Moore, Guy D.",
    title = "{Measuring the broken phase sphaleron rate nonperturbatively}",
    eprint = "hep-ph/9805264",
    archivePrefix = "arXiv",
    reportNumber = "MCGILL-98-7",
    doi = "10.1103/PhysRevD.59.014503",
    journal = "Phys. Rev. D",
    volume = "59",
    pages = "014503",
    year = "1999"
}

@article{Biekotter:2022kgf,
    author = {Biek\"otter, Thomas and Heinemeyer, Sven and No, Jos\'e Miguel and Olea-Romacho, Mar\'{i}a Olalla and Weiglein, Georg},
    title = "{The trap in the early Universe: impact on the interplay between gravitational waves and LHC physics in the 2HDM}",
    eprint = "2208.14466",
    archivePrefix = "arXiv",
    primaryClass = "hep-ph",
    reportNumber = "DESY-22-127, IFT--UAM/CSIC--22-015",
    doi = "10.1088/1475-7516/2023/03/031",
    journal = "JCAP",
    volume = "03",
    pages = "031",
    year = "2023"
}

@article{Abouabid:2021yvw,
    author = {Abouabid, Hamza and Arhrib, Abdesslam and Azevedo, Duarte and Falaki, Jaouad El and Ferreira, Pedro. M. and M\"uhlleitner, Margarete and Santos, Rui},
    title = "{Benchmarking di-Higgs production in various extended Higgs sector models}",
    eprint = "2112.12515",
    archivePrefix = "arXiv",
    primaryClass = "hep-ph",
    doi = "10.1007/JHEP09(2022)011",
    journal = "JHEP",
    volume = "09",
    pages = "011",
    year = "2022"
}

@article{Coimbra:2013qq,
    author = "Coimbra, Rita and Sampaio, Marco O. P. and Santos, Rui",
    title = "{ScannerS: Constraining the phase diagram of a complex scalar singlet at the LHC}",
    eprint = "1301.2599",
    archivePrefix = "arXiv",
    primaryClass = "hep-ph",
    doi = "10.1140/epjc/s10052-013-2428-4",
    journal = "Eur. Phys. J. C",
    volume = "73",
    pages = "2428",
    year = "2013"
}

@article{Bechtle:2008jh,
    author = "Bechtle, Philip and Brein, Oliver and Heinemeyer, Sven and Weiglein, Georg and Williams, Karina E.",
    title = "{HiggsBounds: Confronting Arbitrary Higgs Sectors with Exclusion Bounds from LEP and the Tevatron}",
    eprint = "0811.4169",
    archivePrefix = "arXiv",
    primaryClass = "hep-ph",
    reportNumber = "DCPT-08-172, IPPP-08-86, BONN-TH-2008-17",
    doi = "10.1016/j.cpc.2009.09.003",
    journal = "Comput. Phys. Commun.",
    volume = "181",
    pages = "138--167",
    year = "2010"
}

@article{Bechtle:2013xfa,
    author = "Bechtle, Philip and Heinemeyer, Sven and St\r{a}l, Oscar and Stefaniak, Tim and Weiglein, Georg",
    title = "{$HiggsSignals$: Confronting arbitrary Higgs sectors with measurements at the Tevatron and the LHC}",
    eprint = "1305.1933",
    archivePrefix = "arXiv",
    primaryClass = "hep-ph",
    reportNumber = "BONN-TH-2013-07, DESY-13-078",
    doi = "10.1140/epjc/s10052-013-2711-4",
    journal = "Eur. Phys. J. C",
    volume = "74",
    number = "2",
    pages = "2711",
    year = "2014"
}

@article{Ekstedt:2022bff,
    author = "Ekstedt, Andreas and Schicho, Philipp and Tenkanen, Tuomas V. I.",
    title = "{DRalgo: A package for effective field theory approach for thermal phase transitions}",
    eprint = "2205.08815",
    archivePrefix = "arXiv",
    primaryClass = "hep-ph",
    reportNumber = "HIP-2022-11/TH, NORDITA 2022-030",
    doi = "10.1016/j.cpc.2023.108725",
    journal = "Comput. Phys. Commun.",
    volume = "288",
    pages = "108725",
    year = "2023"
}

@article{Arnold:1992rz,
    author = "Arnold, Peter Brockway and Espinosa, Olivier",
    title = "{The Effective potential and first order phase transitions: Beyond leading-order}",
    eprint = "hep-ph/9212235",
    archivePrefix = "arXiv",
    reportNumber = "UW-PT-92-18, USM-TH-60",
    doi = "10.1103/PhysRevD.47.3546",
    journal = "Phys. Rev. D",
    volume = "47",
    pages = "3546",
    year = "1993",
    note = "[Erratum: Phys.Rev.D 50, 6662 (1994)]"
}

@article{Parwani:1991gq,
    author = "Parwani, Rajesh R.",
    title = "{Resummation in a hot scalar field theory}",
    eprint = "hep-ph/9204216",
    archivePrefix = "arXiv",
    reportNumber = "ITP-SB-91-64",
    doi = "10.1103/PhysRevD.45.4695",
    journal = "Phys. Rev. D",
    volume = "45",
    pages = "4695",
    year = "1992",
    note = "[Erratum: Phys.Rev.D 48, 5965 (1993)]"
}

@article{Lee:1973iz,
    author = "Lee, T. D.",
    editor = "Feinberg, G.",
    title = "{A Theory of Spontaneous T Violation}",
    doi = "10.1103/PhysRevD.8.1226",
    journal = "Phys. Rev. D",
    volume = "8",
    pages = "1226--1239",
    year = "1973"
}

@article{Branco:2011iw,
    author = "Branco, G. C. and Ferreira, P. M. and Lavoura, L. and Rebelo, M. N. and Sher, Marc and Silva, Joao P.",
    title = "{Theory and phenomenology of two-Higgs-doublet models}",
    eprint = "1106.0034",
    archivePrefix = "arXiv",
    primaryClass = "hep-ph",
    doi = "10.1016/j.physrep.2012.02.002",
    journal = "Phys. Rept.",
    volume = "516",
    pages = "1--102",
    year = "2012"
}

@article{ATLAS:2025eii,
    author = "Aad, Georges and others",
    collaboration = "ATLAS, CMS",
    title = "{Highlights of the HL-LHC physics projections by ATLAS and CMS}",
    eprint = "2504.00672",
    archivePrefix = "arXiv",
    primaryClass = "hep-ex",
    reportNumber = "ATL-PHYS-PUB-2025-018, CMS-HIG-25-002",
    month = "4",
    year = "2025"
}

@article{Maura:2025rcv,
    author = "Maura, Victor and Stefanek, Ben A. and You, Tevong",
    title = "{The Higgs Self-Coupling at FCC-ee}",
    eprint = "2503.13719",
    archivePrefix = "arXiv",
    primaryClass = "hep-ph",
    reportNumber = "KCL-PH-TH/2025-05",
    doi = "10.1103/wjcy-1qk6",
    journal = "Phys. Rev. Lett.",
    volume = "135",
    number = "14",
    pages = "141802",
    year = "2025"
}

@article{Selvaggi:2025kmd,
    editor = "Selvaggi, M. and Blondel, A. and Eysermans, J.",
    collaboration = "FCC",
    title = "{Prospects in electroweak, Higgs and Top physics at FCC}",
    doi = "10.17181/n78xk-qcv56",
    month = "3",
    year = "2025"
}

@article{Bahl:2026aou,
    author = "Bahl, Henning and Braathen, Johannes and Gabelmann, Martin and Heinemeyer, Sven and Serdula, Kateryna Radchenko and Verduras Schaeidt, Alain and Weiglein, Georg",
    title = "{Sensitivity to new physics: single-Higgs couplings vs. the trilinear Higgs coupling}",
    eprint = "2603.26633",
    archivePrefix = "arXiv",
    primaryClass = "hep-ph",
    reportNumber = "DESY-26-025, IFT--UAM/CSIC-26-017, FR-PHENO-2026-005",
    month = "3",
    year = "2026"
}

@article{Anisha:2025zbc,
    author = {Anisha and Arco, Francisco and Di Noi, Stefano and Englert, Christoph and M{\"u}hlleitner, Margarete},
    title = "{Z and Higgs factory implications of two Higgs doublets with first-order phase transitions}",
    eprint = "2506.18555",
    archivePrefix = "arXiv",
    primaryClass = "hep-ph",
    reportNumber = "KA-TP-17-2025, DESY-25-085",
    doi = "10.1007/JHEP10(2025)179",
    journal = "JHEP",
    volume = "10",
    pages = "179",
    year = "2025"
}

@article{Planck:2018vyg,
    author = "Aghanim, N. and others",
    collaboration = "Planck",
    title = "{Planck 2018 results. VI. Cosmological parameters}",
    eprint = "1807.06209",
    archivePrefix = "arXiv",
    primaryClass = "astro-ph.CO",
    doi = "10.1051/0004-6361/201833910",
    journal = "Astron. Astrophys.",
    volume = "641",
    pages = "A6",
    year = "2020",
    note = "[Erratum: Astron.Astrophys. 652, C4 (2021)]"
}

@article{Ekstedt:2024fyq,
    author = "Ekstedt, Andreas and Gould, Oliver and Hirvonen, Joonas and Laurent, Benoit and Niemi, Lauri and Schicho, Philipp and van de Vis, Jorinde",
    title = "{How fast does the WallGo? A package for computing wall velocities in first-order phase transitions}",
    eprint = "2411.04970",
    archivePrefix = "arXiv",
    primaryClass = "hep-ph",
    reportNumber = "CERN-TH-2024-174, DESY-24-162, HIP-2024-21/TH",
    doi = "10.1007/JHEP04(2025)101",
    journal = "JHEP",
    volume = "04",
    pages = "101",
    year = "2025"
}

@article{Branchina:2025jou,
    author = "Branchina, Carlo and Conaci, Angela and Delle Rose, Luigi and De Curtis, Stefania",
    title = "{Electroweak phase transition and bubble wall velocity in local thermal equilibrium}",
    eprint = "2504.21213",
    archivePrefix = "arXiv",
    primaryClass = "hep-ph",
    doi = "10.1103/lkht-1nv1",
    journal = "Phys. Rev. D",
    volume = "112",
    number = "9",
    pages = "095008",
    year = "2025"
}

@article{Branchina:2025adj,
    author = "Branchina, Carlo and Conaci, Angela and Delle Rose, Luigi and De Curtis, Stefania",
    title = "{Bubble wall velocity with out-of-equilibrium corrections}",
    eprint = "2510.21942",
    archivePrefix = "arXiv",
    primaryClass = "hep-ph",
    doi = "10.1103/nmkw-7kgk",
    journal = "Phys. Rev. D",
    volume = "113",
    number = "3",
    pages = "035024",
    year = "2026"
}

@article{DeCurtis:2022hlx,
    author = "De Curtis, Stefania and Rose, Luigi Delle and Guiggiani, Andrea and Muyor, {\'A}ngel Gil and Panico, Giuliano",
    title = "{Bubble wall dynamics at the electroweak phase transition}",
    eprint = "2201.08220",
    archivePrefix = "arXiv",
    primaryClass = "hep-ph",
    doi = "10.1007/JHEP03(2022)163",
    journal = "JHEP",
    volume = "03",
    pages = "163",
    year = "2022"
}

@article{DeCurtis:2023hil,
    author = "De Curtis, Stefania and Delle Rose, Luigi and Guiggiani, Andrea and Gil Muyor, {\'A}ngel and Panico, Giuliano",
    title = "{Collision integrals for cosmological phase transitions}",
    eprint = "2303.05846",
    archivePrefix = "arXiv",
    primaryClass = "hep-ph",
    doi = "10.1007/JHEP05(2023)194",
    journal = "JHEP",
    volume = "05",
    pages = "194",
    year = "2023"
}

@article{DeCurtis:2024hvh,
    author = "De Curtis, Stefania and Delle Rose, Luigi and Guiggiani, Andrea and Gil Muyor, {\'A}ngel and Panico, Giuliano",
    title = "{Non-linearities in cosmological bubble wall dynamics}",
    eprint = "2401.13522",
    archivePrefix = "arXiv",
    primaryClass = "hep-ph",
    doi = "10.1007/JHEP05(2024)009",
    journal = "JHEP",
    volume = "05",
    pages = "009",
    year = "2024"
}

@article{10.1162/106365601750190398,
    author = {Hansen, Nikolaus and Ostermeier, Andreas},
    title = "{Completely Derandomized Self-Adaptation in Evolution Strategies}",
    journal = {Evolutionary Computation},
    volume = {9},
    number = {2},
    pages = {159-195},
    year = {2001},
    month = {06},
    issn = {1063-6560},
    doi = {10.1162/106365601750190398},
}

@article{hansen2023cmaevolutionstrategytutorial,
      title="{The CMA Evolution Strategy: A Tutorial}", 
      author={Nikolaus Hansen},
      eprint={1604.00772},
      archivePrefix={arXiv},
      primaryClass={cs.LG},
}

@article{deSouza:2022uhk,
    author = "de Souza, Fernando Abreu and Crispim Rom{\~a}o, Miguel and Castro, Nuno Filipe and Nikjoo, Mehraveh and Porod, Werner",
    title = "{Exploring parameter spaces with artificial intelligence and machine learning black-box optimization algorithms}",
    eprint = "2206.09223",
    archivePrefix = "arXiv",
    primaryClass = "hep-ph",
    doi = "10.1103/PhysRevD.107.035004",
    journal = "Phys. Rev. D",
    volume = "107",
    number = "3",
    pages = "035004",
    year = "2023"
}

@article{Romao:2024gjx,
    author = "Rom{\~a}o, Jorge Crispim and Crispim Rom{\~a}o, Miguel",
    title = "{Combining evolutionary strategies and novelty detection to go beyond the alignment limit of the Z3 3HDM}",
    eprint = "2402.07661",
    archivePrefix = "arXiv",
    primaryClass = "hep-ph",
    reportNumber = "IPPP/24/04, CFTP/24-002",
    doi = "10.1103/PhysRevD.109.095040",
    journal = "Phys. Rev. D",
    volume = "109",
    number = "9",
    pages = "095040",
    year = "2024"
}

@article{deSouza:2025bpl,
    author = "de Souza, Fernando Abreu and Boto, Rafael and Crispim Rom{\~a}o, Miguel and Figueiredo, Pedro N. and Rom{\~a}o, Jorge C. and Silva, Jo{\~a}o P.",
    title = "{Unearthing large pseudoscalar Yukawa couplings with machine learning}",
    eprint = "2505.10625",
    archivePrefix = "arXiv",
    primaryClass = "hep-ph",
    reportNumber = "IPPP/25/28",
    doi = "10.1007/JHEP07(2025)268",
    journal = "JHEP",
    volume = "07",
    pages = "268",
    year = "2025"
}

@article{Boto:2025ovp,
    author = "Boto, Rafael and Matos, Jo{\~a}o A. C. and Rom{\~a}o, Jorge C. and Silva, Jo{\~a}o P.",
    title = "{Surveying the complex three Higgs doublet model with Machine Learning}",
    eprint = "2510.02445",
    archivePrefix = "arXiv",
    primaryClass = "hep-ph",
    doi = "10.1007/JHEP03(2026)182",
    journal = "JHEP",
    volume = "03",
    pages = "182",
    year = "2026"
}

@article{Boto:2026gzj,
    author = {Boto, Rafael and Elyaouti, Karim and Fontes, Duarte and Gon{\c{c}}alves, Maria and M{\"u}hlleitner, Margarete and Rom{\~a}o, Jorge C. and Santos, Rui and Silva, Jo{\~a}o P.},
    title = "{Reassessing CP Violation in the C2HDM with Machine Learning}",
    eprint = "2601.15227",
    archivePrefix = "arXiv",
    primaryClass = "hep-ph",
    reportNumber = "KA-TP-03-2026, P3H-26-005",
    month = "1",
    year = "2026"
}

@article{deSouza:2026sta,
    author = "de Souza, Fernando Abreu and Boto, Rafael and Crispim Rom{\~a}o, Miguel and de Figueiredo, Pedro N. and Rom{\~a}o, Jorge C.",
    title = "{Machine Learning insights on the Z3 3HDM with Dark Matter}",
    eprint = "2603.00254",
    archivePrefix = "arXiv",
    primaryClass = "hep-ph",
    reportNumber = "KA-TP-06-2026, IPPP/26/19, KA-TP-06-2026, IPPP/26/19",
    month = "2",
    year = "2026"
}

@article{CMS:2026nuu,
    author = "Aad, Georges and others",
    collaboration = "CMS, ATLAS",
    title = "{Combination of ATLAS and CMS searches for Higgs boson pair production at $\sqrt{s} = 13$ TeV}",
    eprint = "2602.23991",
    archivePrefix = "arXiv",
    primaryClass = "hep-ex",
    reportNumber = "CERN-EP-2026-011",
    month = "2",
    year = "2026"
}

@article{Dorsch:2017nza,
    author = "Dorsch, G. C. and Huber, S. J. and Mimasu, K. and No, J. M.",
    title = "{The Higgs Vacuum Uplifted: Revisiting the Electroweak Phase Transition with a Second Higgs Doublet}",
    eprint = "1705.09186",
    archivePrefix = "arXiv",
    primaryClass = "hep-ph",
    reportNumber = "CP3-17-15, DESY-17-076, KCL-PH-TH-2017-27",
    doi = "10.1007/JHEP12(2017)086",
    journal = "JHEP",
    volume = "12",
    pages = "086",
    year = "2017"
}

@article{Andersen:2017ika,
    author = "Andersen, Jens O. and Gorda, Tyler and Helset, Andreas and Niemi, Lauri and Tenkanen, Tuomas V. I. and Tranberg, Anders and Vuorinen, Aleksi and Weir, David J.",
    title = "{Nonperturbative Analysis of the Electroweak Phase Transition in the Two Higgs Doublet Model}",
    eprint = "1711.09849",
    archivePrefix = "arXiv",
    primaryClass = "hep-ph",
    reportNumber = "HIP-2017-26/TH, HIP-2017-26-TH",
    doi = "10.1103/PhysRevLett.121.191802",
    journal = "Phys. Rev. Lett.",
    volume = "121",
    number = "19",
    pages = "191802",
    year = "2018"
}

@article{Bernon:2017jgv,
    author = "Bernon, J{\'e}r{\'e}my and Bian, Ligong and Jiang, Yun",
    title = "{A new insight into the phase transition in the early Universe with two Higgs doublets}",
    eprint = "1712.08430",
    archivePrefix = "arXiv",
    primaryClass = "hep-ph",
    doi = "10.1007/JHEP05(2018)151",
    journal = "JHEP",
    volume = "05",
    pages = "151",
    year = "2018"
}

@article{Goncalves:2021egx,
    author = "Gon{\c{c}}alves, Dorival and Kaladharan, Ajay and Wu, Yongcheng",
    title = "{Electroweak phase transition in the 2HDM: Collider and gravitational wave complementarity}",
    eprint = "2108.05356",
    archivePrefix = "arXiv",
    primaryClass = "hep-ph",
    doi = "10.1103/PhysRevD.105.095041",
    journal = "Phys. Rev. D",
    volume = "105",
    number = "9",
    pages = "095041",
    year = "2022"
}

@article{Bhatnagar:2025jhh,
    author = "Bhatnagar, Ansh and Croon, Djuna and Schicho, Philipp",
    title = "{Interpreting the 95 GeV resonance in the Two Higgs Doublet Model. Implications for the electroweak phase transition}",
    eprint = "2506.20716",
    archivePrefix = "arXiv",
    primaryClass = "hep-ph",
    reportNumber = "IPPP/25/39",
    doi = "10.1007/JHEP03(2026)014",
    journal = "JHEP",
    volume = "03",
    pages = "014",
    year = "2026"
}

@article{Haller:2018nnx,
    author = {Haller, Johannes and Hoecker, Andreas and Kogler, Roman and M{\"o}nig, Klaus and Peiffer, Thomas and Stelzer, J{\"o}rg},
    title = "{Update of the global electroweak fit and constraints on two-Higgs-doublet models}",
    eprint = "1803.01853",
    archivePrefix = "arXiv",
    primaryClass = "hep-ph",
    doi = "10.1140/epjc/s10052-018-6131-3",
    journal = "Eur. Phys. J. C",
    volume = "78",
    number = "8",
    pages = "675",
    year = "2018"
}

@article{HFLAV:2016hnz,
    author = "Amhis, Y. and others",
    collaboration = "HFLAV",
    title = "{Averages of $b$-hadron, $c$-hadron, and $\tau$-lepton properties as of summer 2016}",
    eprint = "1612.07233",
    archivePrefix = "arXiv",
    primaryClass = "hep-ex",
    reportNumber = "FERMILAB-PUB-16-611-ND",
    doi = "10.1140/epjc/s10052-017-5058-4",
    journal = "Eur. Phys. J. C",
    volume = "77",
    number = "12",
    pages = "895",
    year = "2017"
}

@article{CMS:2014xfa,
    author = "Khachatryan, Vardan and others",
    collaboration = "CMS, LHCb",
    title = "{Observation of the rare $B^0_s\to\mu^+\mu^-$ decay from the combined analysis of CMS and LHCb data}",
    eprint = "1411.4413",
    archivePrefix = "arXiv",
    primaryClass = "hep-ex",
    reportNumber = "CERN-PH-EP-2014-220, CMS-BPH-13-007, LHCB-PAPER-2014-049",
    doi = "10.1038/nature14474",
    journal = "Nature",
    volume = "522",
    pages = "68--72",
    year = "2015"
}

@article{LHCb:2017rmj,
    author = "Aaij, Roel and others",
    collaboration = "LHCb",
    title = "{Measurement of the $B^0_s\to\mu^+\mu^-$ branching fraction and effective lifetime and search for $B^0\to\mu^+\mu^-$ decays}",
    eprint = "1703.05747",
    archivePrefix = "arXiv",
    primaryClass = "hep-ex",
    reportNumber = "CERN-EP-2017-041, LHCB-PAPER-2017-001",
    doi = "10.1103/PhysRevLett.118.191801",
    journal = "Phys. Rev. Lett.",
    volume = "118",
    number = "19",
    pages = "191801",
    year = "2017"
}

@article{Kajantie:1995kf,
    author = "Kajantie, K. and Laine, M. and Rummukainen, K. and Shaposhnikov, Mikhail E.",
    title = "{The Electroweak phase transition: A Nonperturbative analysis}",
    eprint = "hep-lat/9510020",
    archivePrefix = "arXiv",
    reportNumber = "CERN-TH-95-263, HD-THEP-95-44, HU-TFT-95-57, IUHET-318",
    doi = "10.1016/0550-3213(96)00052-1",
    journal = "Nucl. Phys. B",
    volume = "466",
    pages = "189--258",
    year = "1996"
}

@article{Kajantie:1996qd,
    author = "Kajantie, K. and Laine, M. and Rummukainen, K. and Shaposhnikov, Mikhail E.",
    title = "{A Nonperturbative analysis of the finite T phase transition in SU(2) x U(1) electroweak theory}",
    eprint = "hep-lat/9612006",
    archivePrefix = "arXiv",
    reportNumber = "BI-TP-96-54, CERN-TH-96-334A, HD-THEP-96-48",
    doi = "10.1016/S0550-3213(97)00164-8",
    journal = "Nucl. Phys. B",
    volume = "493",
    pages = "413--438",
    year = "1997"
}

@article{Karsch:1996yh,
    author = "Karsch, F. and Neuhaus, T. and Patkos, A. and Rank, J.",
    editor = "Bernard, C. and Golterman, M. and Ogilvie, M. and Potvin, J.",
    title = "{Critical Higgs mass and temperature dependence of gauge boson masses in the SU(2) gauge Higgs model}",
    eprint = "hep-lat/9608087",
    archivePrefix = "arXiv",
    reportNumber = "FSU-SCRI-96C-79",
    doi = "10.1016/S0920-5632(96)00736-0",
    journal = "Nucl. Phys. B Proc. Suppl.",
    volume = "53",
    pages = "623--625",
    year = "1997"
}

@article{Gurtler:1997hr,
    author = "Gurtler, M. and Ilgenfritz, Ernst-Michael and Schiller, A.",
    title = "{Where the electroweak phase transition ends}",
    eprint = "hep-lat/9704013",
    archivePrefix = "arXiv",
    reportNumber = "UL-NTZ-10-97, HUB-EP-97-24, DESY-97-086",
    doi = "10.1103/PhysRevD.56.3888",
    journal = "Phys. Rev. D",
    volume = "56",
    pages = "3888--3895",
    year = "1997"
}

@article{Rummukainen:1998as,
    author = "Rummukainen, K. and Tsypin, M. and Kajantie, K. and Laine, M. and Shaposhnikov, Mikhail E.",
    title = "{The Universality class of the electroweak theory}",
    eprint = "hep-lat/9805013",
    archivePrefix = "arXiv",
    reportNumber = "CERN-TH-98-08, NORDITA-98-30-HE",
    doi = "10.1016/S0550-3213(98)00494-5",
    journal = "Nucl. Phys. B",
    volume = "532",
    pages = "283--314",
    year = "1998"
}

@article{Aoki:1999fi,
    author = "Aoki, Y. and Csikor, F. and Fodor, Z. and Ukawa, A.",
    title = "{The Endpoint of the first order phase transition of the SU(2) gauge Higgs model on a four-dimensional isotropic lattice}",
    eprint = "hep-lat/9901021",
    archivePrefix = "arXiv",
    reportNumber = "ITP-BUDAPEST-547, UTCCP-P-60, UTHEP-397",
    doi = "10.1103/PhysRevD.60.013001",
    journal = "Phys. Rev. D",
    volume = "60",
    pages = "013001",
    year = "1999"
}

@article{DOnofrio:2015gop,
    author = "D'Onofrio, Michela and Rummukainen, Kari",
    title = "{Standard model cross-over on the lattice}",
    eprint = "1508.07161",
    archivePrefix = "arXiv",
    primaryClass = "hep-ph",
    reportNumber = "HIP-2015-30-TH",
    doi = "10.1103/PhysRevD.93.025003",
    journal = "Phys. Rev. D",
    volume = "93",
    number = "2",
    pages = "025003",
    year = "2016"
}

@article{Brdar:2025hxw,
    author = "Brdar, Vedran and Finetti, Marco and Matteini, Marco and Morais, Ant{\'o}nio P. and Nemev{\v{s}}ek, Miha",
    title = "{$\text{PT2GWFinder}$ : A package for cosmological first-order phase transitions and gravitational waves}",
    eprint = "2505.04744",
    archivePrefix = "arXiv",
    primaryClass = "hep-ph",
    doi = "10.1016/j.cpc.2026.110119",
    journal = "Comput. Phys. Commun.",
    volume = "323",
    pages = "110119",
    year = "2026"
}

@article{Costa:2025pew,
    author = "Costa, Francesco and Hoefken Zink, Jaime and Lucente, Michele and Pascoli, Silvia and Rosauro-Alcaraz, Salvador",
    title = "{ELENA: a software for fast and precise computation of first order phase transitions and gravitational waves production in particle physics models}",
    eprint = "2510.00289",
    archivePrefix = "arXiv",
    primaryClass = "hep-ph",
    doi = "10.1140/epjc/s10052-026-16003-5",
    journal = "Eur. Phys. J. C",
    volume = "86",
    number = "7",
    pages = "851",
    year = "2026"
}

@article{Ekstedt:2023sqc,
    author = "Ekstedt, Andreas and Gould, Oliver and Hirvonen, Joonas",
    title = "{BubbleDet: a Python package to compute functional determinants for bubble nucleation}",
    eprint = "2308.15652",
    archivePrefix = "arXiv",
    primaryClass = "hep-ph",
    doi = "10.1007/JHEP12(2023)056",
    journal = "JHEP",
    volume = "12",
    pages = "056",
    year = "2023"
}

@article{Muhlleitner:2026uqd,
    author = {M{\"u}hlleitner, Margarete and Plotnikov, Johann and Santos, Rui and Viana, Jo{\~a}o},
    title = "{A Deep Dive into Baryon Asymmetry -- the C2HDM}",
    eprint = "2606.04229",
    archivePrefix = "arXiv",
    primaryClass = "hep-ph",
    reportNumber = "KA-TP-11-2026",
    month = "6",
    year = "2026"
}

@article{Balui:2025kat,
    author = "Balui, Debanjan and Chakrabortty, Joydeep and Dey, Debmalya and Mohanty, Subhendra",
    title = "{Gauge invariant effective potential}",
    eprint = "2502.17156",
    archivePrefix = "arXiv",
    primaryClass = "hep-th",
    doi = "10.1103/PhysRevD.111.085032",
    journal = "Phys. Rev. D",
    volume = "111",
    number = "8",
    pages = "085032",
    year = "2025"
}

@article{Balui:2025yvd,
    author = "Balui, Debanjan and Biswas, Tisa and Chakrabortty, Joydeep and Dey, Debmalya and Englert, Christoph and Mohanty, Subhendra",
    title = "{Gauge choices, infrared pitfalls, and thermal effects in effective potentials}",
    eprint = "2507.22706",
    archivePrefix = "arXiv",
    primaryClass = "hep-th",
    doi = "10.1103/drsd-wfns",
    journal = "Phys. Rev. D",
    volume = "112",
    number = "5",
    pages = "056022",
    year = "2025"
}

@article{Balui:2026ghs,
    author = "Balui, Debanjan and Chakrabortty, Joydeep and Englert, Christoph and Mohanty, Subhendra and Tushar",
    title = "{Background fields meet the heat kernel: Gauge invariance and RGEs without diagrams}",
    eprint = "2604.05972",
    archivePrefix = "arXiv",
    primaryClass = "hep-th",
    doi = "10.1103/mcpq-jckb",
    journal = "Phys. Rev. D",
    volume = "113",
    number = "11",
    pages = "116040",
    year = "2026"
}

@article{Caprini:2019pxz,
    author = "Caprini, Chiara and Figueroa, Daniel G. and Flauger, Raphael and Nardini, Germano and Peloso, Marco and Pieroni, Mauro and Ricciardone, Angelo and Tasinato, Gianmassimo",
    title = "{Reconstructing the spectral shape of a stochastic gravitational wave background with LISA}",
    eprint = "1906.09244",
    archivePrefix = "arXiv",
    primaryClass = "astro-ph.CO",
    reportNumber = "LISA-CosWG-19-02",
    doi = "10.1088/1475-7516/2019/11/017",
    journal = "JCAP",
    volume = "11",
    pages = "017",
    year = "2019"
}

\end{document}